\documentclass[11pt]{article}

\usepackage{graphicx}
\usepackage{epsfig}
\usepackage{bbm}
\usepackage{amsmath, amssymb}
\usepackage{cite}
\usepackage{subfig}
\usepackage{float}
\usepackage{braket}
\usepackage{resizegather}
\usepackage{multirow}
\usepackage{tikz}
\usepackage{longtable}

\usepackage[margin=2.5cm]{geometry}
\numberwithin{equation}{section}
\usepackage{hyperref}

\usepackage{color}

\newcommand{\IP}{\mathbb{P}}

\newcommand{\IZ}{\mathbb{Z}}
\newcommand{\cO}{{\cal O}}
\newcommand{\cN}{{\cal N}}

\newcommand{\cA}{{\cal A}}
\newcommand{\cB}{{\cal B}}
\newcommand{\cC}{{\cal C}}

\newcommand{\cK}{{\cal K}}
\newcommand{\cV}{{\cal V}}

\def\cjn1{{\cA, \cC^*\otimes \wedge^j \cN^*}}
\def\bjn1{{\cA, \cB^*\otimes \wedge^j \cN^*}}
\def\vjn1{{\cA, \cV^*\otimes \wedge^j \cN^*}}
\def\cjn2{{\cA, \cC\otimes \wedge^j \cN^*}}
\def\bjn2{{\cA, \cB\otimes \wedge^j \cN^*}}
\def\vjn2{{\cA, \cV\otimes \wedge^j \cN^*}}

\newcommand{\varstr}[2]{\vrule height #1 depth #2 width0pt}

\newcommand{\cicy}[2]{\begin{matrix} #1\end{matrix}\!\left[\begin{matrix}#2 \end{matrix}\right]}

\begin{document}
\begin{flushright}
$ $\\
\end{flushright}
\vspace{8mm}
\begin{center}
{
\Large {\bf Heterotic String Model Building \\ with Monad Bundles and Reinforcement Learning}\\[12pt]
\vspace{1cm}
\normalsize
{\bf{Andrei Constantin}\footnote{andrei.constantin@physics.ox.ac.uk}},  
{\bf{Thomas R.~Harvey}\footnote{thomas.harvey@physics.ox.ac.uk}},
{\bf{Andre Lukas}\footnote{andre.lukas@physics.ox.ac.uk}},
\bigskip}\\[0pt]
\vspace{0.23cm}
{\it 
Rudolf Peierls Centre for Theoretical Physics, University of Oxford\\
Parks Road, Oxford OX1 3PU, UK
}\\[2ex]
\end{center}
\vspace{0.5cm}

\begin{abstract}\noindent
We use reinforcement learning as a means of constructing string compactifications with prescribed properties. Specifically, we study heterotic $SO(10)$ GUT models on Calabi-Yau three-folds with monad bundles, in search of phenomenologically promising examples. Due to the vast number of bundles and the sparseness of viable choices, methods based on systematic scanning are not suitable for this class of models. By focusing on two specific manifolds with Picard numbers two and three, we show that reinforcement learning can be used successfully to explore monad bundles. Training can be accomplished with minimal computing resources and leads to highly efficient policy networks. They produce phenomenologically promising states for nearly 100\% of episodes and within a small number of steps. In this way, hundreds of new candidate standard models are found. 
\end{abstract}

\setcounter{footnote}{0}
\setcounter{tocdepth}{2}
\clearpage
\tableofcontents

\newpage
\section{Introduction}\label{secIntro}

The earliest, and one of the most promising, proposals for connecting string theory to low-energy physics has been the $E_8 \times E_8$ heterotic string setup compactified to four-dimensions on smooth Calabi-Yau (CY) three-folds~\cite{PhysRevLett.54.502, Candelas:1985en}. Over the last couple of decades this line of research has moved forward from the position of constructing models `by hand' to increasingly more systematic automated searches, facilitated by the gradual assimilation of new mathematical techniques into the analysis of string models, as well as by substantial advancements in computational power. In the early days of string theory it seemed plausible to believe that the number of solutions resembling our world at the crudest level of analysis is small enough to allow the identification of the correct string vacuum `by eye'. Today we know that there may be up to $10^{723}$ heterotic string compactifications that lead to consistent models with the correct low-energy gauge group and particle content \cite{Constantin:2018xkj}. This huge number of `good solutions' is not a problem in itself: as more phenomenological constraints are being imposed, the large exponent is bound to fall dramatically. However, the real challenge is how to access this wealth of models within the even larger space of consistent string theory compactifications. 

The initial heterotic model building efforts focused on the standard embedding of the background gauge connection into the spin connection of the Calabi-Yau three-fold. These efforts produced a handful of $(2,2)$ supersymmetric models leading to three-generations and an $E_6$ GUT group in the four-dimensional effective theory \cite{Greene:1986bm, Greene:1986jb, Braun:2009qy, Braun:2011ni}.
The low number of `good solutions' within this class is no surprise: one typically looks for Calabi-Yau three-folds with a small Euler number (in absolute value) that moreover admit freely acting discrete symmetries needed in order to break the GUT group down to the observable $SU(3)\times SU(2)\times U(1)$. Given that there are only a few hundred known Calabi-Yau three-folds admitting such symmetries \cite{Candelas:2008wb, Braun:2010vc, Candelas:2010ve, Braun:2017juz, Candelas:2015amz, Candelas:2016fdy, Larfors:2020weh}, the real surprise is that any three-generation models at all could be found in this way. 

The realisation that $(0,2)$ Calabi-Yau models provide true solutions of the heterotic string opened up a much wider class of compactifications in which one could also obtain $SO(10)$ and $SU(5)$ as effective gauge groups \cite{Distler:1986wm, Distler:1987ee}. Such compactifications are specified by triplets $(X,V,\tilde V)$ where $X$ is a Calabi-Yau three-fold and $V$, $\tilde V$ are holomorphic stable bundles whose connections define the $E_8\times E_8$ gauge background. While the number of available choices for $X$ remains relatively limited, the number of possibilities for $V$ and $\tilde{V}$ is virtually unbounded. Various constructions of holomorphic stable bundles have been used over the years, including the spectral cover construction over elliptically fibered Calabi-Yau three-folds \cite{Friedman:1997yq, Friedman:1997ih, Donagi:1998xe, Andreas:1999ty, Donagi:1999gc, Donagi:1999ez, Donagi:2000zf, Donagi:2000zs, Braun:2005ux, Braun:2005bw, Braun:2005nv, Blumenhagen:2006ux, Blumenhagen:2006wj, Gabella:2008id, Anderson:2019agu}, monad bundles \cite{Distler:1987ee, Kachru:1995em, Anderson:2008ex, Anderson:2008uw, Anderson:2009mh, He:2009wi}, extension bundles \cite{Bouchard:2005ag, Blumenhagen:2006ux, Blumenhagen:2006wj}, as well as direct sums of line bundles \cite{Blumenhagen:2005ga, Blumenhagen:2006ux, Blumenhagen:2006wj, Anderson:2011ns, Anderson:2012yf, Anderson:2013xka, He:2013ofa, Buchbinder:2013dna, Buchbinder:2014qda, Buchbinder:2014sya, Anderson:2014hia, Larfors:2020weh}, the latter construction leading to the largest dataset of string $SU(5)$ GUT models available to date. Each of these constructions has its own virtues: bundles obtained through the spectral cover construction can be directly used in the study of heterotic/F-theory duality, while monad and extension sequences provide an accessible construction of non-abelian bundles. 
The main virtue of line bundle sums resides in their `split' nature: many of the consistency and phenomenological constraints can be imposed line bundle by line bundle, making this class searchable by systematic methods. In this manner, in Ref.~\cite{Anderson:2013xka} an exhaustive search for $SU(5)$ GUT models has been accomplished for Calabi-Yau three-folds with non-trivial fundamental group and a Picard number smaller than $7$, the search being extended in Ref.~\cite{Constantin:2018xkj} to manifolds of Picard number equal to $7$. While the space of line bundle sums of a fixed rank over a given manifold is unbounded, it was noticed that phenomenologically viable models  correspond to line bundle sums where all entires are relatively small integers, an observation which effectively renders the search space finite, though typically very large. To provide concrete numbers, the search space involved in Ref.~\cite{Anderson:2013xka} was of order $10^{40}$ bundles, leading to an order of $10^5$ three-generation models with $SU(5)$ GUT group.

In the present work we focus on a class of bundles known to be a fruitful source of phenomenologically attractive models for which, however, systematic scanning methods cannot be used in any comprehensive way: monad bundles. The problem is that for monads there are no clever tricks which allow decomposing the geometry into several pieces and applying the constraints individually to each piece, as in the case of line bundle sums. On the other hand, the size of the search space gets prohibitively large even for manifolds of relatively small Picard number. To see this, recall that a monad bundle $V$ on a complex manifold $X$ can be constructed from two sums of holomorphic line bundles $B$ and $C$, with ranks $r_B={\rm rk}(B)$ and $r_C={\rm rk}(C)$, respectively, via the short exact sequence
	\begin{equation*}\label{eqnMonadSeqenceIntro}
		0 \rightarrow V \rightarrow B \stackrel{f}{\rightarrow} C \rightarrow 0, 
	\end{equation*}
and assume that the bundle morphism $f$ is sufficiently generic. Each line bundle in $B$ and~$C$ is specified by its first Chern class, which in a basis of $H^2(X)$ corresponds to an integer list of length equal to $h=h^{1,1}(X)$, the Picard number of $X$. This means that $V$ is specified by $h(r_B+r_C)$ integers. In fact, for reasons explained below, we will be looking for bundles satisfying $c_1(V)=0$, a constraint that cuts down the number of integer parameters to $h(r_B+r_C-1-1)$. If we let these integers to run between $-4$ and $5$, which turns out to be the range where most of the good models lie, the size of the search space is of order
\begin{equation}\label{eq:SizeMonadSpace}
10^{h(r_B+r_C-1)}\; .
\end{equation}
Now ${\rm rk}(V) =r_B-r_C$ and for $SO(10)$ models ${\rm rk}(V)=4$, while for $SU(5)$ models we need ${\rm rk}(V)=5$. Since $r_C \geq 1$, this implies that the exponent is at least $5h$. The computational time required to perform even the most basic checks for a monad bundle being of order of a few mili-seconds on a standard machine, this implies that for any manifold with $h> 2$ a systematic and comprehensive search is not possible (or just about possible in the case $h=2$). 

Supervised learning methods are not a viable alternative in this problem as they run essentially into the same issues. Supervised learning could be applied by first obtaining a large enough set of good models (for example through a random search) which is then passed as training data to a suitably designed neural network. Once trained, the neural network may be quicker than a standard check-list validation in classifying unseen models as viable or not. In fact the authors of Ref.~\cite{Deen:2020dlf} have shown that a fully connected feed-forward network can be successfully designed in this way to deal  with models constructed on line bundle sums. The same was shown to hold true for an auto-encoder unsupervised learning architecture, which could be used to reliably identify the `standard model' property. 
However, the core issue of how to deal with the exponentially large size of the space of unseen models remains as problematic as in the case of systematic searches. 

A promising alternative\footnote{Other alternatives may include genetic algorithms and Markov chain Monte Carlo methods such as Metropolis, which turned out to be successful in the search for type IIB flux vacua undertaken in Ref.~\cite{Krippendorf:2021uxu}.} for approaching this problem seems to be reinforcement learning (RL), a method that has proved to be spectacularly efficient in solving qualitatively similar problems within unfathomably large spaces, as demonstrated by the AlphaZero Go-player \cite{AlphaGoZero}. RL has been successfully implemented to generate type IIA intersecting brane configurations that lead to standard-like models in Ref.~\cite{Halverson:2019tkf} and in the exploration of the landscape of type IIB flux vacua in \cite{Krippendorf:2021uxu}. 
Moreover, in Ref.~\cite{Larfors:2020ugo} the line bundle sum results of Ref.~\cite{Anderson:2013xka} could be recovered using RL, the important lesson being that, while maintaining comparable levels of comprehensiveness, RL-based searches perform substantially better on large Picard number manifolds as compared to standard systematic scans. Beyond string model building, RL has been shown in Ref.~\cite{Harvey:2021oue} to be a successful method for generating realistic Froggatt-Nielsen models that could explain the observed quark mass hierarchy. It has also been used in the study of knot theory for finding sequences of unknotting actions \cite{Gukov:2020qaj}.

We will discuss RL in more detail below; here we only sketch the main ideas. RL involves an artificial intelligence agent exploring the space of possible states of a given system (the space of potential solutions for a given problem), called {\itshape environment}. Each state is associated with a numerical value reflecting how well it fits the properties sought from target solutions. Often the {\itshape value function} is defined to be semi-negative such that the target states correspond to the zeros of this function. 

The agent is provided with a cleverly constructed set of penalties and rewards which guide its progression (set of {\itshape actions}) towards states that resemble more and more the features sought from the desired states. It self-trains without any prior knowledge of the environment, a feature that distinguishes RL from supervised learning. The scan is divided into multiple {\itshape episodes} involving a fixed number of maximal states. Typically, the initial state of an episode is randomly chosen and the episode ends either when a target state is found or when the maximal episode length is reached. The progression of states within an episode is dictated by the current {\itshape policy}, which is encoded in a neural network. This network is updated iteratively with data obtained from the agent's exploration of the environment. Training is usually stopped once the loss becomes small enough and it can be considered successful if desirable states are reached quickly from virtually any starting point. An estimate of what should be considered a ``quick" approach can be obtained as follows. If the space of solutions is a $d$-dimensional lattice hypercube with $l$ sites in each direction, the length of the diagonal is
\begin{equation}\label{eq:straightpath}
 \mbox{longest straight path }\sim d^{1/2}l~.
\end{equation}
Episodes reaching desirable states after a number of steps comparable to (or smaller than) this length should be considered efficient. To encourage short episodes, it is customary to introduce a penalty on the episode length, giving the agent an incentive to find terminal states that are as close as possible to the original random starting point. On the other hand, if the search space contains sizeable `gaps' with no target states, the episode length should be large enough so that the agent can move out of these regions within an episode. Often the distribution of target states in the search space is not known, which makes the episode length an important hyper-parameter that needs adjustment.
\begin{figure}[ht]
\centering
\includegraphics[width=0.24\textwidth]{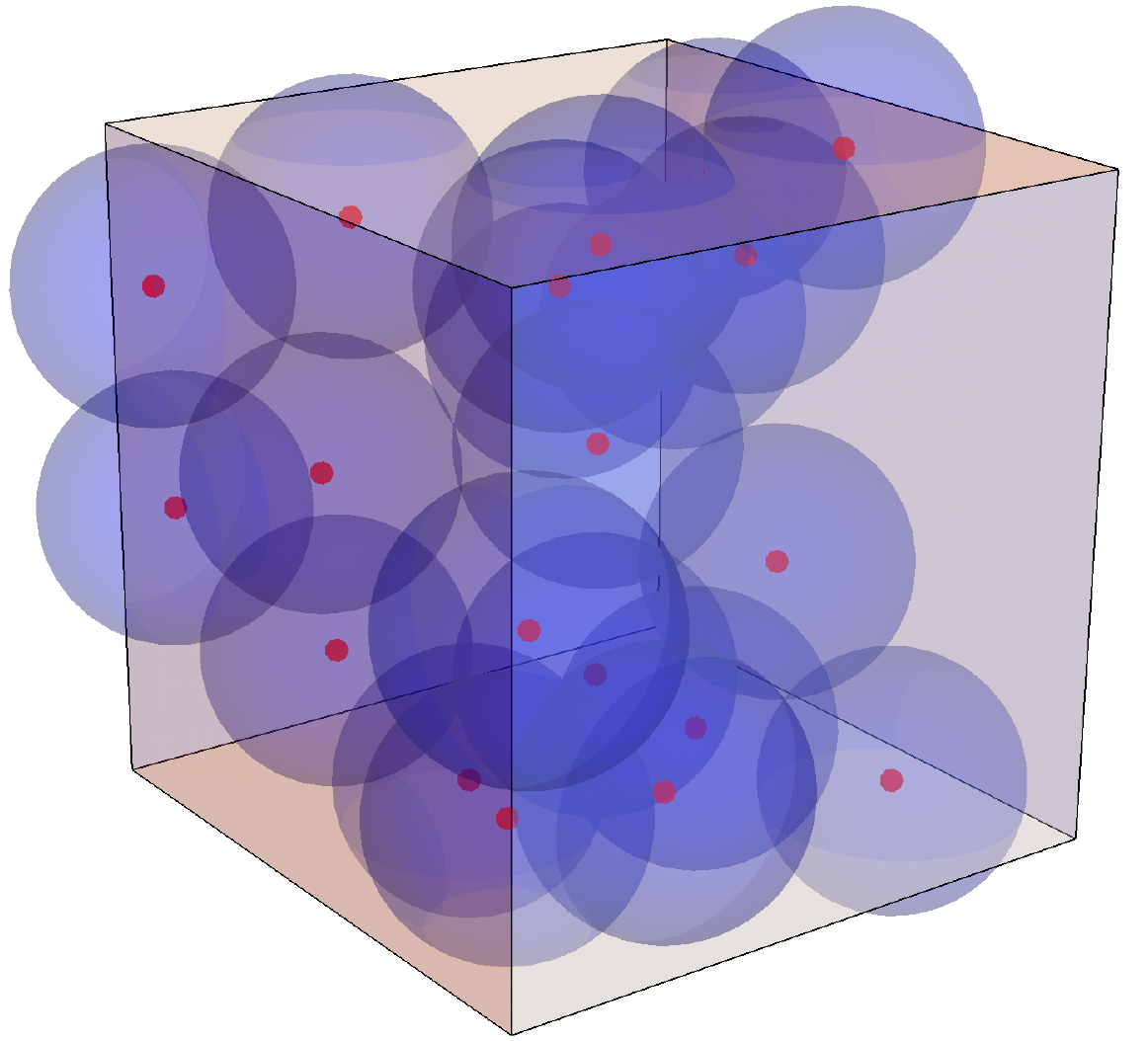}
\caption{\sf An idealised picture of the search space and target states.}
\label{Fig:SearchSpace}
\end{figure}

Ideally, after sufficiently many training episodes, the AI agent `knows' enough about the landscape (the value function) to (1) reach a terminal state for virtually any initial random point and (2) reach any specific target state within an episode provided that the initial random point is close enough. For the purpose of illustration, Figure~\ref{Fig:SearchSpace} shows a situation where the target states (the red points) are uniformly distributed. If the initial random point of an episode falls within a blue ball (basin of attraction), the corresponding target state will be found within that episode. The basins of attraction cover the entire space, which means that the number of episodes needed to obtain all the target states is, in principle, comparable to the total number of target states available in the search space. Of course, in practice the basins of attraction are not spheres; they roughly correspond to level hypersurfaces of the value function, though their exact shape depends very much on the training history. However, the main idea remains essentially the same: the target states get `thickened', acting as attractor points within the corresponding attractor basins.  

For string theory model building this can be a real game changer. We know from the study of line bundle sums that the number of topologically distinct bundles that lead to $SU(5)$ models with the correct particle spectrum is, roughly, $10^h$ on a CY three-fold $X$ with Picard number $h=h^{1,1}(X)$. Of course, this may not generalise to other constructions, but for the sake of the argument we take it as an estimate figure for the size of the solution space. In the light of the above discussion the computational time needed in an RL search will (roughly) scale in the same way, that is the number $10^h$ of target states. This represents a substantial reduction in time from what is required in any systematic scan, as given by the estimate \eqref{eq:SizeMonadSpace}. Moreover, one can envisage including more phenomenological constraints in the scan which are bound to reduce the size of the solution space. While in any systematic search including more checks necessarily leads to an increase in the computational time, the expectation is that for RL searches this may actually lead to a further substantial decrease in time, depending on how much the size of the solution space gets reduced. 

The rest of the paper is structured as follows. In Section \ref{secHeterotic} we review the main features of $E_8\times E_8$ heterotic model building with monad bundles. In Section \ref{secExample} we present a new heterotic standard model based on a monad bundle that was found using RL, including the details of the various checks that have been performed. A reader who is less interested in the mathematical and string theoretic background, but would rather like to focus on the machine learning aspects can skip straight to Section \ref{secRL} where we review RL and apply it to a simple proof-of-concept search for line bundles with a fixed Euler characteristic. In Section \ref{secMonadRL} we apply RL to the problem of searching for heterotic standard models on CY three-folds with low Picard number carrying monad bundles. 

\section{Heterotic model building with monad bundles}\label{secHeterotic}
The low-energy field theory limit of the $E_8\times E_8$ heterotic string is $10$-dimensional $\mathcal N =1$ supergravity, coupled to $E_8\times E_8$ super-Yang-Mills theory. The background geometry is taken to be a direct product $\mathcal M _{10} = \mathbb R^{1,3} \times X$, where $X$ is a compact six-dimensional manifold. Throughout this paper $X$ will be a Calabi-Yau three-fold, which facilitates $\mathcal N =1$ supersymmetry in four dimensions \cite{Candelas:1985en}. The gauge background is specified by the connections on two poly-stable holomorphic $E_8$ vector bundles $V$ and $\tilde V$ on $X$. 

\subsection{Heterotic bundles: general constraints}
In the following we provide a brief summary of the constraints on the triple $(X,V,\tilde{V})$ which defines a compactification, both from consistency and from phenomenological requirements.\\[2mm]
\noindent{\bfseries Vanishing first Chern class.}
The gauge group in the four-dimensional theory is given by the commutant of the structure group of $V$ in $E_8$. In order to obtain one of the standard GUT groups, $E_6$, $SO(10)$ and $SU(5)$, $V$ must be an $SU(n)$-bundle with $n=3,4,5$, respectively. This implies $c_1(V)=0$. The bundle $\tilde V$ gives rise to hidden matter which only interacts gravitationally with the observable sector associated to $V$. For our purposes, we do not need to construct $\tilde{V}$ explicitly.\\[2mm]
\noindent{\bfseries Anomaly cancellation condition.} The bundles and the manifold are related by a topological constraint that guarantees the vanishing of gravitational and gauge anomalies 
\begin{equation}\label{eqAnomalyCancelationm}
  c_2(TX) - c_2(V) - c_2(\tilde V) = [\mathcal C],
\end{equation}
where $[\mathcal C] \in H^4 (X)$ is the class of an effective holomorphic curve wrapped by five-branes which span across the four un-compactified dimensions.  Of course, the term $c_2(\tilde V)$ can be dropped for trivial hidden bundles. In practice, we will only explicitly construct the Calabi-Yau three-fold $X$ and the bundle $V$ and demand that
\begin{equation}
   c_2(TX) - c_2(V) \in\mbox{ Mori cone of }X\; .
\end{equation}
This condition guarantees that there is an anomaly-free and supersymmetric completion of the model by including five branes wrapping a holomorphic curve $\mathcal{C}$ with class $[\mathcal{C}]=c_2(TX) - c_2(V)$.\\[2mm]
\noindent{\bfseries Bundle Stability.} In order to preserve $\mathcal N =1$ supersymmetry in four dimensions, both bundles need to be  slope poly-stable and have vanishing slopes. Checking bundle stability is hard, in general. We review the main strategy below; for a more detailed exposition see Ref.~\cite{Anderson:2008ex}. The slope of a coherent sheaf $\mathcal{F}$ on a K\"ahler manifold $X$ with K\"ahler form~$J$ is defined as 
\begin{equation}
\mu_J(\mathcal{F}) = \frac{1}{{\rm rk}(\mathcal{F})}\int_{X}  c_1(\mathcal{F})\wedge J^{{\rm dim}(X)-1}~.
\end{equation}
By definition, $V$ is slope-stable iff for any sub-sheaf $\mathcal{F}$ with $0<{\rm rk}(\mathcal{F})<{\rm rk}(V)$ the inequality $\mu_J(\mathcal{F})<\mu_J(V)$ holds. In our case $c_1(V)=0$ so that $\mu_J(V)=0$ for any K\"ahler form $J$. Hence, $V$ is stable if $\mu_J(\mathcal{F})<0$ for all sub-sheaves $\mathcal{F}\subset V$ with $0<{\rm rk}(\mathcal{F})<{\rm rk}(V)$.

It is not easy to consider all sub-sheaves $\mathcal{F}\subset V$. Fortunately, the problem can be simplified by studying instead all line sub-bundles $L=\Lambda^r\mathcal{F}\subset\Lambda^r V$, where $r={\rm rk}(\mathcal{F})$, of the anti-symmetric powers of $V$. In practice, this can be done by computing cohomologies. A line bundle $L$ injects into $\Lambda^r V$ iff $h^0(X,L^*\otimes\Lambda^rV)\neq 0$ and in this way we can find the set $\mathcal{I}$ of potentially destabilising line bundles. (In practice, it is usually difficult to check all line bundles and only a finite subset can be tested for membership in $\mathcal{I}$.) The question is then whether the subset
\begin{equation}\label{stabreg}
 \{J\in\mathcal{K}(X)\,|\, \mu_J(L)<0\;\forall L\in\mathcal{I}\}
\end{equation}
of the K\"ahler cone $\mathcal{K}(X)$ where all these injecting line bundle have negative slope is non-empty. If it is, then there is a locus in K\"ahler moduli space where the compactification is supersymmetric and the model is accepted. Otherwise, it is rejected.

The trivial bundle $\mathcal{O}_X$ has a vanishing slope for all K\"ahler forms. If it injects into any of the wedge powers $\Lambda^rV$ it will destabilise $V$ in the entire K\"ahler cone. Therefore, a necessary condition for stability (also known as Hoppe's criterion) is
\begin{equation}\label{hoppe}
h^0(X,\wedge^r V) = 0\;,\quad \text{for all } r = 1, 2,\ldots, {\rm rk}(V)-1.
\end{equation}
{\bfseries Low-energy gauge symmetry.}
In this paper we focus on $SO(10)$ models, for which we require $SU(4)$-bundles. The standard method to break the GUT symmetry to the standard model group (times an additional $U(1)_{B-L}$ factor) is to quotient the CY manifold $X$ by a freely-acting discrete symmetry $\Gamma$ and consider the non-simply connected quotient CY $\hat{X}=X/\Gamma$. On $\hat{X}$ it is then possible to include a $\Gamma$ Wilson-line in the direction of weak hypercharge to accomplish the GUT breaking. Of course the bundle $V\rightarrow X$ has to descend to a bundle $\hat{V}\rightarrow\hat{X}$ for this construction to make sense and this is equivalent to saying that $V$ has a $\Gamma$-equivariant structure. For specific constructions (see below), the existence of such a structure may well impose additional constraints on $V$ which have to be checked. There are also group-theoretical constraints on the discrete symmetry which allow for a GUT breaking to the standard model group. In was shown in Ref.~\cite{Braun:2004xv} that for $SO(10)$ breaking the discrete symmetry has to be at least as large as  $\mathbb Z_3\times \mathbb Z_3$. Not many CY three-folds $X$ with such large freely-acting symmetry groups are known, so the choice of manifold is quite restrictive for models based on $SO(10)$ GUTs.\\[2mm] 
\noindent{\bfseries Low-energy spectrum.}
Finally, the bundle $V$ needs to be compatible with a number of spectrum requirements. The four-dimensional particle spectrum can be found by looking at the breaking pattern of the adjoint $\bf{248}$ representation of $E_8$ under $E_8\rightarrow SU(4)\times SO(10)$,
\begin{equation}
\begin{array}{ccccccccccc}
\textbf{248}_{E_8} ~\rightarrow &\big{[}\, (\textbf{1},\textbf{45}) &\oplus&(\textbf{4},\textbf{16})& \oplus\!\!\!\!\!&(\overline{\textbf{4}},\overline{\textbf{16}}) &\!\!\!\!\!\oplus&(\textbf{6},\textbf{10}) &\oplus&(\textbf{15},\textbf{1})&\!\!\!\!\!\big{]}_{SU(4)\times SO(10)}\\[2pt]
& \text{gauge}&&\text{families}&&\text{anti-families}&&\text{Higgs}&& \text{bundle}&\\[-2pt]
& \text{bosons}&& && &&&& \text{moduli}&
\end{array}
\end{equation}
The relevant numbers of $SO(10)$-multiplets are computed as dimensions of certain cohomologies:
\begin{equation}
		\label{eqParticleSpectrum}
		\begin{aligned}
			n_{\mathbf{16}} &= h^1(X,V)~,\qquad
			~~~n_{\overline{\mathbf{16}}} =h^1(X,V^\star) = h^2(X,V), \\
			n_{\mathbf{10}} &= h^1(X,\wedge^2 V)~, \qquad
			n_{\mathbf{1}} = h^1(X,V \otimes V^\star).
		\end{aligned}
\end{equation}
The net number of chiral families is then given by the negative of the Euler characteristic
\begin{equation}
	\label{eqIndex}
	n_{\mathbf{16}}- n_{\overline{\mathbf{16}}} = -\mbox{ind}(V) = h^1(X,V) - h^2(X,V) \stackrel{!}{=} 3|\Gamma|\; .
\end{equation}
To match the required three families after taking the quotient by $\Gamma$ it must equal $3|\Gamma|$ in the ``upstairs" GUT model. Models with anti-families are unattractive (although not necessarily unacceptable) and we require $n_{\overline{\mathbf{16}}}=0$. Of course, the number of Higgs multiplets should be at least one. However, we do not require $n_{\mathbf{10}}\geq 1$, as past experience indicates this condition may only be satisfied for specific sub-loci of the complex structure moduli space. 

\subsection{CICY three-folds.} The class of complete intersection Calabi-Yau (CICY) three-folds $X$ contains many of the known examples with large discrete symmetry groups and is, therefore, a good starting point for the construction of models based on $SO(10)$ GUTs. CICYs are defined as the common zero locus of several multi-homogeneous polynomials in the coordinates of a product space  ${\cal A}=\IP^{n_1}\times\dots\times\IP^{n_m}$. The multi-degrees of the defining polynomials can be recorded as the columns of a matrix, known as the configuration matrix, of the form
\begin{equation*}
\label{conf}
\cicy{\IP^{n_1}\\[4pt] \vdots\\[4pt] \IP^{n_m}}{q^1_1&\cdots&q^1_R\\[4pt] \vdots&\cdots&\vdots\\[4pt]q^m_1&\hdots&q^m_R}^{h^{1,1}(X),~h^{2,1}(X)}
\end{equation*}
where $h^{1,1}(X)$ and $h^{2,1}(X)$ are the two non-trivial Hodge numbers of $X$, independent of the coefficients in the defining equations (assuming the complete intersection is smooth). The Calabi-Yau condition $c_1(TX)=0$ corresponds to the condition that the sum of the degrees in each row of the configuration matrix equals the dimension of the corresponding projective space plus one. The embedding of a CICY three-fold is called {\itshape favourable} if $H^2(X)$ descends from $H^2(\cA)$, that is all holomorphic line bundles on $X$ can be obtained from holomorphic line bundles on $\cA$ by restriction. Since the construction of monad bundles involves line bundles, the notion of favourability is relevant here. Favourable embeddings are particularly useful as many of the algebraic computations on $X$ can be traced back to computations on the ambient space.

All CICY three-folds are simply connected. A number of these admit, for certain choices of coefficients in the defining equations, freely acting discrete symmetries that are large enough to be useful for $SO(10)$ symmetry breaking. These can be identified from the classification of Ref.~\cite{Braun:2010vc} and, restricting to favourable CICY and Abelian groups $\Gamma$, they are listed in Table~\ref{tab:CYs}.
\begin{table}
\begin{center}
\begin{tabular}{|c||l|l|}\hline
$\Gamma$&CICY configurations&\\\hline\hline
\varstr{24pt}{8pt}$\IZ_3\times\IZ_3$&$X_{7878} = \cicy{\IP^{5}}{3&3}^{1,73}$&$X_{7808}=\cicy{\IP^{2}\\[4pt] \IP^{5}}{0&1&1&1\\[4pt] 3&1&1&1}^{2,56}$\\
\varstr{34pt}{8pt}&$X_{7669}= \cicy{\IP^{2}\\[4pt] \IP^{2}\\[4pt] \IP^{2}}{1&1&1\\[4pt] 1&1&1\\[4pt] 1&1&1}^{3,48}$&$X_{7240}=\cicy{\IP^{2}\\[4pt] \IP^{2}\\[4pt] \IP^{5}}{1&1&1&0&0&0\\[4pt] 0&0&0&1&1&1\\[4pt] 1&1&1&1&1&1}^{3,39}$\\
\varstr{28pt}{17pt}&$X_{7884} = \cicy{\IP^{2}\\[4pt] \IP^{2}}{3\\[4pt] 3}^{2,83}$&\\\hline
\varstr{41pt}{35pt}$\IZ_4\times\IZ_4$&$X_{7861}= \cicy{\IP^{7}}{2&2&2&2}^{1,65}$&$X_{7862}= \cicy{\IP^{1}\\[4pt] \IP^{1}\\[4pt]\IP^{1}\\[4pt] \IP^{1}}{2\\[4pt] 2\\[4pt] 2\\[4pt] 2}^{4,68}$\\\hline
\varstr{14pt}{8pt}$\IZ_5\times\IZ_5$&$X_{7890}=\cicy{\IP^{4}}{5}^{1,101}$&\\\hline
\varstr{14pt}{8pt}$\IZ_8\times \IZ_4$&$X_{7861}= \cicy{\IP^{7}}{2&2&2&2}^{1,65}$&\\
\varstr{0pt}{8pt}$\IZ_4\times\IZ_4\times \IZ_2$&&\\\hline
\end{tabular}
\caption{\sf Favourable CICYs with sufficiently large symmetry groups for $SO(10)$ symmetry breaking.}\label{tab:CYs}
\end{center}
\end{table}

\subsection{Line bundle cohomology formulae}\label{sec:lbcohformulae}
Computing bundle cohomology is the key to determining the spectrum of elementary particles as well as to checking bundle stability. Usually the methods for computing cohomology are algorithmic and involve patching together local data to infer global features. However, it has recently been noticed that for many classes of manifolds of interest in string theory, line bundle cohomology dimensions are described by simple formulae, involving a decomposition of the Picard group into disjoint regions, in each of which the cohomology function is polynomial or very close to polynomial \cite{Constantin:2018otr, Buchbinder:2013dna, Constantin:2018hvl, Larfors:2019sie, Brodie:2019pnz,Klaewer:2018sfl, Brodie:2019dfx, Brodie:2019ozt, Brodie:2020wkd}. This pattern has been observed for the zeroth as well as all higher cohomologies. 

Line bundles $L=\cO_X({\bf k})$ on $X$ are labelled by integer vectors ${\bf k} = (k_1,k_2,\ldots ,k_h)$, where $h=h^{1,1}(X)$, relative to an integral basis $(J_1,\ldots,J_h)$ of $H^2(X)$ such that 
\begin{equation*}
c_1(\cO_X({\bf k})) = \sum_{i=1}^hk_i J_i~.
\end{equation*}
The line bundle index can be computed from the formula
\begin{equation}\label{indgen}
 {\rm ind}(L)=\sum_{q=0}^3(-1)^qh^q(L)=\frac{1}{6}c_1(L)^3+\frac{1}{12}c_2(TX)c_1(L)=\frac{1}{6}d_{ijl}k^ik^jk^l+\frac{1}{12}c_{2i}(TX)k^i\; ,
\end{equation}
where $d_{ijk}$ are the triple intersection numbers (completely symmetric in all three indices) and $c_{2i}(TX)$ is the second Chern class of the CY manifold, relative to the dual basis of $4$-forms.

For the manifold $X_{7884}$ with $h=2$ we can choose a basis $(J_1,J_2)$ of K\"ahler cone generators, so that the K\"ahler cone $\mathcal{K}(X)$ is the positive quadrant in those coordinates. In this basis the non-zero triple intersection numbers and second Chern class are given by
\begin{equation}
 d_{112}=d_{122}=3\;,\qquad c_{2i}(TX)=(36,36)\; ,
\end{equation}
so that the index formula~\eqref{indgen} specialises to
\begin{equation}\label{eq:bicubicind}
 {\rm ind}(X,\cO_X(k_1,k_2))=\frac{3}{2}(k_1+k_2)(2+k_1k_2)\; .
\end{equation}
The effective cone coincides with $\overline{\mathcal{K}(X)}$, so all line bundles outside this region have no sections. The formula for $h^0(X,{\cal O}_X(k))$ is summarised in the table below.
\begin{equation}
\begin{tabular}{|c|c|}\hline
 {\rm region }&$h^0(X,\cO_X(k_1,k_2))$ \\[4pt]\hline
\varstr{14pt}{8pt}$k_1>0,~k_2>0$ &~ ${\rm ind}(X,\cO_X(k_1,k_2))=\frac{3}{2}(k_1+k_2)(2+k_1k_2)$ \\[2pt]
\varstr{4pt}{8pt}$k_1>0,~k_2=0$ & ${\rm ind}(\IP^2,\cO_{\IP^2}(k_1))=\frac{1}{2}(1+k_1)(2+k_1)$\\[2pt]
\varstr{4pt}{8pt}$k_1=0,~k_2>0$ & ${\rm ind}(\IP^2,\cO_{\IP^2}(k_2))=\frac{1}{2}(1+k_2)(2+k_2)$\\[2pt]
$k_1=k_2=0$ & $1$\\[2pt]
otherwise&$0$\\\hline
\end{tabular}
\label{eq:7887_formulae}
\end{equation}
For the manifold $X_{7669}$ with $h=3$ we can also choose a basis $(J_1,J_2,J_3)$ of K\"ahler cone generators so that the K\"ahler cone $\cK(X)$ is the positive octant. Its non-zero intersection numbers and the second Chern class are given by
\begin{equation}
 d_{iij}=3\;\;\forall i\neq j\;,\quad d_{123}=6\;,\qquad c_{2i}(TX)=(36,36,36)\; ,
\end{equation} 
which results in the index formula
\begin{equation}\label{eq:7669_index}
  {\rm ind}(L)=3(k_1+k_2+k_3)+\frac{3}{2}(k_1^2k_2+k_1k_2^2+k_1^2k_3+k_1k_3^2+k_2^2k_3+k_2k_3^2)+6k_1k_2k_3\; .
\end{equation}  
The effective cone for this manifold consists of an infinite number of additional K\"ahler cones, adjacent to the three boundaries of $\cK(X)$, which corresponds to bi-rationally equivalent and isomorphic Calabi-Yau three-folds related to $X$ by sequences of flops (see Refs.~\cite{Brodie:2020fiq, Brodie:2021ain, Brodie:2021prep}). These additional cones are obtained from the Kahler cone by the action of a symmetry generated by
\begin{equation*}
M_1 = \left(\begin{array}{rrr}-1&0&0\\2&1&0\\2&0&1\end{array}\right)~,\qquad
M_2 = \left(\begin{array}{rrr}1&2&~~0\\0&-1&~~0\\0&2&~~1\end{array}\right)~,\qquad
M_3 = \left(\begin{array}{rrr}1&~~0&2\\0&~~1&2\\0&~~0&-1\end{array}\right)~.
\end{equation*}
Consequently, any effective line bundle $L$ is related to a nef line bundle $L'$ by a finite number of transformations
\begin{equation*}
c_1({L'})=M_{i_1}M_{i_2}\ldots M_{i_k}c_1(L) \in \overline{\cK(X)}~.
\end{equation*}
with the above matrices. Since the number of global sections of a line bundle is invariant under a flop, it follows that 
\begin{equation*}
h^0(X,L)=h^0(X,L')= {\rm ind}(X,L')\;, 
\end{equation*}
where the index can be computed from Eq.~\eqref{eq:7669_index}.

\subsection{Monad bundles} \label{sec:monads}
Given a Calabi-Yau three-fold $X$, monad bundles can be defined by a short exact sequence
\begin{equation}\label{MonadSeqence}
0 \longrightarrow V \longrightarrow B \stackrel{f}{\longrightarrow} C \longrightarrow 0\;,\qquad
B = \bigoplus_{i=1}^{r_B} \mathcal O_X(\mathbf b_i)\;,\quad C = \bigoplus_{a=1}^{r_C} \mathcal O_X(\mathbf c_a)
\end{equation}
where $B$ and $C$ are line bundle sums with ranks $r_B = {\rm rk}(B)$ and  $r_C = {\rm rk}(C)$, respectively. The map $f$ is a bundle homomorphism and by exactness 
\begin{equation} \label{Vprop}
V\cong{\rm ker}(f)\;,\quad {\rm coker}(f)=0\;,\quad {\rm ch}(V) = {\rm ch}(B)- {\rm ch}(C)\;.
\end{equation}
\noindent{\bfseries Bundleness.} The monad construction leads to an additional consistency condition which needs to be checked. Namely, for $V$ to be a vector bundle rather than a sheaf, its rank must be constant and equal to $r_B-r_C$. For this to happen, the degeneracy locus of $f$, the locus where the rank of $f$ is less than maximal, must be empty. We will take the dimension of the degeneracy locus of $f$, which can be $0,1,2$ or $3$, as a measure of how badly $V$ fails to be a bundle. 

In this paper we consider monad bundles over favourable CICYs $X\subset \cA=\IP^{n_1}\times\dots\times\IP^{n_m}$, hence both $B$ and $C$ are restrictions to $X$ of line bundle sums
\begin{equation}
 \tilde{B} = \bigoplus_{i=1}^{r_B} \mathcal O_\cA(\mathbf b_i)\;,\qquad \tilde{C} = \bigoplus_{a=1}^{r_C} \mathcal O_\cA(\mathbf c_a)
\end{equation}
on $\mathcal{A}$. The monad map $f$ is then the restriction of a map $\tilde{f}:\tilde{B}\rightarrow\tilde{C}$ which can be written as an $r_C\times r_B$ matrix with entries
\begin{equation}\label{ftilde}
 \tilde{f}_{ai}\in H^0(\mathcal{A},\mathcal{O}_\mathcal{A}({\bf c}_a-{\bf b}_i))\; ,
\end{equation} 
 that is, either polynomials of multi-degree ${\bf c}_a-{\bf b}_i$ or zero if any component of ${\bf c}_a-{\bf b}_i$ is negative. The degeneracy locus of $\tilde{f}$ defines a variety in $\mathcal{A}$ and, for bundleness of $V$, we should demand that this variety does not intersect $X$. For sufficiently generic choices of polynomials this is satisfied provided the co-dimension of the degeneracy locus in $\mathcal{A}$ is at least four. The dimension of the degeneracy locus of $\tilde f$ and $f$ are related by 
\begin{equation}
d_{\rm deg} (f) = d_{\rm deg} (\tilde f) - ({\rm dim}\,\cA-3)~.
\end{equation}
 
Naively, the computation of the dimension of the (generic) degeneracy locus proceeds as follows. Replace $\tilde f$ by a numerical matrix with a random entry for each non-trivial polynomial $\tilde f_{ai}$. If the rank of this matrix is non-maximal then $\tilde f$ degenerates everywhere, hence $d_{\rm deg} (\tilde f) = {\rm dim}\,\cA$ and $d_{\rm deg} (f) = 3$. If the rank is maximal, proceed further by replacing any one of the non-trivial entries by $0$ and checking each time if the rank has dropped. If for any such replacement the rank drops, then $d_{\rm deg} (f) = 2$. Otherwise replace two non-trivial entries by $0$ at a time and if the rank drops, $d_{\rm deg} (f) = 1$. If that's still not the case, then replace three non-trivial entries by $0$ and if now the rank drops, then $d_{\rm deg} (f) = 0$. If this did not happen, the degeneracy locus is empty and  $V$ is a bundle. 

There are, however, a number of subtleties in setting various entries of $\tilde f$ to zero which arise from the fact that the entries are polynomials, rather than numbers. Two (or more) polynomials that depend only on the coordinates of a single $\IP^1$ cannot be set to $0$ simultaneously. The same is true for three (or more) polynomials that depend only on the coordinates of a single $\IP^2$ or the coordinates of two $\IP^1$ spaces. 
\noindent{\bfseries Chern classes.} In view of Eq.~\eqref{Vprop} the Chern classes of an $SU(4)$ monad bundle on a Calabi-Yau threefold $X$ can be computed from the following expressions:
\begin{equation}\label{eq:monchern}
\begin{array}{rcl}
 {\rm rk}(V)&=&{\rm rk}(B)-{\rm rk}(C)\stackrel{!}{=}4\\
c_1^k(V)&=& \displaystyle c_1^k(B)-c_1^k(C)=\sum_{i=1}^{r_B}b_i^k- \sum_{a=1}^{r_C} c_a^k\stackrel{!}{=}0\\
 c_{2k}(V)&=&\displaystyle  {\rm ch}_{2k}(C)-{\rm ch}_{2k}(B)=\frac{1}{2}d_{klm}\left(\sum_{a=1}^{r_C}c_a^lc_a^m-\sum_{i=1}^{r_B}b_i^lb_i^m\right)\stackrel{!}{\leq}c_{2k}(TX)\\[4pt]
 {\rm ind}(V)&=& \displaystyle \sum_{q=0}^3(-1)^q h^q(X,V)=\frac{1}{2}c_3(V)={\rm ch}_3(B)-{\rm ch}_3(C)\\[4pt]
 &=& \displaystyle \frac{1}{6}d_{klm}\left(\sum_{i=1}^{r_B}b_i^kb_i^lb_i^m-\sum_{a=1}^{r_C}c_a^kc_a^lc_a^m\right)\stackrel{!}{=}-3|\Gamma|~.
\end{array} 
\end{equation}
For the two manifolds of interest the triple intersection numbers and the second Chern class of the tangent bundle have been provided in Section~\ref{sec:lbcohformulae}. On the right-hand sides of Eqs.~\eqref{eq:monchern} we have indicated the desired value for the respective Chern class.

\noindent{\bfseries Low-energy spectrum.} The chiral asymmetry between ${\bf 16}$ and $\overline{\bf 16}$ multiplets is given by the index, that is,
\begin{equation}
 n_{\bf 16}-n_{\overline{\bf 16}}=h^1(X,V)-h^2(X,V)=-{\rm ind}(V)\; ,
\end{equation} 
since Hoppe's criterion for stable bundles implies that $h^0(X,V)=h^3(X,V)=0$.

In order to compute the number of $\mathbf{16}$ and $\overline{\mathbf{16}}$ $SO(10)$-multiplets seperately, the first and second cohomologies of $V$ are required. These can be obtained from the long exact sequence in cohomology 
\begin{equation}\label{LongMonadSequence}
\begin{aligned}
0 & \rightarrow ~H^0(X,V)~ \rightarrow H^0(X,B) \rightarrow H^0(X,C) \rightarrow \\
& \rightarrow  \boxed{H^1(X,V) }\rightarrow H^1(X,B) \rightarrow H^1(X,C) \rightarrow  \\
& \rightarrow \boxed{H^2(X,V)} \rightarrow H^2(X,B) \rightarrow H^2(X,C) \rightarrow \\
& \rightarrow ~H^3(X,V)~ \rightarrow H^3(X,B) \rightarrow H^3(X,C) \rightarrow 0\; ,
\end{aligned}
\end{equation}
which is associated with the monad sequence \eqref{MonadSeqence}. From the same sequence $h^0(X,V)$ and $h^3(X,V)$ can be computed. Both of these need to vanish if $V$ is stable, according to Hoppe's criterion \eqref{hoppe}. In general, such cohomology computations are difficult and require two pieces of information: (1) the maps $H^{i}(X,B)\rightarrow H^i(X,C)$ induced by the bundle morphism $f$ and (2) the co-boundary maps $H^{i}(X,C)\rightarrow H^{i+1}(X,V)$. Both of these issues have been dealt with and implemented in the CICY package \cite{cicypackage}, which allows us to compute the required cohomologies even when the long exact sequence in cohomology does not split.

The Higgs field arises from the cohomology $H^1(X,\wedge^2 V)$ which can be obtained from the second wedge power sequence
\begin{equation}\label{wedge2seq}
 0\rightarrow \wedge^2V\rightarrow \wedge^2 B\rightarrow B\otimes C\rightarrow S^2C\rightarrow 0\; ,
\end{equation}
of the monad sequence \eqref{MonadSeqence}. This can be split up into two short exact sequences 
\begin{equation}\label{SplitSequence}
 0\rightarrow \wedge^2V\rightarrow \wedge^2 B\rightarrow K \rightarrow 0~,\qquad 0\rightarrow K  \rightarrow   B\otimes C\rightarrow S^2C\rightarrow 0~.
\end{equation}
The long exact sequence in cohomology associated with the second sequence provides the cohomology of $K$, which can then be fed into the long exact sequence associated with the first sequence to obtain the cohomology of $\wedge^2 V$. These computations are non-trivial for the same reasons as above; working out the induced maps in the first long exact sequence in cohomology may be even harder since the cohomology of $K$ can be a complicated sum of kernels and co-kernels.

\noindent{\bfseries Equivariance.} As discussed earlier, the bundle $V\rightarrow X$ needs to admit a $\Gamma$-equivariant structure for it to descend to a bundle $\hat{V}\rightarrow\hat{X}$ on the quotient manifold $\hat{X}=X/\Gamma$. For a monad bundle a sufficient condition for this is that the constituent bundles $B$ and $C$ admit $\Gamma$-equivariant structures. If all line bundles in $B$ (and in $C$) are different from each other then each line bundle needs to carry a $\Gamma$-equivariant structure individually and a (strong) necessary check for this to be the case is that the indices ${\rm ind}({\cal O}_X({\bf b}_i))$ and ${\rm ind}({\cal O}_X({\bf c}_a))$ are each divisible by the group order $|\Gamma|$. Things are slightly more complicated if $B$ or $C$ contain repeated line bundles since sums of same line bundles may admit an equivariant structure even though the individual line bundle does not. Suppose $B$ (or $C$) contains $m$ copies of the line bundle $L$. Then we should check whether $L^{\oplus m}$ admits a $\Gamma$-equivariant structure and a necessary condition is that $m\,{\rm ind}(L)$ is divisible by $|\Gamma|$.

\subsection{A new example on the bicubic}\label{secExample}
Before discussing the RL set-up for heterotic models based on monad bundles, it is instructive to consider an example for such a model. In fact, the following monad bundle on the bicubic CY (number $7884$ from Table~\ref{tab:CYs}) was found by the RL system presented in the following sections. The line bundles involved in the definition of this monad contain both negative and positive entries and, to our knowledge, it is the first model of this kind with the correct particle spectrum. (See Ref.~\cite{Anderson:2009mh} for a bicubic standard model based on a semi-positive monad.)

Let $X$ denote a generic bicubic three-fold $X$ admitting a freely acting discrete symmetry $\Gamma=\IZ_3\times \IZ_3$. From Table~\ref{tab:CYs}, its configuration matrix is
\begin{equation}\label{eq:bicubicCM}
\cicy{\IP^{2}\\[4pt] \IP^{2}}{3\\[4pt] 3}^{2,83}~.
\end{equation}
On this manifold, we define a rank four monad bundle $V$ by the sequence~\eqref{MonadSeqence} with constituent line bundle sums
\begin{equation}
B= {\cal O}_X (-1, 1)^3 \oplus {\cal O}_X (2, 0)^3\;,\quad
C= {\cal O}_X (1, 1) \oplus {\cal O}_X (2,2)\; . 
\end{equation}
As discussed earlier, the monad map $f$ is the restriction of a polynomial map $\tilde{f}$ whose multi-degrees can be determined from the line bundle integers as in Eq.~\eqref{ftilde}. For the above choice of $B$ and $C$ this leads to the matrix
\begin{equation}\label{eq:MonadMapEx1}
\tilde f = \left(\begin{array}{llllll}
 f^{1,1}_{(2,0)}& f^{1,2}_{(2,0)}&f^{1,3}_{(2,0)}& 0&0&0\\[8pt]
 f^{2,1}_{(3,1)}& f^{2,2}_{(3,1)}& f^{2,3}_{(3,1)}&f^{2,4}_{(0,2)}&f^{2,5}_{(0,2)}&f^{2,6}_{(0,2)}
 \end{array}\right)\,,
\end{equation} 
where the subscripts indicate the multi-degrees of the otherwise generic polynomials and the superscripts are bookkeeping labels. For instance, $f^{1,1}_{(2,0)}$ is of degree 2 in the coordinates of first ${\mathbb P}^2$ space, and does not depend on the coordinates of the second  ${\mathbb P}^2$ space.\\[2mm]
\noindent{\bfseries Bundleness.} The matrix associated with the monad map $f$ is potentially rank-changing over $X$ since the first row in \eqref{eq:MonadMapEx1} contains only three non-trivial polynomials which could, in principle, intersect $X$ in a number of points. However, this is not the case. All three polynomials depend only on the coordinates of the first projective space, but the intersection of three sufficiently generic polynomials in $\IP^2$ vanishes.\\[2mm]
\noindent{\bfseries Vanishing first Chern class.} The vanishing $c_1(V)=0$ follows easily from Eq.~\eqref{eq:monchern}.\\[2mm]
\noindent{\bfseries Anomaly cancellation condition.} Since $c_{2k}(TX)=(36,36)$ and $c_{2k}(V)=(27,9)$ the anomaly cancellation condition can be satisfied with a five-brane wrapping the holomorphic curve $\mathcal{C}$ with class $[\cC]=9C_1+27C_2$, where $C_1$ and $C_2$ are the curve classes that are dual to the cohomology classes $J_1$ and $J_2$.\\[2mm]
\noindent {\bfseries Low-energy spectrum.} The long exact sequence \eqref{LongMonadSequence} associated with the monad sequence gives the cohomology of $V$. The cohomology dimensions of the line bundle sums $B$ and $C$ are straightforward from the formulae in Section~\ref{sec:lbcohformulae} and are given in the following table.
\begin{equation*}
\begin{array}{c|ccc}
q& h^q(X,V)& h^q(X,B)& h^q(X,C)\\
\hline
0& ? & 18 & 45\\
1& ? & 0 & 0\\
2& 0 & 0 & 0\\
3& 0 & 0 & 0\\
\end{array}
\end{equation*}
The long exact sequence \eqref{LongMonadSequence} gives $H^0(X,V)={\rm ker}\left( H^0(X,B)\rightarrow H^0(X,C)\right)$. But this kernel is trivial: all the global section of $B$ come from global sections of $\cO_X(2,0)$, which are mapped to global sections of $\cO_x(2,2)$ (the second term in $C$) by multiplication with the polynomials $f^{2,4}_{(0,2)}$, $f^{2,5}_{(0,2)}$ and $f^{2,6}_{(0,2)}$. Since these polynomials are generic, the map is injective and it follows that 
\begin{equation*}
h^\bullet(X,V) = (0,27,0,0)\; .
\end{equation*}
This result has a number of implications. First, since $h^0(X,V)=h^3(X,V)=0$ two of the three conditions for Hoppe's stability criterion in Eq.~\eqref{hoppe} are satisfied. Secondly, $h^1(X,V)=27=3|\Gamma|$ is the correct number which leads to three families after dividing by $\Gamma$. And finally, $h^2(X,V)=0$ implies the absence of anti-families.

To determine the number of $\mathbf{10}$-multiplets, containing the Higgs field, we have to study the second wedge power sequence~\eqref{wedge2seq} which is split into two short exact sequences as in Eq.~\eqref{SplitSequence}. The cohomology dimensions in the two associated long exact sequences can again be computed from the formulae in Section~\ref{sec:lbcohformulae} and this leads to the following tables.
\begin{equation*}
\begin{array}{c|ccc}
q& h^q(X,\wedge^2 V)& h^q(X,\wedge^2 B)& h^q(X,K)\\
\hline
0& ? & 126 & ?\\
1& ? & 9 & ?\\
2& ? & 0 & 0\\
3& 0 & 0 & 0\\
\end{array}\hspace{21pt}
\begin{array}{c|ccc}
q& h^q(X,K)& h^q(X,B\otimes C)& h^q(X,S^2C)\\
\hline
0& ? & 468 & 351\\
1& ? & 0 & 0\\
2& 0 & 0 & 0\\
3& 0 & 0 & 0\\
\end{array}
\end{equation*}
In order to compute the cohomology of $K$, we need: (1) the bundle morphism $B\otimes C\rightarrow S^2 C$ induced by the morphism in the monad sequence and (2) the maps induced in cohomology by this bundle morphism. The result is
\begin{equation*}
h^\bullet(X,K) = (135,18,0,0)~.
\end{equation*}
This information then feeds into the long exact sequence in cohomology associated with the first short exact sequence. Working though the various induced maps gives the final result
\begin{equation*}
h^\bullet(X,\wedge^2 V) = (0,9+c,9+c,0)\; .
\end{equation*}
where $0\leq c\leq 9$ is the dimension of ${\rm ker}\left( H^1(X,\wedge^2 B)\rightarrow H^1 (X,K)\right)$, which we were not able to compute explicitly. 
Hence, we have at least nine ${\bf 10}$ multiplets upstairs with a chance of retaining one Higgs pair in the downstairs model. The details depend on the choice of equivariant structure which we will not study in detail. We also note that $h^0(X,\Lambda^2 V)=0$, so that all three conditions of Hoppe's criterion~\eqref{hoppe} are satisfied.\\[2mm]
\noindent{\bfseries Bundle stability} With Hoppe's criterion satisfied the bundle $V$ has already passed an important stability test. To improve on this, we have to determine the set $\mathcal{I}$ of all line bundles which inject into $V$, $\Lambda^2 V$ and $V^*$ and then use Eq.~\eqref{stabreg} to check if there exists a locus in the K\"ahler cone where all these line bundles have a negative slope. In practice, this has to be carried out for line bundles ${\cal O}_X(k_1,k_2)$ with entries in a finite range which we take to be $-5\leq k_i\leq 5$. From these line bundles the following subset injects into $V$ or $V^*$:
\begin{equation*}
\mathcal{I}\supset\{\cO(1,k_2),\;\cO(2,k_2)\,|\, k_2=-2,-3,-4,-5\}
\end{equation*}
%
It can be easily checked that there exist K\"ahler forms $J=t^1J_1+t^2J_2$ for which the slope
$ \mu_J(\cO_X({\bf k}))=d_{ijk}t^it^jk^l$
is negative for all the above line bundles. Indeed, any K\"ahler form in the region $s_2>s_1>0$ where $s_i = d_{ijk}t^j t^k$ renders $V$ stable with respect to the sub-bundles associated with the above line bundles. 

\section{Reinforcement Learning}\label{secRL}

\subsection{Generalities}
Reinforcement learning (RL) is an artificial intelligence method positioned between supervised and unsupervised learning. Both RL and supervised learning rely on labelled data which is used to train a neural network. However, unlike supervised learning, RL does not rely on pre-compiled training data but, rather, this data is incrementally generated by exploring an environment during the course of the training process. In this way, data sampling can be guided and refined by the neural network, using rewards and penalties to aid the navigation of the environment. In the following we review the basic structure of RL, mainly to fix notation and terminology. For more in-depth accounts see, for example, Refs.~\cite{sutton2018reinforcement, Ruehle:2020jrk}.

The mathematical framework for RL is provided by Markov decision processes (MDPs) which are given by a tuple $(\mathcal{S},\mathcal{A},\mathcal{P},\gamma,\mathcal{R})$. Here, $\mathcal{S}$ is the {\it environment} which consists of certain {\it states} $s\in\mathcal{S}$. For our applications, the environment consists of a class of string compactifications, with states typically represented by integer matrices. The set $\mathcal{A}$ contains {\it actions} $a:\mathcal{S}\rightarrow\mathcal{S}$ and the function $\mathcal{P}$ provides probabilities $\mathbb{P}(S=s'|S=s,A=a)$ for an action $a$ to change a state $s$ to a state $s'$. In our case, actions will be deterministic, that is, an action $a\in\mathcal{A}$ will convert a state $s$ into a certain other state $s'$ with probability one. In practice, they involve changing one of the entries of the integer matrices representing states by $\pm 1$. Finally, $\gamma\in[0,1]$ is the {\it reward factor} which will enter the definition of the {\it return} given below and $\mathcal{R}:\mathcal{S}\times\mathcal{A}\rightarrow\mathbb{R}$ is the {\it reward function} which provides the reward/penalty $\mathcal{R}(s,a)$ for an action $a$ carried out on a state $s$. For our applications, the reward function will measure whether the action has produced a more or less desirable string model. Usually, a certain subset of states $\mathcal{T}\subset\mathcal{S}$, called {\it terminal states}, is singled out. In our case, the terminal states will be those string models which have all required properties for a candidate string standard model.

A sequence
\[
 s_0\stackrel{a_0,\,r_0}{\xrightarrow{\hspace*{8mm}}} s_1\stackrel{a_1,\,r_1}{\xrightarrow{\hspace*{8mm}}}s_2\stackrel{a_2,\,r_2}{\xrightarrow{\hspace*{8mm}}}s_3\cdots
\]
of states $s_t$, connected by actions $a_t$ and with rewards $r_t=\mathcal{R}(s_t,a_t)$, where $t=0,1,2,\ldots$, is called an {\it episode}. An {\it agent} explores the environment carrying out such episodes, thereby producing data triplets $(s_t,a_t,r_t)$. An episode terminates if it reaches a terminal state or else after a maximal pre-set number of steps $t_{\rm max}$. The {\it return} $G_t$ of each state $s_t$ in an episode is defined as
\begin{equation}\label{returndef}
 G_t=\sum_{k\geq 0} \gamma^k r_{t+k}\; ,
\end{equation} 
where a small discount factor $\gamma\ll 1$ leads to a short-term return and a discount factor close to~$1$ leads to a long-term return. Episodes are carried out from a random starting state $s_0$, which is determined from a given probability distribution on the state space $\mathcal{S}$. In our case this will be either a flat distribution or a distribution somewhat peaked on small integer entries. A {\it policy} $\pi$ provides probabilities $\pi(\alpha|s)=\mathbb{P}(A_t=a|S_t=s)$ for applying an action $a$ to a state $s$. Given a policy, the {\it state value function} $V_\pi$ and the {\it state-action value function} $Q_\pi$ are defined as the expectation values of the return, that is,
\begin{equation}
 V_\pi (s)=\mathbb{E}(G_t|S_t=s)\;,\qquad Q_\pi(s,a)=\mathbb{E}(G_t|S_t=s,A_t=a)\; .
\end{equation} 
The goal of the RL system is to maximise these functions over the space of policies $\pi$ and this can be accomplished by a number of different RL algorithm, which differ by which of the functions $\pi$, $V_\pi$ and $A_\pi$ are realised as a neural network.

In this paper, we consider policy-based approaches, that is, the policy $\pi$ is realised by a neural network $\pi_\theta$ with parameters $\theta$ which guides the episodes
\begin{equation}\label{eppi}
 s_0\stackrel{\pi_\theta}{\xrightarrow{\hspace*{8mm}}} s_1\stackrel{\pi_\theta}{\xrightarrow{\hspace*{8mm}}}s_2\stackrel{\pi_\theta}{\xrightarrow{\hspace*{8mm}}}s_3\cdots \; .
\end{equation}
For the simplest algorithm, known as REINFORCE, $\pi_\theta$ is the only neural network of the system. From the {\it policy-gradient theorem} this network is trained on the loss function
\begin{equation}\label{losspn}
 L(\theta)=Q_\pi(s,a)\ln (\pi_\theta(s,a))\; ,
\end{equation} 
where $Q_\pi(s,a)$ can, in practice, be replaced by the return $G$ of the state $s$. An actor-critic algorithm relies on two neural networks, the policy network $\pi_\theta$, as above, and a network $V_\eta$ with parameters~$\eta$ which represents the value function (state or state-action). The policy network is trained on the loss~\eqref{losspn}, as before, while the value network $V_\eta$ is trained on a mean square loss. The presence of the value network allows replacing the formula~\eqref{returndef} for the return by the {\it TD-return}
\begin{equation}\label{tdreturn}
 G_t=r_t+\gamma V_\eta(s_{t+1})\; .
\end{equation}
This means computation of the return does not have to await the end of the episode but can be computed immediately after each action. This also facilitates exploring the environment by multiple agents. Schematically, the training proceeds as follows:
\begin{enumerate}
\item[(1)] Initialise the policy network $\pi_\theta$ and the value network $V_\pi$, if present.
\item[(2)] Collect a batch of data triplets $(s_t,a_t,G_t)$. For REINFORCE this is produced by one agent completing sufficiently many episodes and computing the return from Eq.~\eqref{returndef}. For the actor-critic method multiple agents produce data by carrying our sufficiently many actions in parallel and the return is computed from Eq.~\eqref{tdreturn}.
\item[(3)] Use this batch to update the weights $\theta$ of the policy network $\pi_\theta$, based on the loss~\eqref{losspn}, and the weights $\eta$ of the value network $V_\eta$, if present, based on a mean square loss.
\item[(4)] Repeat from (2) until the loss is sufficiently small so that the policy has converged.
\end{enumerate}

Computational versions of RL systems require two basic components, a simulation of the environment which computes rewards $r$ for a state-action pair $(s,a)$ and a realisation of the agent/neural network(s) following one of the algorithms outlined above. In our case, the environment has been realised as a MATHEMATICA package which computes the reward for a monad string compactification by assessing whether a small modification leads to a more or less desirable model. The agent for both the REINFORCE and the actor-critic algorithms has also been realised as MATHEMATICA package, based on the MATHEMATICA suite of machine learning functions. Before we describe these computational realisations in more detail we discuss a simple toy example, based on single line bundles.

\subsection{A toy example: searching for line bundles with a given index}
In this section, we apply RL to a simple environment which consists of the set of all (single) line bundles (with entries in a certain range) on a given CY manifold. The goal is to identify line bundle with a given target index. This environment is of course much simpler and smaller than the environment of monad bundles, which we will tackle in the next section. However, it is qualitatively similar in that it is based on vector bundles whose topological properties we are attempting to engineer. It provides us with a first confirmation that RL is indeed capable of carrying out topological engineering. Moreover, the environment is small enough that it can easily be scanned systematically so that we have an independent check on the results obtained from RL. Given that we focus on a simple two-dimensional environment which facilitates graphical representation it is also a good opportunity to develop intuition and illustrate the workings of an RL system.

Our setting is a fixed CY three-fold $X$ with Picard number $h=h^{1,1}(X)$ together with its line bundles $L=\mathcal{O}_{X}({\bf k})$, labelled by $h$-dimensional integer vectors ${\bf k}=(k^1,\ldots ,k^h)$. Let us be more precise about how this mathematical setting is mapped to the ingredients of a MDP. The environment is given by $h$-dimensional integer vectors
 \begin{equation}
  \mathcal S = \{{\bf k}\in \IZ^h\,:\,|k_i|\leq k_{\rm max}\}\; ,
 \end{equation} 
which represent the line bundles $\cO_X({\bf k})$. To obtain a finite environment, we have imposed an upper bound $k_{\rm max}$ on the absolute values of the entries. The allowed actions involve increasing or decreasing one entry in ${\bf k}$ by one, so 
\begin{equation}
  \mathcal A = \{{\bf k}\mapsto {\bf k} \pm {\bf e}_i\}\; ,
\end{equation}  
where ${\bf e}_i$ is the $i^{\rm th}$ standard unit vector in $h$ dimensions.
The definition of the reward is based on an {\it intrinsic state value}
\begin{equation}
 v({\bf k})=-\frac{10\,|{\rm ind}(\cO_X({\bf k}))-\tau|}{h k_{\rm max}^3}
\end{equation}
which (modulo normalisation factors) measures the negative difference of the index from the target index $\tau$. Based on this intrinsic value, the reward is defined as 
\begin{equation}\label{lbreward}
r_{s\mapsto s'} = 
\left\{\begin{array}{ccl}
(v(s')-v(s))^p&\text{if}& v(s')-v(s)>0\\
r_{\rm offset}&\text{if}&v(s')-v(s)\leq 0
\end{array}\right\}
+ r_{\rm step}+r_ {\rm boundary}+r_{\rm terminal}
\end{equation}
where $p\in\mathbb{R}$ is a suitably chosen power and $r_{\rm offset}<0$ is a penalty for decreasing the intrinsic state value. To favour finding terminal states by short episodes a penalty $r_{\rm step}<0$ is added for each step and actions which lead beyond the boundary set by $k_{\rm max}$ attract a penalty $r_{\rm boundary}<0$. Finally, actions which lead to a terminal state are rewarded with a bonus $r_{\rm terminal}>0$.

\begin{figure}[H]
\centering
\subfloat[]{\includegraphics[width=0.3\linewidth]{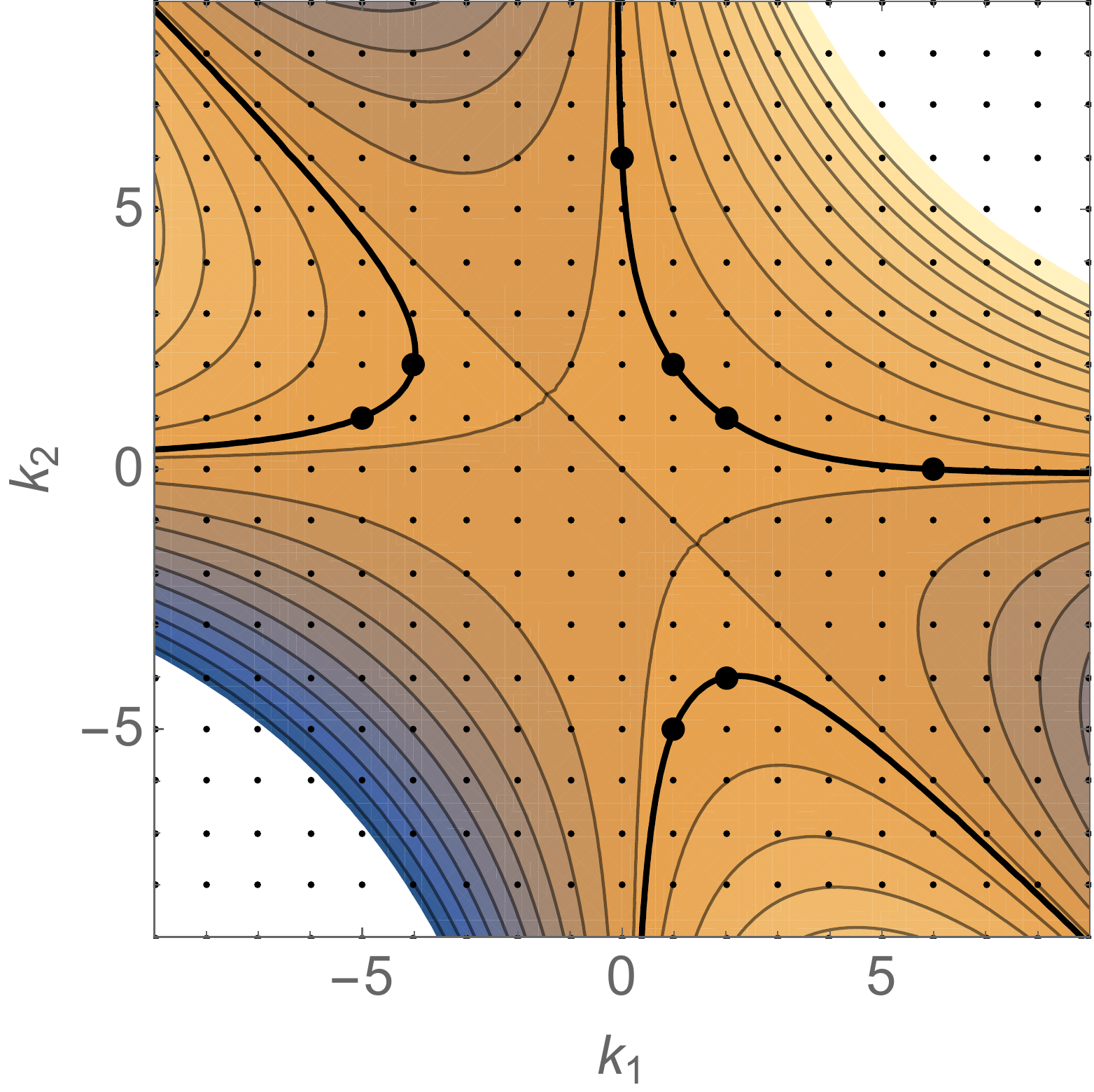}}\qquad
\subfloat[]{\includegraphics[width=0.3\linewidth]{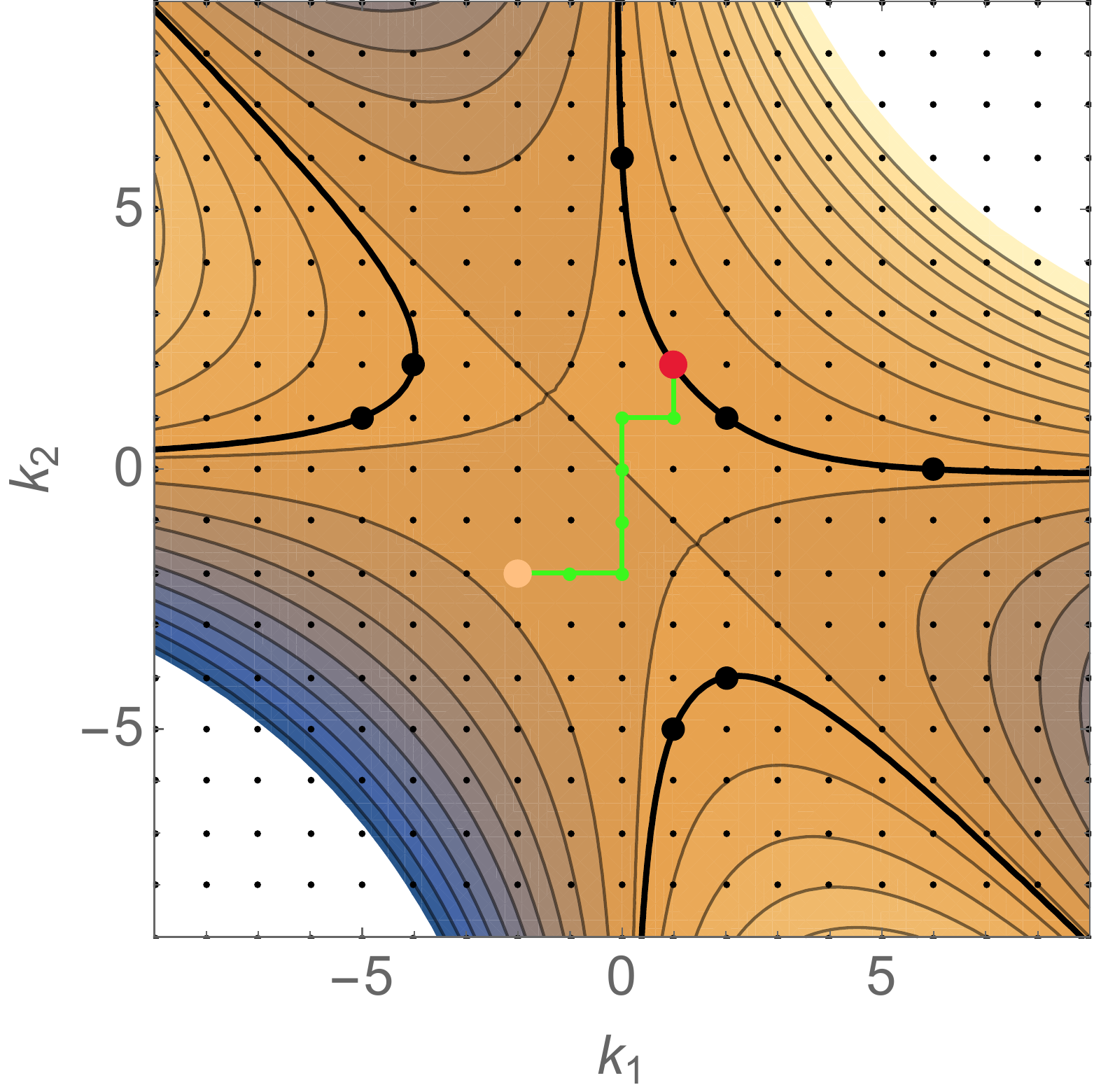}}\qquad
\subfloat[]{\includegraphics[width=0.3\linewidth]{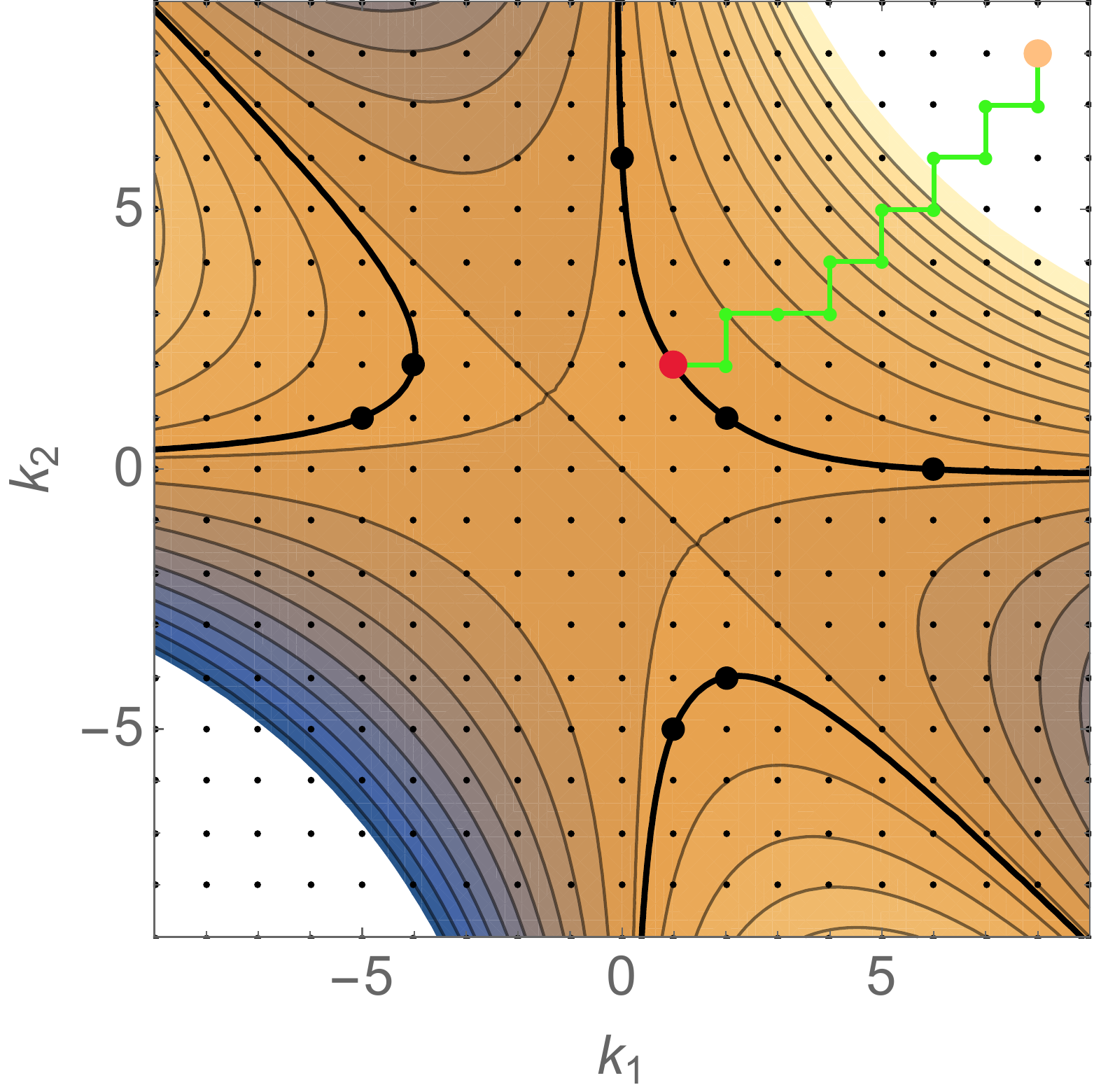}}\\
\caption{\sf (a) Result of a systematic scan for line bundles on the bicubic three-fold, with the large back points representing line bundles with index equal to $\tau=18$ (b) (c) Two sample episodes (green lines) obtained by following an RL-trained policy, starting at the yellow points and ending up in terminal states (red points).}
\label{figLineBundleScan}
\end{figure}

The above set-up can be applied to line bundles on any CY three-fold but, for simplicity, we focus on a generic bicubic CY 
 whose configuration matrix is given in Eq.~\eqref{eq:bicubicCM}. Since the Picard number equals $h=2$ for this manifold, line bundles $L=\mathcal{O}_{X}({\bf k})$ are specified by two dimensional integer vectors ${\bf k}=(k_1,k_2)$ and the explicit formula for the index has been presented in Eq.~\eqref{eq:bicubicind}. We restrict the range of line bundle integers by setting $k_{\rm max}=9$ and we are looking for line bundles with target index $\tau=18$. The parameters which enter the definition of the reward function~\eqref{lbreward} are chosen as
 \[
  p=1\;,\quad r_{\rm offset}=-1\;,\quad r_{\rm step}=0\;,\quad r_{\rm boundary}=-1\;,\quad r_{\rm terminal}=2\; .
 \] 
The environment contains $19^2=361$ states so it is small enough to be scanned systematically. The result of such a scan is shown in  Figure~\ref{figLineBundleScan}(a), where the eight black dots indicate the terminal states, that is, the line bundles with the required target index $\tau=18$.

\begin{figure}[H]
\begin{center}
\begin{tikzpicture}
\draw[->,thick] (0.35,0) -- (0.7,0);
\draw[thick] (0.7,0.25) -- (0.7,-0.25) -- (1.9,-0.25) -- (1.9,0.25) -- (0.7,0.25);
\draw[->,thick] (1.9,0) -- (2.6,0);
\draw[thick] (2.6,0.25) -- (2.6,-0.25) -- (3.8,-0.25) -- (3.8,0.25) -- (2.6,0.25);
\draw[->,thick] (3.8,0) -- (4.5,0);
\draw[thick] (4.5,0.25) -- (4.5,-0.25) -- (5.7,-0.25) -- (5.7,0.25) -- (4.5,0.25);
\draw[->,thick] (5.7,0) -- (6.4,0);
\draw[thick] (6.4,0.25) -- (6.4,-0.25) -- (7.6,-0.25) -- (7.6,0.25) -- (6.4,0.25);
\draw[->,thick] (7.6,0) -- (8.3,0);
\draw[thick] (8.3,0.25) -- (8.3,-0.25) -- (9.5,-0.25) -- (9.5,0.25) -- (8.3,0.25);
\draw[->,thick] (9.5,0) -- (10.2,0);
\draw[thick] (10.2,0.25) -- (10.2,-0.25) -- (11.4,-0.25) -- (11.4,0.25) -- (10.2,0.25);
\draw[->,thick] (11.4,0) -- (12.1,0);
\draw[thick] (12.1,0.25) -- (12.1,-0.25) -- (13.2,-0.25) -- (13.2,0.25) -- (12.1,0.25);
\draw[->,thick] (13.2,0) -- (13.9,0);
\draw[thick] (13.9,0.25) -- (13.9,-0.25) -- (15.1,-0.25) -- (15.1,0.25) -- (13.9,0.25);
\draw[->,thick] (15.1,0) -- (15.45,0);
\node at (1.3,0.02) {\scriptsize affine};
\node at (3.2,0.02) {\scriptsize SELU};
\node at (5.1,0.02) {\scriptsize affine};
\node at (7,0.02) {\scriptsize SELU};
\node at (8.9,0.02) {\scriptsize affine};
\node at (10.8,0.02) {\scriptsize SELU};
\node at (12.7,0.02) {\scriptsize affine};
\node at (14.55,0.02) {\scriptsize softmax};
\node at (0,0.05) {\scriptsize $\mathbb{R}^{d_0}$};
\node at (15.85,0.05) {\scriptsize $\mathbb{R}^{d_1}$};
\node at (2.25,0.25) {\scriptsize $\mathbb{R}^{d}$};
\node at (4.2,0.25) {\scriptsize $\mathbb{R}^{d}$};
\node at (6.1,0.25) {\scriptsize $\mathbb{R}^{d}$};
\node at (8,0.25) {\scriptsize $\mathbb{R}^{d}$};
\node at (9.9,0.25) {\scriptsize $\mathbb{R}^{d}$};
\node at (11.8,0.25) {\scriptsize $\mathbb{R}^{d}$};
\node at (13.6,0.25) {\scriptsize $\mathbb{R}^{d_1}$};
\end{tikzpicture}
\end{center}
\caption{\sf Structure of neural network used as a policy and value network.}\label{fig:nn}
\end{figure}
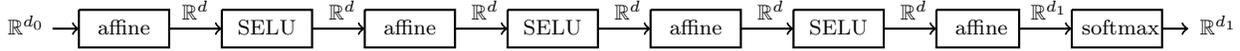

To realise the REINFORCE algorithm we need to supply a neural network $\pi_\theta$ which represents the policy. Its input are the two-dimensional vectors ${\bf k}=(k_1,k_2)$ and its output is a four-dimensional vector which provides the probabilities for the four possible actions. We use a standard feed-forward neural network with the eight-layer architecture shown in Fig~\ref{fig:nn}, with input and output dimensions $d_0=2$ and $d_1=4$ and width $d=16$. 

Here ``affine" refers to an affine layer performing the transformation ${\bf x}\mapsto W{\bf x}+{\bf b}$ with weight $W$ and bias ${\bf b}$, The layers ``SELU" refer to the standard scaled exponential linear unit activation function 
\[
\rho(x)=\begin{cases}1.0507 x~,& x\geq 0\\ 1.7581 (e^x -1)~, &x<0 \end{cases}\; ,
\]
while ``softmax" is a softmax (normalised exponential) layer defined by
\[
 \sigma:\mathbb R^n\rightarrow [0,1]^n~,\qquad \sigma({\bf x})_i = \frac{e^{x_i}}{\sum_{i=1}^n e^{x_i}}\; .
\]
It ensures that the output components are positive and sum up to one so they can be interpreted as probabilities. 

To train this neural network, the agent is coupled to the line bundle environment which is explored in episodes of maximal  length $t_{\rm max}=16$ and with a discount factor $\gamma=0.98$. Training data is supplied in batches of size $32$ (two full episodes) and for stochastic gradient descent we use the ADAM optimiser with learning rate $1/5000$.
\begin{figure}[ht]
\centering
  \subfloat[\sf Loss vs batch number.]{\raisebox{5mm}{\includegraphics[width=0.44\textwidth]{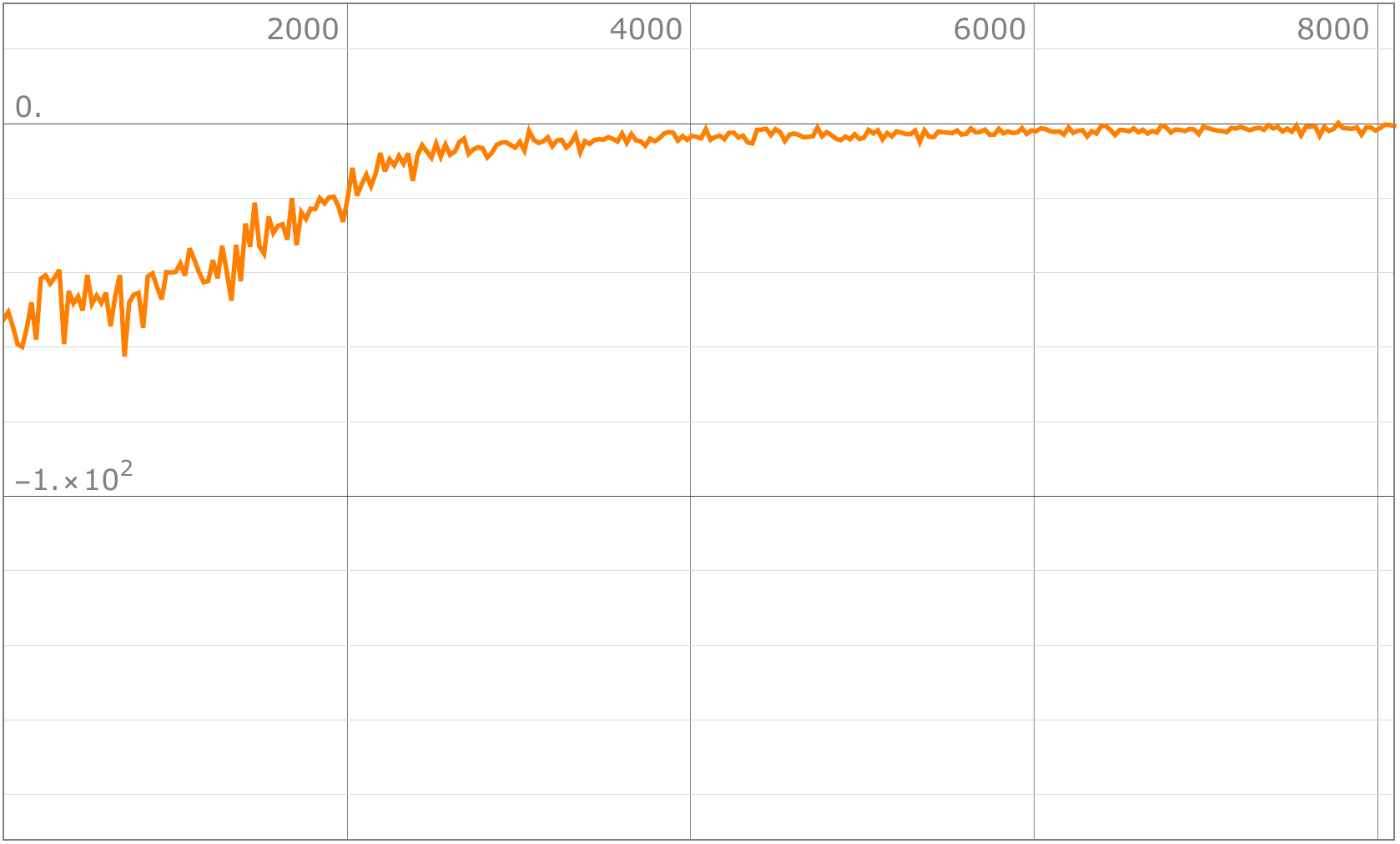}}}\qquad
  \subfloat[\sf Fraction of terminal episodes vs episode number.]{\includegraphics[width=0.47\textwidth]{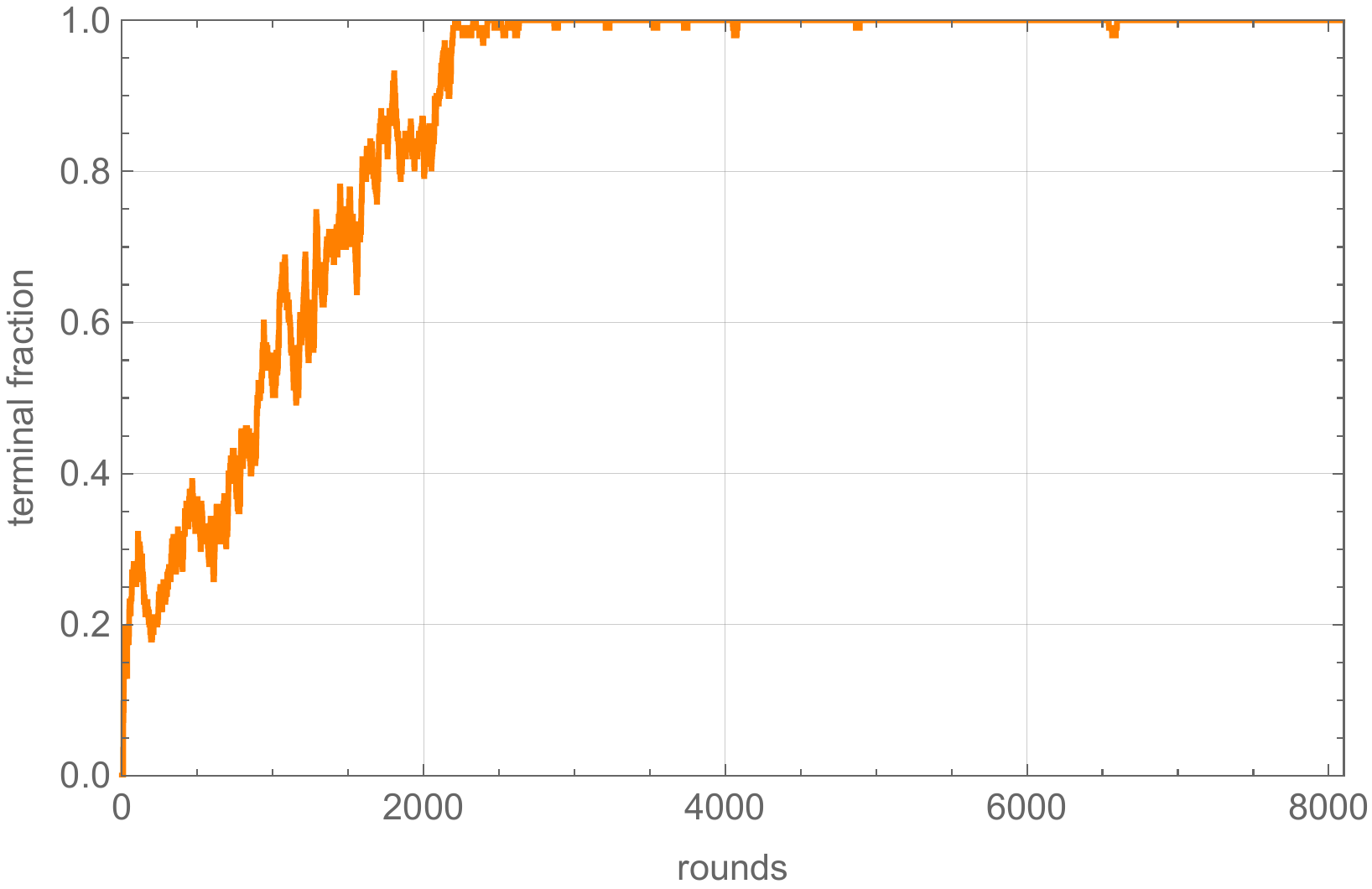}}\qquad
\caption{\sf Training metrics for line bundle environment on bicubic with a target index of $18$.}
        \label{figLineBundleTrainingData}
\end{figure}
Training is accomplished on a single CPU and only takes a few minutes. The training measurements are shown in Fig.~\ref{figLineBundleTrainingData}. 
Since the environment is quite small every state is sampled multiple times during training (this will be different for the much larger monad environment discussed in the next section), so it is not surprising that all eight terminal states are found.
More impressively, after about 200 training rounds, when the loss goes to zero, the fraction of terminal episodes approaches 1 as is evident from Fig.~\ref{figLineBundleTrainingData} (b). At the same time, the average episode length decreases to about $6$ steps. This means, the trained network guides episodes from any starting point to a terminal state on an efficient path with average length $6$. Two examples for such paths are shown in Fig.~\ref{figLineBundleScan} (b), (c). Similar results are obtained using the actor critic method.

\begin{figure}[h]
\centering
\includegraphics[width=0.3\textwidth]{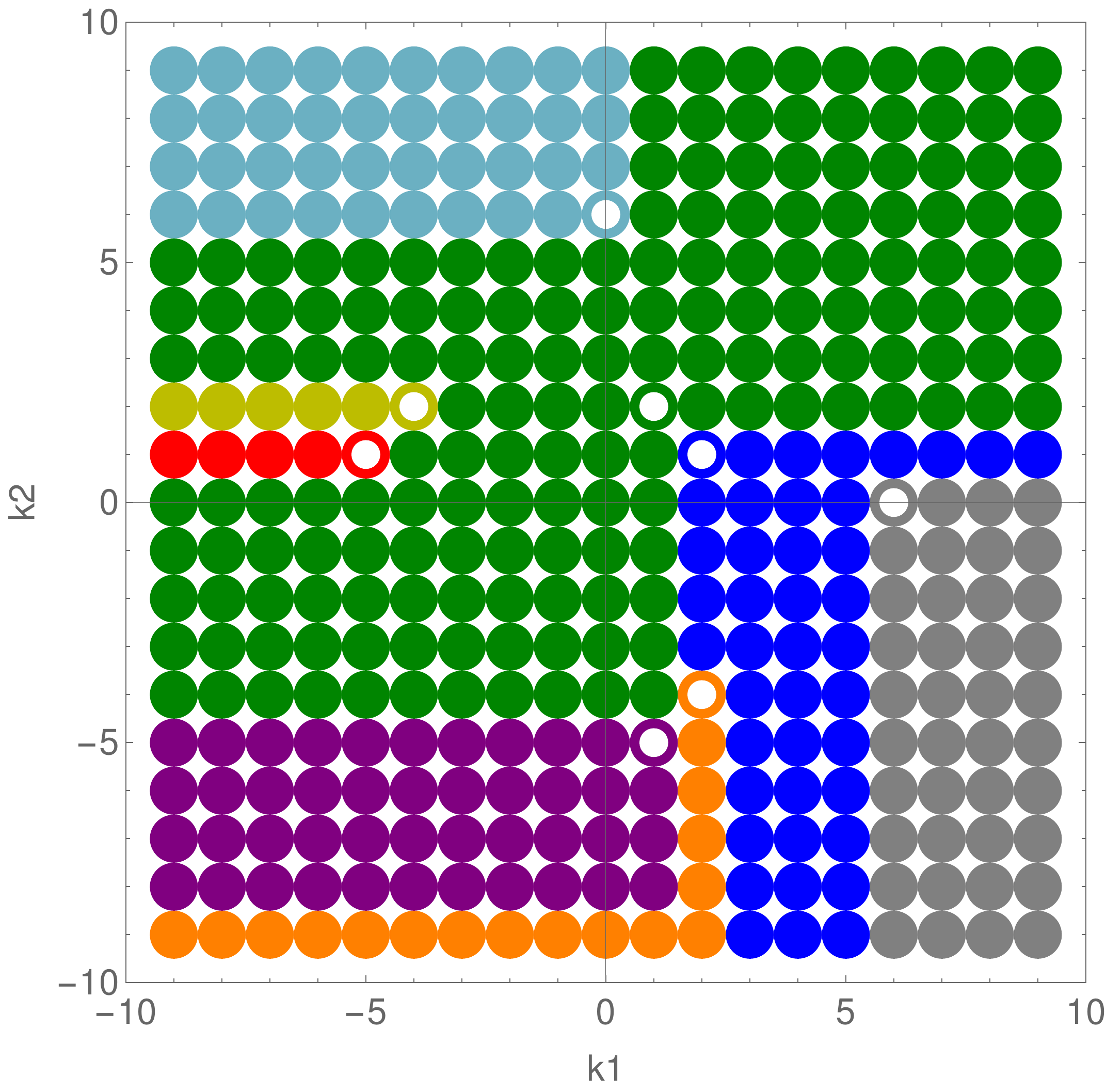}
\caption{\sf The basins of attraction for the terminal states of the line bundle environment, as encoded in the trained policy network. Each colour represents a basin of attraction with the corresponding terminal state indicated by the white circle in one corner of the region.}
\label{figLineBundleDomains}
\end{figure}

Given the small environment size we can approach this more systematically and carry out an episode guided by the training network starting from every state. In this way, we find the basin of attraction for each terminal state and the result is shown in Fig.~\ref{figLineBundleDomains}.
This figure is a more concrete version of the schematic Fig.~\ref{Fig:SearchSpace}. It shows that the basins of attraction can be of different sizes and can have complicated shapes. It is also interesting to note that the network does not always guide to the nearest terminal state. Fig.~\ref{figLineBundleDomains} is an illustration of the phenomenon of ``thickening" of states discussed earlier. We only need to carry out one episode guided by the trained network starting in each domain to find all the terminal states. This compares favourably with a systematic scan of the entire environment. 
	
In summary, training for the line bundle environment is quite successful and leads to a policy network which efficiently guides to the states with the desired target index for all starting points. This provides a first indication that RL is a suitable method for engineering topological quantities.

\section{Learning heterotic monads bundles}\label{secMonadRL}
We now move on to our main interest, namely applying RL to an environment of monad bundles. One of the properties which needs to be incorporated in this context is the index of the bundle and the experience with the previous line bundle environment suggests how to accomplish this.

We concentrate on $SU(4)$ monad bundles on two of the CICY three-folds from Table~\ref{tab:CYs}, namely the bicubic CY, $X_{7884}$, and the triple trilinear CY, $X_{7669}$. Both of these admit freely acting $\IZ_3\times \IZ_3$ discrete symmetries at certain loci in complex structure moduli space and can therefore be used for $SO(10)$ model building. We expect our methods can be applied to all manifolds in Table~\ref{tab:CYs} and quite likely other classes of manifolds and types of bundles as well, but our purpose here is not to be exhaustive. Rather, we would like to show that RL can successfully engineer string models with prescribed properties in a context where systematic scans would fail.

\subsection{The set-up}\label{secSetupMonadRL}
On a CY three-fold $X$ with Picard number $h=h^{1,1}(X)$, our environment consists of monad bundles of the form~\eqref{MonadSeqence} given by two line bundle sums $(B,C)$, with fixed ranks $r_B={\rm rk}(B)$ and $r_C={\rm rk}(C)$ such that $r_B-r_C=4$. We also build in the Chern class condition $c_1(V)=0$, that is, we restrict the environment to pairs $(B,C)$ with $c_1(B)=c_1(C)$. Concretely, we think of these states as $h\times (r_B+r_C)$ integer matrices $(B,C)=({\bf b}_1,\ldots ,{\bf b}_{r_B},{\bf c}_1,\ldots ,{\bf c}_{r_C})=(b_i^k,c_a^k)$, where each column corresponds to a line bundle. In practice, the size of these integers has to be limited to a certain range, so the environment is defined as
\begin{equation}
 \mathcal{S}=\left\{({\bf b}_1,\ldots ,{\bf b}_{r_B},{\bf c}_1,\ldots ,{\bf c}_{r_C})\,|\, b_{\rm min}\leq b_i^k\leq b_{\rm max},\; c_{\rm min}\leq c_a^k\leq c_{\rm max},\ \sum_{i=1}^{r_B}{\bf b}_i=\sum_{a=1}^{r_C}{\bf c}_a\right\}
\end{equation} 
The actions should amount to a ``minimal"modification of a state $(B,C)$ but also have to leave the condition $c_1(B)=c_1(C)$ intact. This can be accomplished by adding $\pm 1$ to one entry of a line bundle in $B$ and perform the same action in the same row for a line bundle in $C$, while keeping all other line bundles unchanged. Hence, the action space is of the form
\begin{equation}
 \mathcal{A}=\{{\bf b}_i\mapsto {\bf b}_i\pm {\bf e}_k,\; {\bf c}_a\mapsto{\bf c}_a\pm {\bf e}_k\,|\, i=1,\ldots ,r_b,\; a=1,\ldots ,r_C,\; k=1,\ldots ,h\}\; ,
\end{equation} 
where ${\bf e}_k$ is the $k^{\rm th}$ standard unit vector in $h$ dimensions.

To define the reward we proceed as we did for the line bundle environment and first define an intrinsic state value $v(B,C)$. In addition to the deviation of the index from the target it has several other contributors whose purpose is to incentivise the desired properties of the monad bundle. They include penalties for violating the anomaly condition, non-bundleness, equivariance and stability. Trivial line bundles appearing in both $B$ and $C$ can be dropped so we penalise their appearance in order to avoid such trivial configurations. Finally, if the monad map is too simple, the bundle $V$ might split, so the structure group is not $SU(4)$ but a non-trivial sub-group thereof. Since we would like to obtain genuine $SU(4)$ bundles such split monad bundles receive an additional penalty. The detailed form of these various contributions is given in Table~\ref{tab:intvaluemonad}.
\begin{table}[!h]
\begin{center}
\begin{tabular}{|l|l|l|}\hline
property&term in $v(B,C)$&comment\\\hline\hline
\varstr{19pt}{8pt}index match&$\displaystyle-\frac{2|{\rm ind}(V)- \tau|}{h M^3}$&$\tau=-3|\Gamma|$ is the target index,\\
&&${\rm ind}(V)$ computed from Eq.~\eqref{eq:monchern}\\\hline
\varstr{16pt}{8pt}anomaly&$\displaystyle\frac{1}{hM^2}\sum_{i=1}^h {\rm min}\left(c_{2i}(TX)-c_{2i}(V),0\right)$&no penalty if anomaly condition satisfied,\\
&&$c_{2i}(V)$ computed from Eq.~\eqref{eq:monchern}\\\hline
\varstr{12pt}{7pt} bundleness&$-(d_{\rm deg}+1)$&$d_{\rm deg}=\;$dimension of degeneracy locus\\
&&as discussed in Sec.~\ref{sec:monads}; if the degeneracy \\
&& locus is empty, $d_{\rm deg}$ is to be taken as $-1$
\\\hline
\varstr{12pt}{7pt} split bundle&$-n_{\rm split}$&$n_{\rm split}=\;$number of splits in $V$\\\hline
\varstr{14pt}{7pt} equivariance&$\displaystyle -\sum_{U\subset B,C}{\rm mod}({\rm ind}(U),|\Gamma|)$&$U$ runs over all line bundles in $B,C$\\
&&or blocks of same line bundles,\\
&&as discussed in Sec.~\ref{sec:monads}\\\hline
\varstr{12pt}{7pt}  trivial bundle&$-n_{\rm trivial}$&$n_{\rm trivial}=\;$number of trivial line bundles\\\hline
\varstr{21pt}{7pt} stability $V$&$\displaystyle- \frac{{\rm max}(0,h^0(X,B)-h^0(X,C))}{hM^3}$&tests Hoppe's criterion for $V$,\\
&&cohomologies from formulae in Sec.~\ref{sec:lbcohformulae}\\\hline
\varstr{21pt}{7pt} stability $V^*$&$\displaystyle - \frac{{\rm max}(0,h^0(X,B^*)-h^0(X,C^*))}{hM^3}$&tests Hoppe's criterion for $V^*$,\\
&&cohomologies from formulae in Sec.~\ref{sec:lbcohformulae}\\\hline
\end{tabular}
\caption{\sf Contributions to the intrinsic value for the monad environment. The intrinsic value $v(B,C)$ is the sum of all eight terms and $M={\rm max}(b_{\rm max},c_{\rm max})$.}\label{tab:intvaluemonad}
\end{center}
\end{table}
In terms of this intrinsic state value function, the reward is then defined exactly as for the line bundle environments, that is, by Eq.~\eqref{lbreward}. The various parameters in this formula are chosen as
\[
 p=1.2\;,\quad r_{\rm offset}=-2\;,\quad r_{\rm step}=-1\;,\quad r_{\rm boundary}=-2\;,\quad r_{\rm terminal}=10\; .
\] 

Ideally, the value function should include more and more sophisticated properties of $V$, such as, for example, the full particle spectrum (rather than just the chiral asymmetry) and a detailed stability check. However, including these properties would require carrying out cohomology calculations during training. Given that these calculations are currently based on commutative algebra methods this is not feasible, as it would lead to an unpredictable slow-down of the training process. Progress in this direction can be made if analytical formulae for monad cohomology can be derived, in analogy with the formulae for line bundle cohomology discussed in Section~\ref{sec:lbcohformulae}. This may well be possible but, at present, such formulae are not known. Hence, for the time being, we limit ourselves to training on the properties listed in Table~\ref{tab:intvaluemonad}. More detailed calculations of the spectrum and stability checks will only be carried out after training and for the terminal states found by the RL system. 

The monad environment has considerable degeneracy. Permutations of the line bundles in $B$ and $C$ of course do not change the monad bundle, so we have a permutation symmetry $S_{r_B}\times S_{r_C}$. Also, depending on the underlying manifold, there can be an additional discrete symmetry $H\subset S_h$ which permutes the rows of $(B,C)$, so the total symmetry group is
\begin{equation}\label{eq:redsymm}
 H\times S_{r_B}\times S_{r_C}
\end{equation} 
In fact, for the bicubic CY, we have $H=S_2$ and for the triple trilinear CY, $H=S_3$.

Finally, we should add a comment on how we sample the initial states for episodes. Naively, one might choose a flat distribution on the environment to choose these states. However, model building experience shows that successful models tend to have entries which are relatively small (typically $0$, $\pm 1$ or $\pm 2$). Since we are trying to match relatively small numbers, such as the number of families, this is perhaps not surprising. For this reason it is helpful to choose a distribution which favours small entries $k$ in initial states $(B,C)$ and we have opted for 
\[
  P(k)\sim\frac{1}{1+|k|^2}\; .
\]  

This environment is realised as a MATHEMATICA package which is coupled to either the REINFORCE or the actor-critic package. The policy network (as well as the value network in the actor-critic case) are fully connected neural networks of the type shown in Fig.~\ref{fig:nn}. The input dimension is $d_0=h(r_B+r_C-1)$, the number of independent entries in a state $(B,C)$ subject to the contraint $c_1(B)=c_1(C)$, and the output dimension is $d_1=2h(r_B+r_C)$, the size of the action space. For the network width we have chosen $d=64$. We use the ADAM optimiser with a learning rate of $1/3500$. The maximal episode length is $t_{\rm max}=32$, batch sizes are $64$ and the discount factor is set to $\gamma=0.98$ for REINFORCE and to $\gamma=0.6$ for actor-critic.

\subsection{Results on the bicubic with $(r_B,r_C)=(6,2)$}\label{secResBC62}
Our first example is for the bicubic CY with ranks $(r_B,r_C)=(6,2)$ and entries in the range
\begin{equation}\label{eq:bcrange}
 -3=b_{\rm min}\leq b_i^k\leq b_{\rm max}=5\;,\qquad 0=c_{\rm min}\leq c_a^k\leq c_{\rm max}=5\; ,
\end{equation} 
which amounts to an environment with about $10^{14}$ states. This is already quite sizeable and, given that monad bundles do not allow for simplifications such as checks carried out for each line bundle, a systematic scan of this environment is not feasible. Moreover, terminal states are very rare; for instance, by randomly sampling $10^9$ models, no terminal state are typically found.

\begin{figure}
\centering
        \subfloat[\sf Loss vs batch number.]{\includegraphics[width=0.3\linewidth]{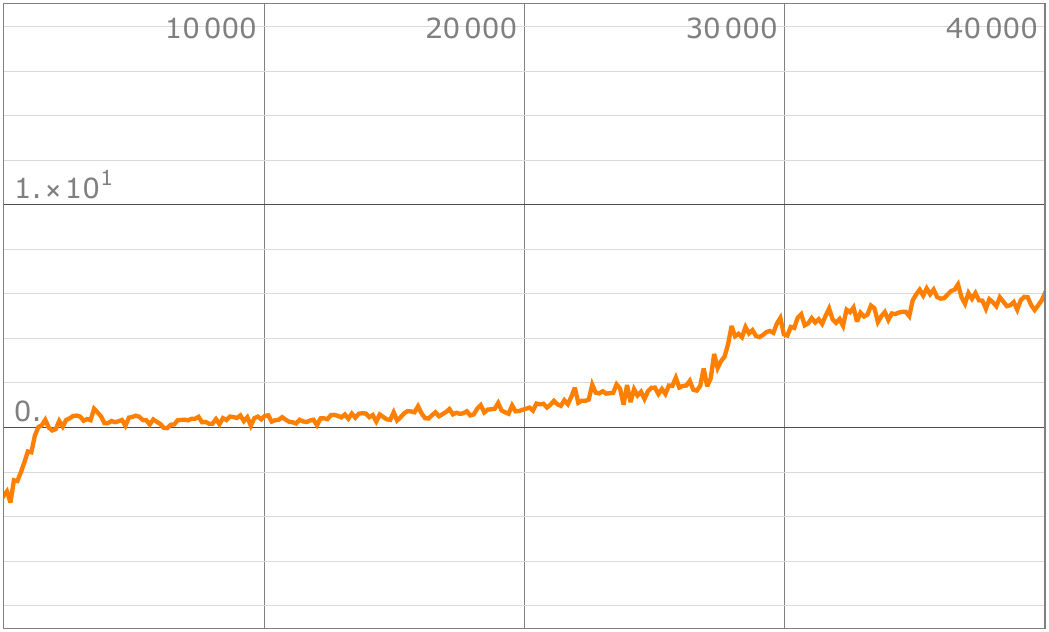}}\qquad
        \subfloat[\sf Policy loss vs batch number.]{\includegraphics[width=0.3\linewidth]{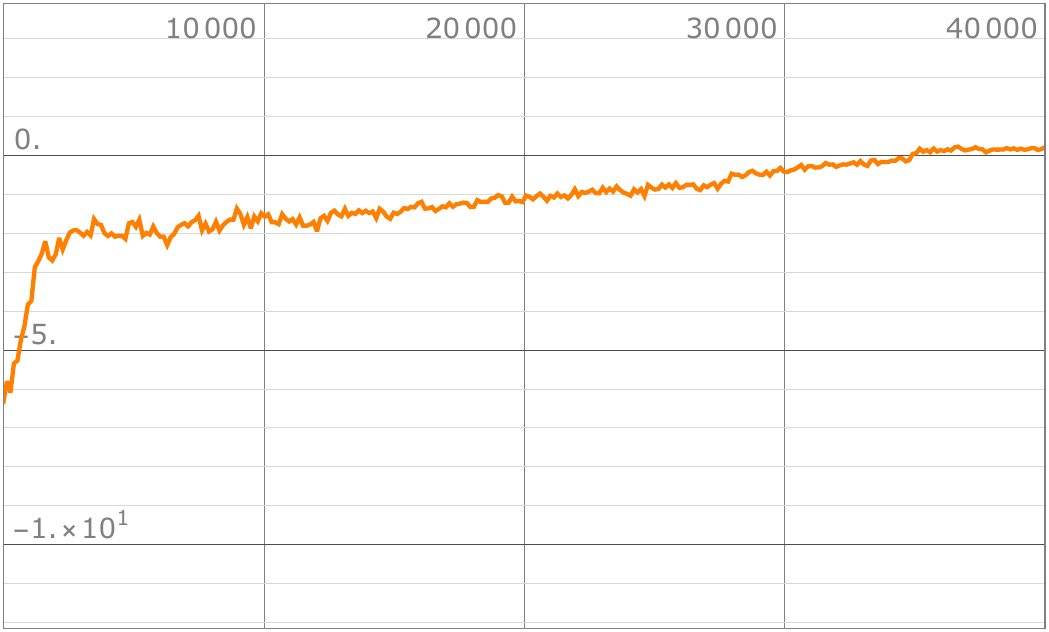}}\qquad
        \subfloat[\sf Value loss vs batch number.]{\includegraphics[width=0.3\textwidth]{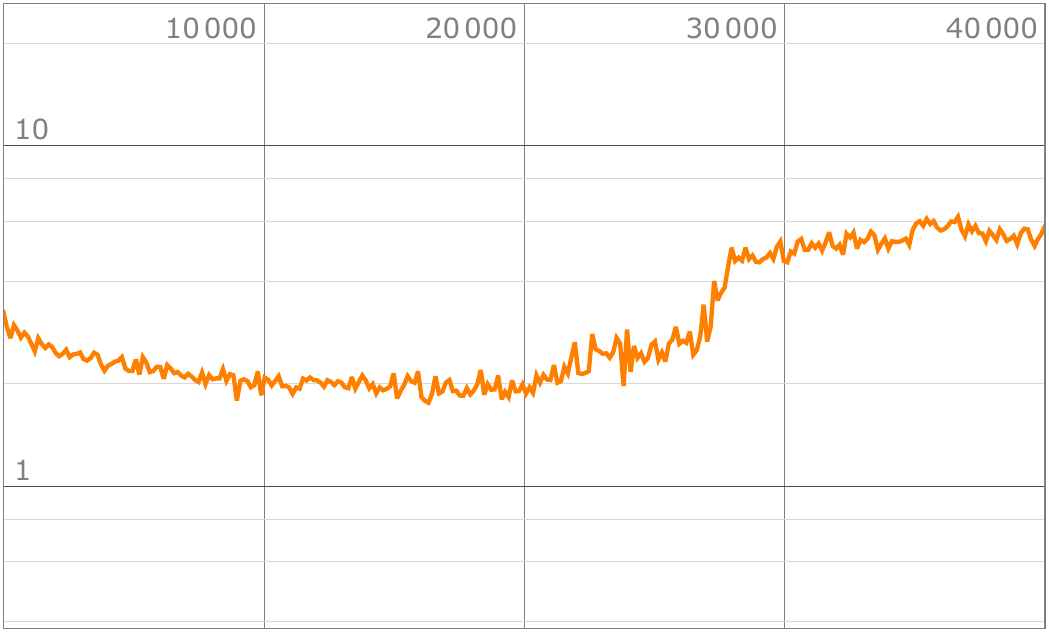}}\\%
        \subfloat[\sf TD return vs batch number.]{\includegraphics[width=0.3\textwidth]{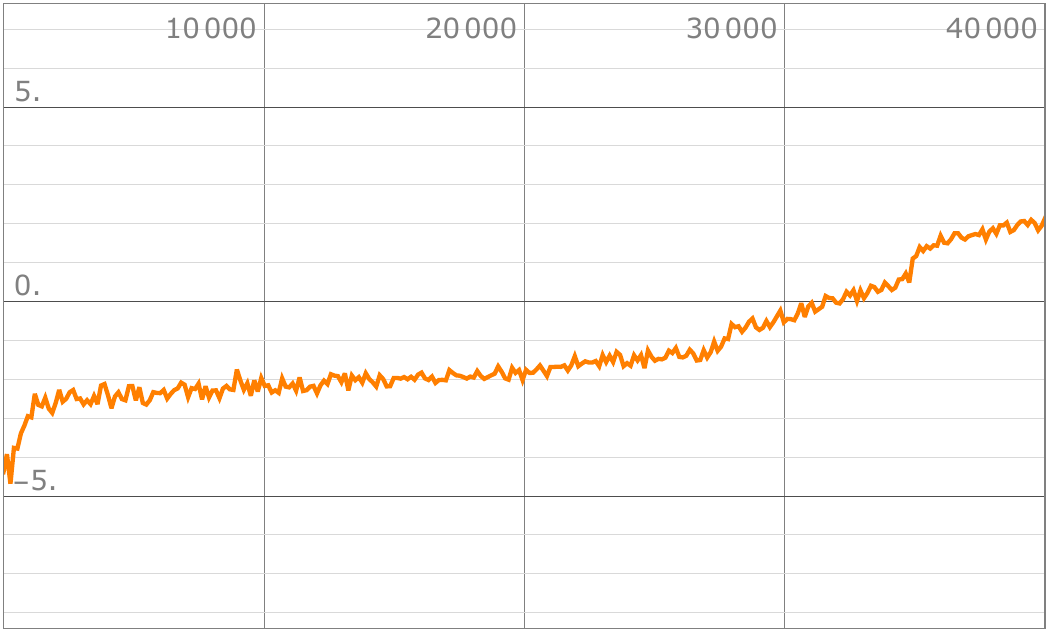}}\qquad
        \subfloat[\sf Number of terminal states vs episode number.]{\includegraphics[width=0.3\textwidth]{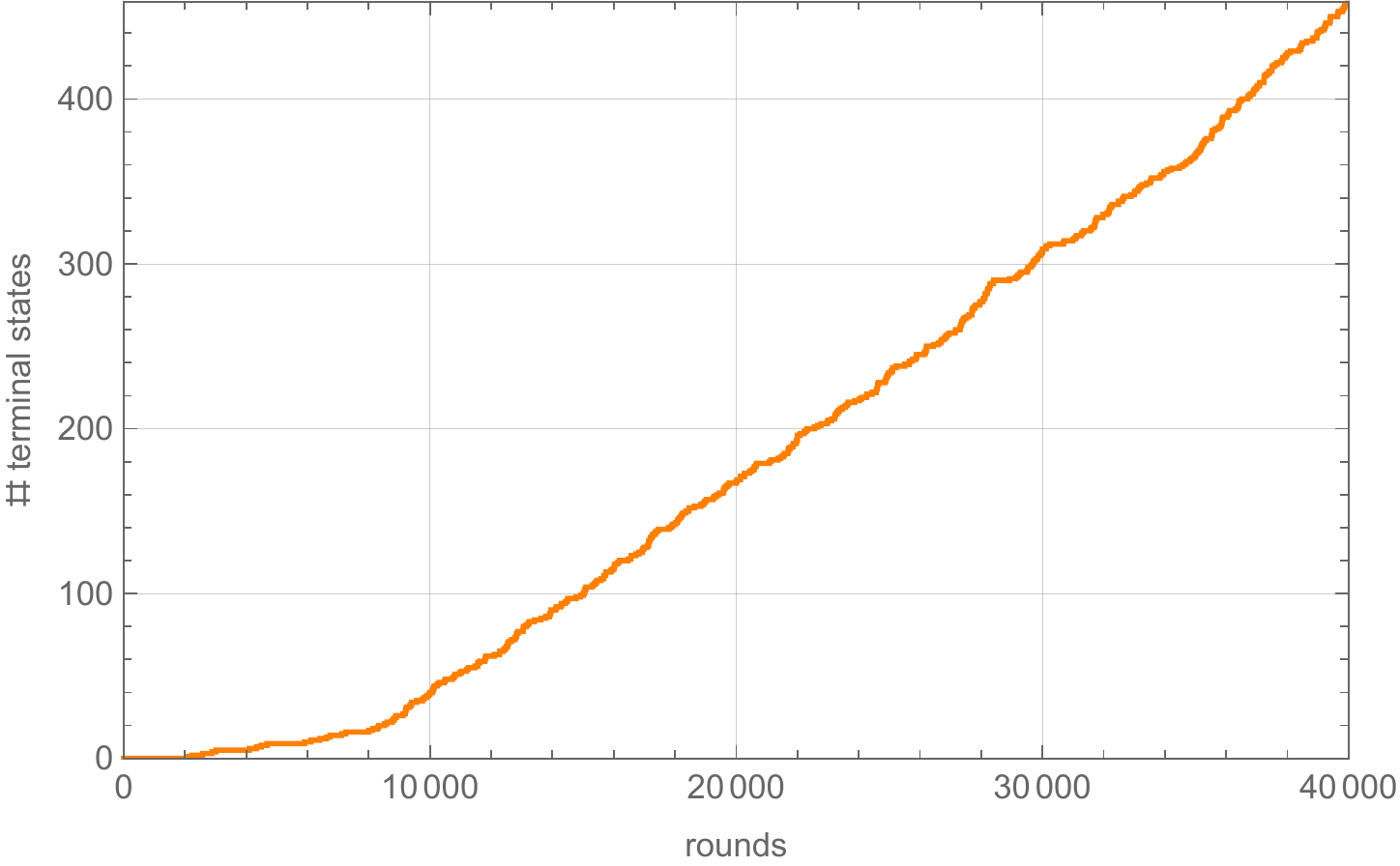}}\qquad
        \subfloat[\sf Terminal fraction vs episode number.]{\includegraphics[width=0.3\textwidth]{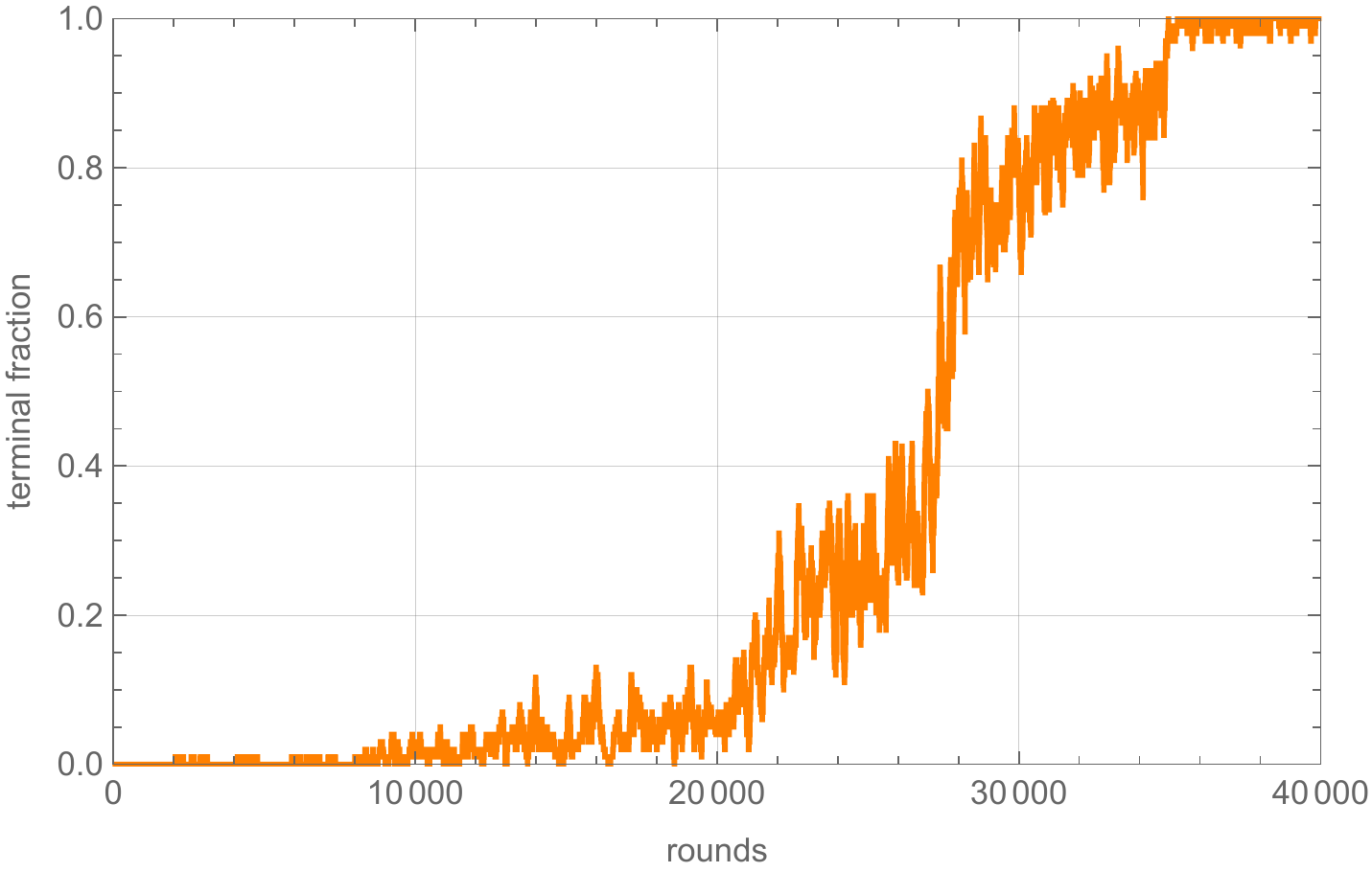}}%
        \caption{\sf Training metrics for the bicubic monad environment with $(r_B,r_C)=(6,2)$.}
        \label{figMonadBicubic62Training}
\end{figure}

In the following we present the results we have obtained with the actor-critic algorithm but results for REINFORCE are, in fact, quite similar.
The measurements taken during $40000$ rounds of training (about an hour on a single CPU) are shown in Fig.~\ref{figMonadBicubic62Training}. The most impressive indicator is Fig.~\ref{figMonadBicubic62Training}(f) which shows the fraction of terminal episodes. At about $30000$ rounds this fraction quickly rises to a value close to $1$, showing that every starting state is guided to a terminal state and it turns out that this happens within $19$ steps on average.  We note that this is achieved by sampling only a tiny fraction of about $\sim 10^{-8}$ of the environment's states. During training a few hundred terminal states are found (see Fig.~\ref{figMonadBicubic62Training}(e)). After removing redundancies due to the symmetry~\eqref{eq:redsymm} this number reduces to $59$ terminal states.

\begin{figure}[h]
\centering
\includegraphics[width=0.7\textwidth]{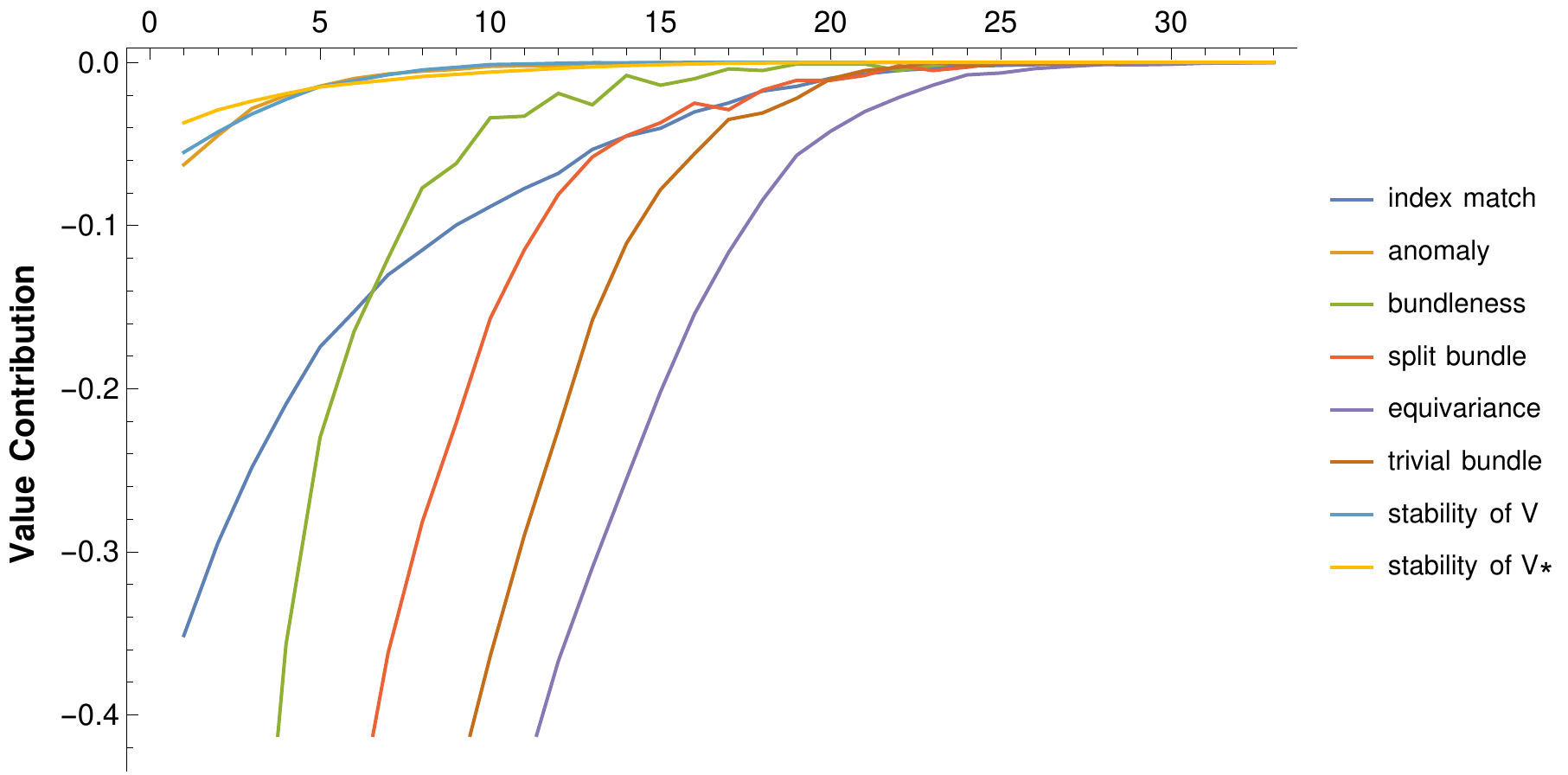}
\caption{\sf The different contribtutions to the intrinsic value for $(r_b,r_c)=(6,2)$ bicubic models. This data is averaged over 1000 termianl states using the trained network.}
\label{figValueContributions}
\end{figure}

The trained network can be used to search for further terminal states. Running about $1000$ episodes from random starting points and guided by the trained network two more terminal states are found (after removing redundancies) which brings the total to 61. The episodes generated by the trained network have, in general, a standard order in which the various requirements for a terminal state are satisfied. Specifically the network tries to satisfy the anomaly cancelation and the rudimentary stability conditions first. The network consistently adjust for the equivariance and the index conditions last (though sometimes joint last). This may be considered an indication to how a model builder may try to construct such a model with pen and paper. This is demonstrated in Fig.~\ref{figValueContributions}, by considering $1000$ terminal states generated using the trained network.

The $61$ monad bundle models found in this RL scan have been further checked. We found that $47$ of these have $h^1(X,V)=27$ families before taking the $\IZ_3\times \IZ_3$-quotient and no anti-families, that is, $h^2(X,V)=0$. Further stability checks, testing the injection of line bundles with entries in the range $-6,\ldots ,6$ into $V$ and $V^*$, shows that many of these models are unstable. However, $18$ models survive these fairly extensive checks. These models, which are listed in Appendix~\ref{appbicubic62}, include the new model presented in Section~\ref{secExample} as well as the semi-positive monad with $B = \cO_X(0,1)^{\bigoplus 3} \oplus\cO_X(1,0)^{\bigoplus 3}$ and $C = \cO_X(1,1) \oplus\cO_X(2,2)$  found in Ref.~\cite{Anderson:2009mh}. In fact, the latter model is the only semi-positive monad found by the network, so it is likely the only semi-positive standard model with $(r_B,r_C)=(6,2)$ on the bicubic.
	
\subsection{Results on the bicubic with $(r_B,r_C)=(7,3)$}\label{secResBC73}
Raising our ambition moderately higher, we consider next a monad environment on the bicubic with larger ranks, $(r_B,r_C)=(7,3)$, but with a range of integer entries as in Eq.~\eqref{eq:bcrange}, leading to about $10^{16}$ states. Both REINFORCE and actor-critic algorithms lead to similar results and, for definiteness, we present the latter.

\begin{figure}
 \centering
        \subfloat[\sf Loss vs batch number.]{\includegraphics[width=0.3\linewidth]{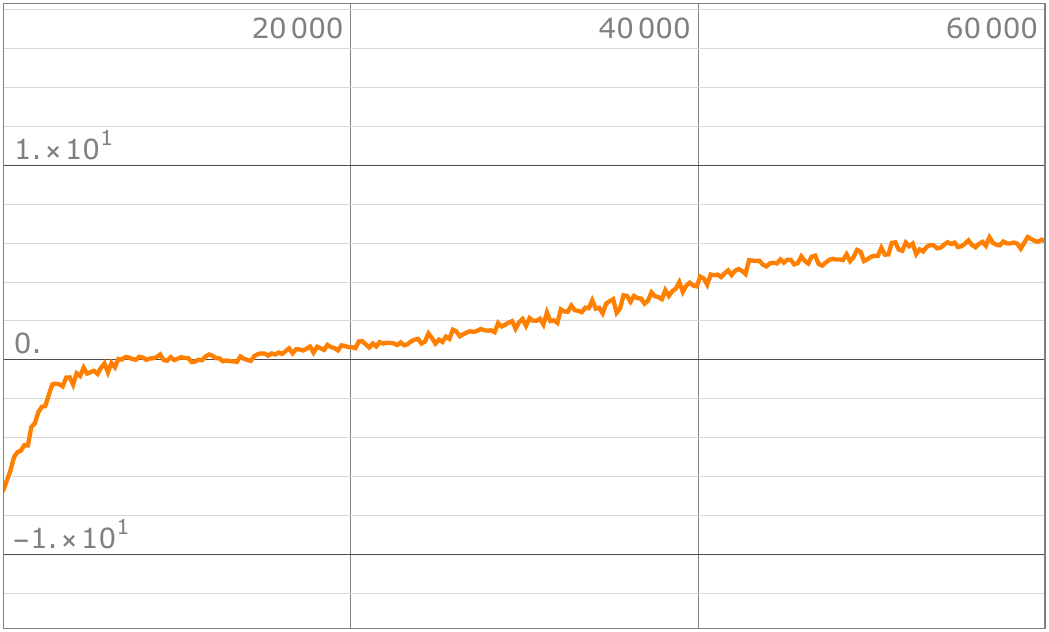}}\qquad
        \subfloat[\sf Policy loss vs batch number.]{\includegraphics[width=0.3\linewidth]{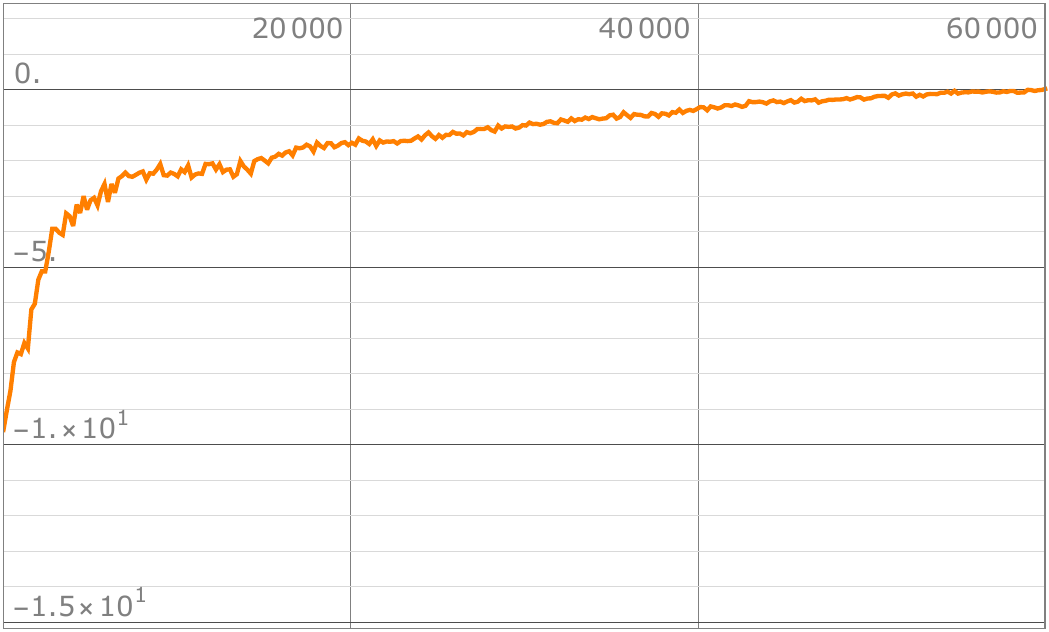}}\qquad
        \subfloat[\sf Value loss vs batch number.]{\includegraphics[width=0.3\textwidth]{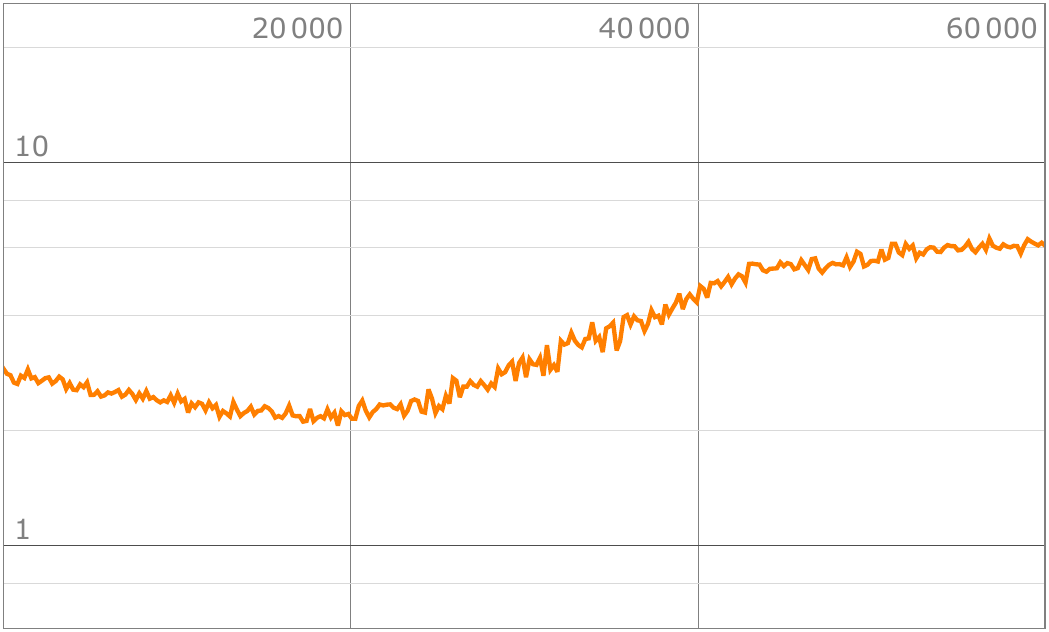}}\\ 
        \subfloat[\sf TD return vs batch number.]{\includegraphics[width=0.3\textwidth]{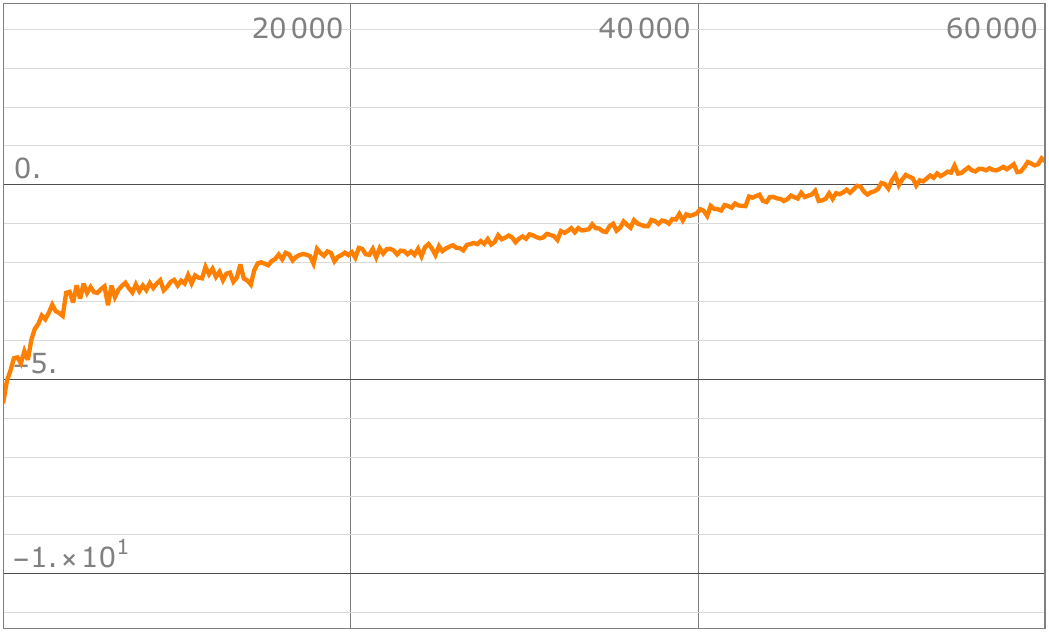}}\qquad
        \subfloat[\sf Number of terminal states vs episode number.]{\includegraphics[width=0.3\textwidth]{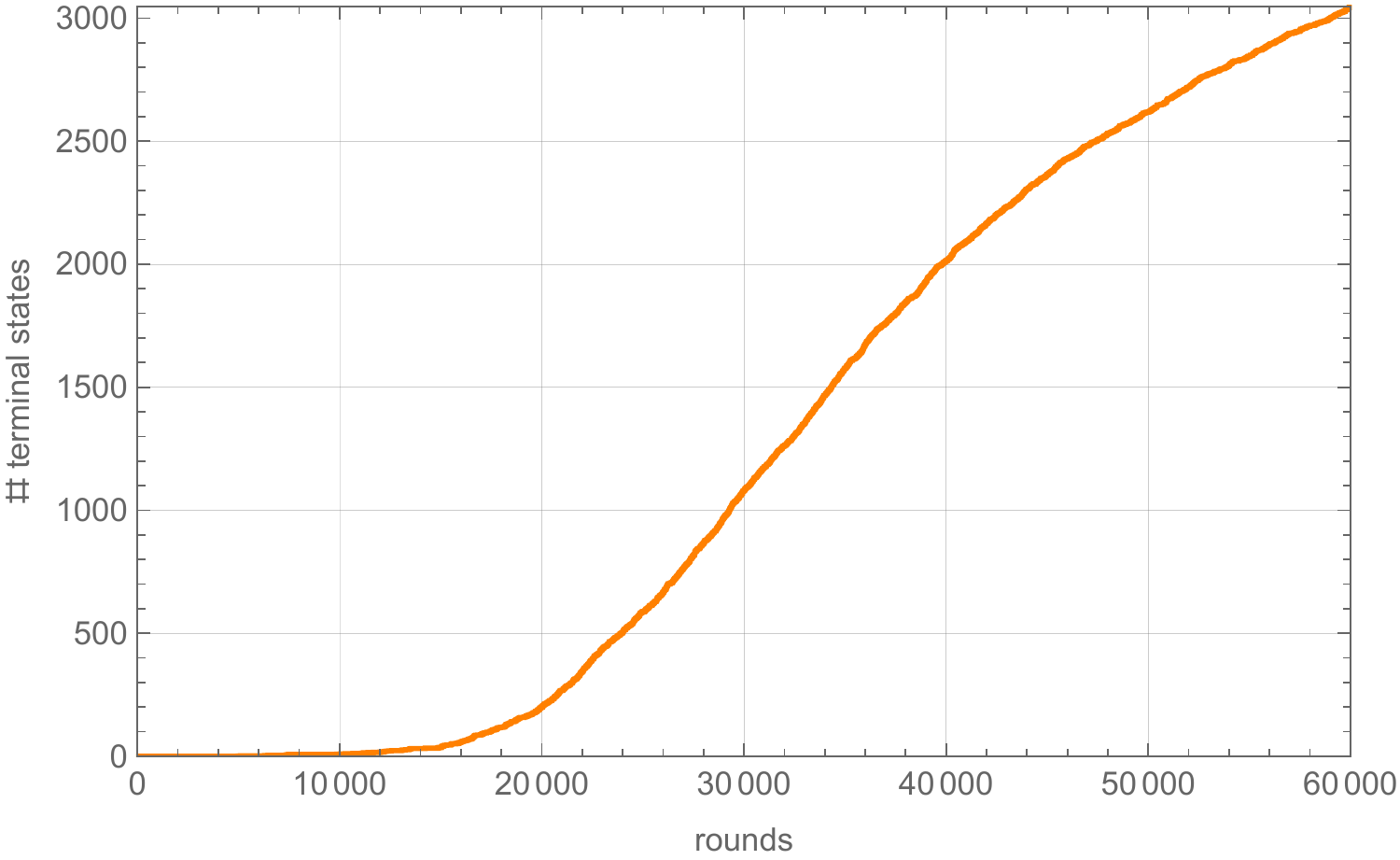}}\qquad
        \subfloat[\sf Terminal fraction vs episode number.]{\includegraphics[width=0.3\textwidth]{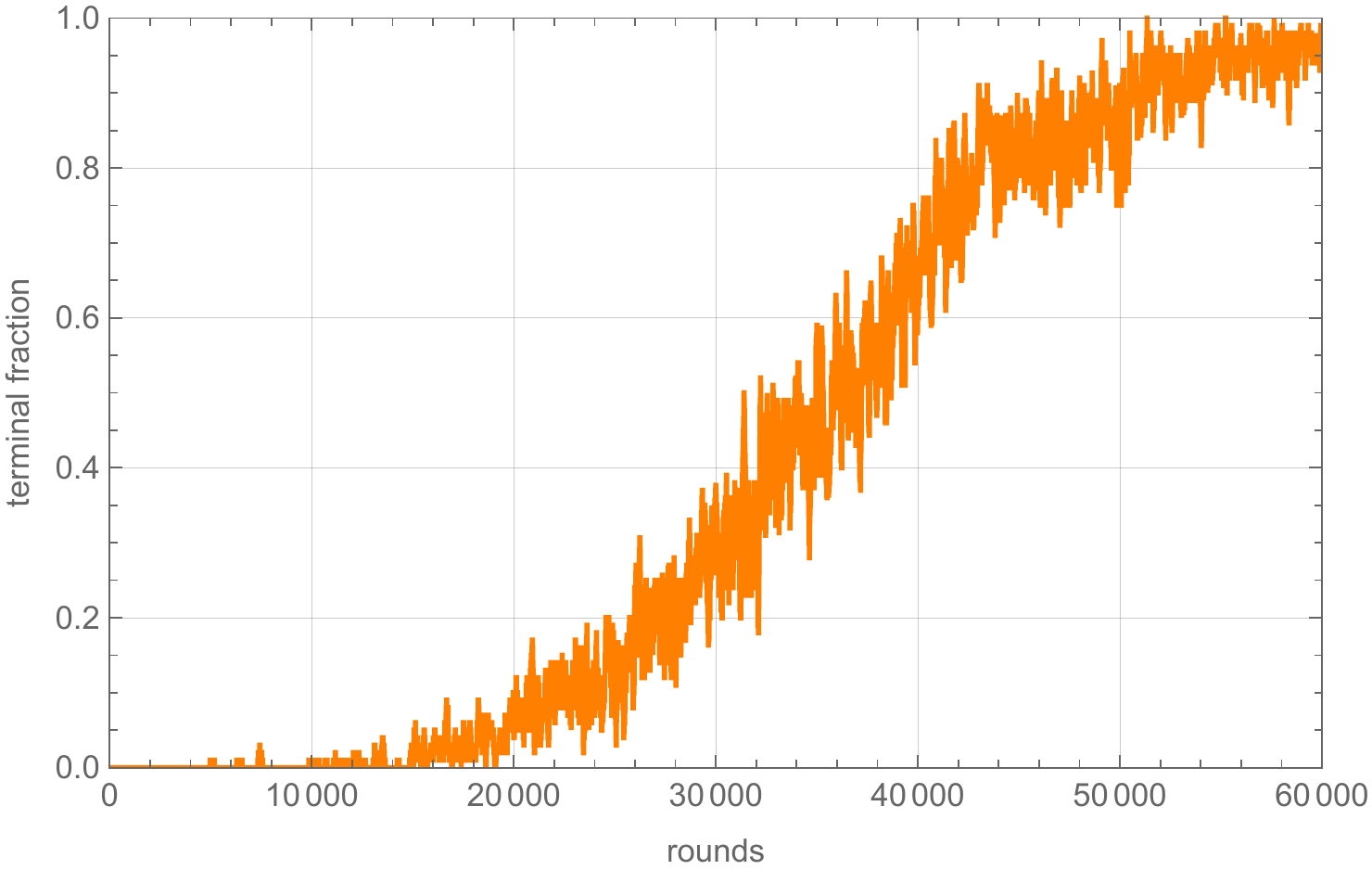}}
        \caption{\sf Training metrics for the bicubic monad environment with $(r_B,r_C)=(7,3)$.}
        \label{figMonadBicubic73Training}
\end{figure}

The network has been trained for $60000$ rounds, sampling only a tiny fraction of about $10^{-10}$ of the environment's states. Training takes about two hours on a single CPU and training measurements are shown in Fig.~\ref{figMonadBicubic73Training}.
As in the previous case, we observe a transition to a policy with a terminal fraction close to $1$ during the training process (Fig.~\ref{figMonadBicubic73Training}(f)), with the average episode length decreasing to about $20.5$. Fig.~\ref{figMonadBicubic73Training} shows that about $3000$ terminal states are found during training. After removing redundancies, this number reduces to $141$, of which $129$ have $27$ families and no anti-families. It is worth mentioning that $101$ of these models have a common line bundle in $B$ and $C$ and are, hence, equivalent to models with $(r_B,r_C)=(6,2)$. A further $5$ models have two common line bundles in $B$ and $C$ are are equivalent to models with $(r_B,r_C)=(5,1)$.

As before, the trained network can be used to find more terminal states. After running $1000$ episodes guided by the trained network, removing redundancies and checking the cohomology for the terminal states, one more model is found, bringing the total number to $130$. A stability check, looking at injection of line bundles with entries between $\pm 5$ into $V$ and $V^\star$, rejects $61$ of these models as unstable, leaving $69$ models which pass. These models are listed in Appendix~\ref{appbicubic73}.
	
\subsection{Results on the triple trilinear manifold with $(r_B,r_C)=(6,2)$}\label{secResTTL62}
Finally, we consider monads on the triple tri-linear CY $X_{7669}$ (see Table~\ref{tab:CYs}) with line bundle sum ranks $(r_B,r_C)=(6,2)$ and entries bounded as before, see Eq.~\eqref{eq:bcrange}. Since the Picard number is $h=3$, the environment is much larger than for the previous example  and contains about $10^{20}$ states. As in the previous case, the manifold admits a freely acting $\IZ_3\times\IZ_3$ symmetry, hence we are looking for models with $27$ families. 

\begin{figure}
\centering
        \subfloat[\sf Loss vs batch number.]{\includegraphics[width=0.3\linewidth]{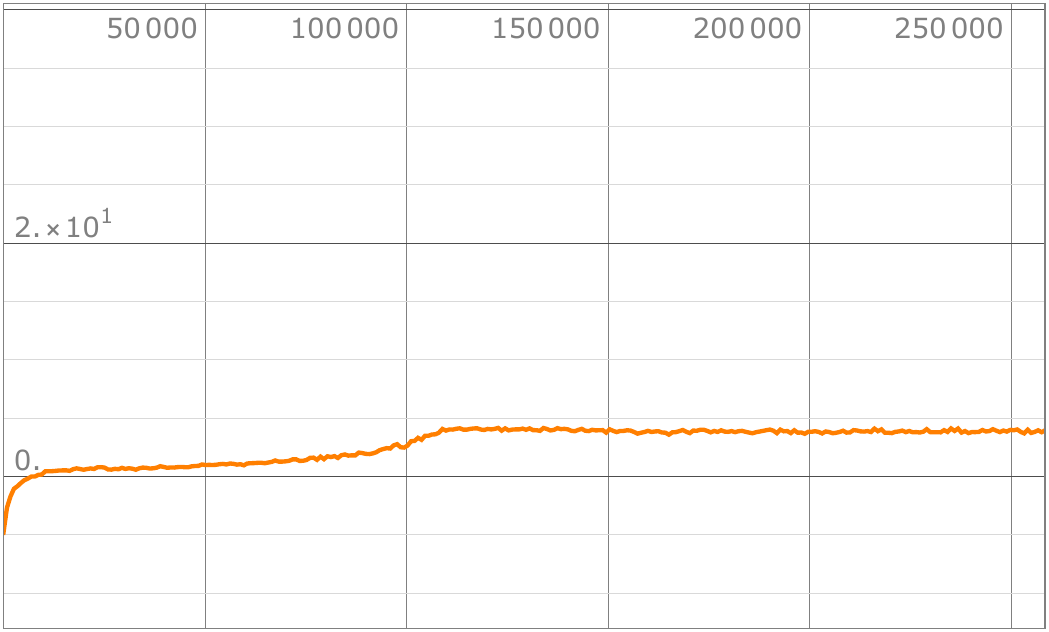}}\qquad
        \subfloat[\sf Policy loss vs batch number.]{\includegraphics[width=0.3\linewidth]{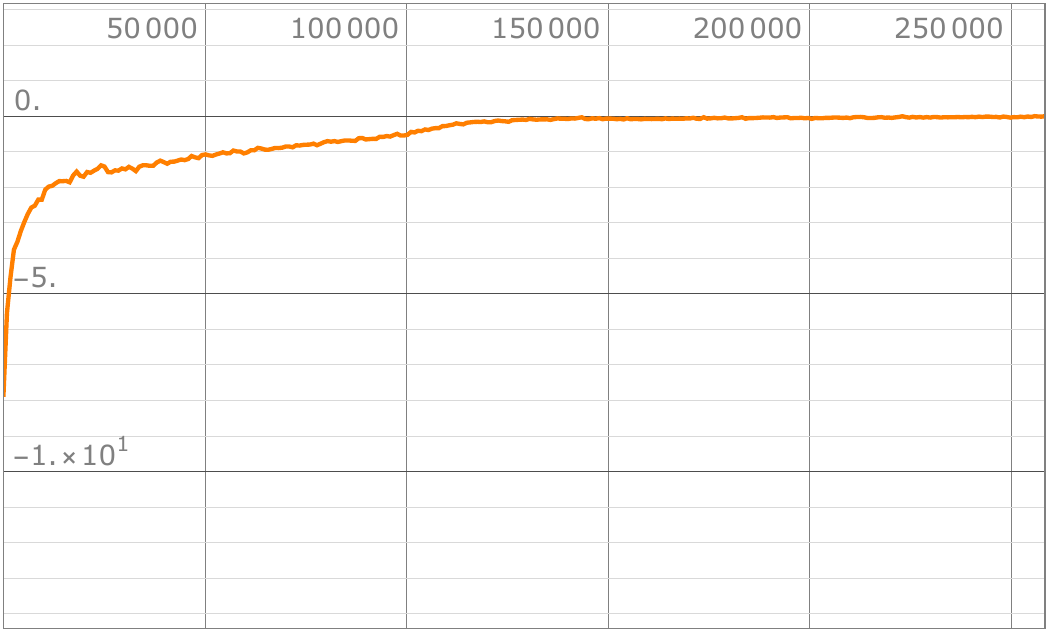}}\qquad
        \subfloat[\sf Value loss vs batch number.]{\includegraphics[width=0.3\textwidth]{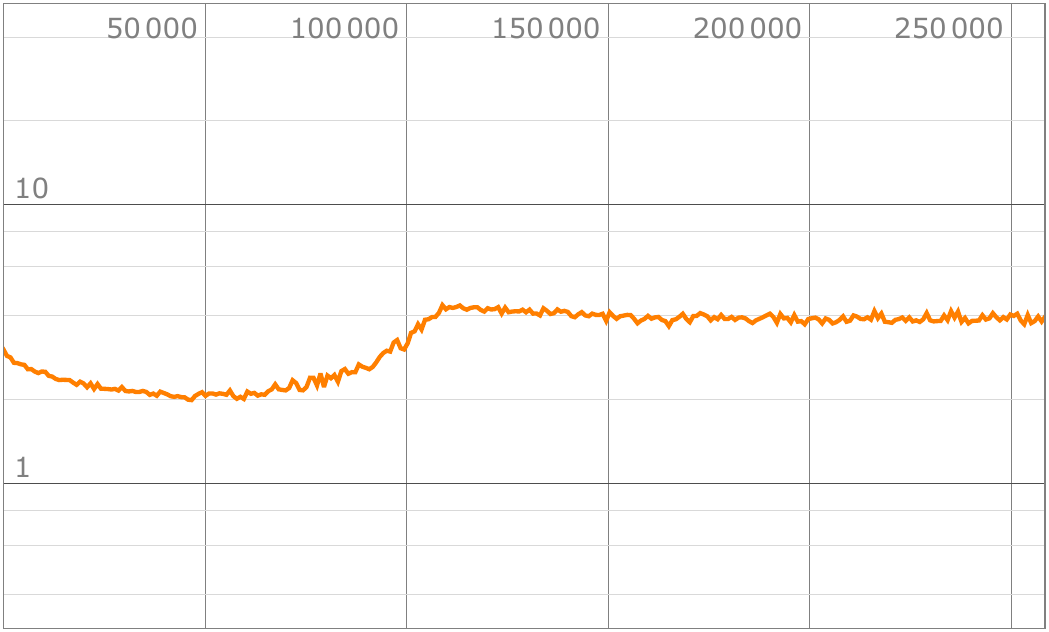}}\\%
        \subfloat[\sf TD return vs batch number.]{\includegraphics[width=0.3\textwidth]{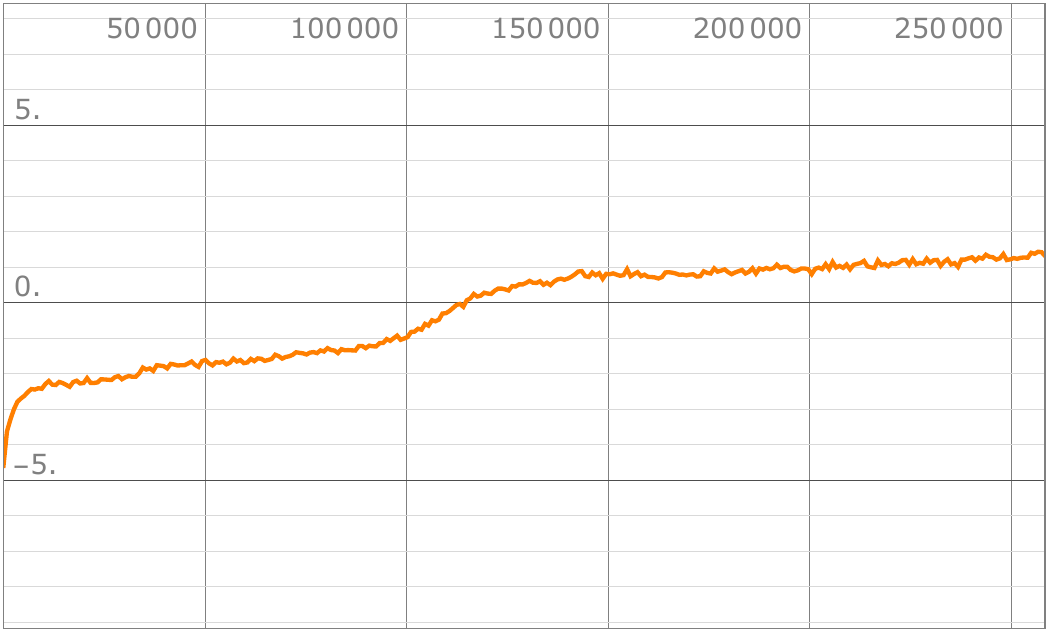}}\qquad
        \subfloat[\sf Number of terminal states vs episode number.]{\includegraphics[width=0.3\textwidth]{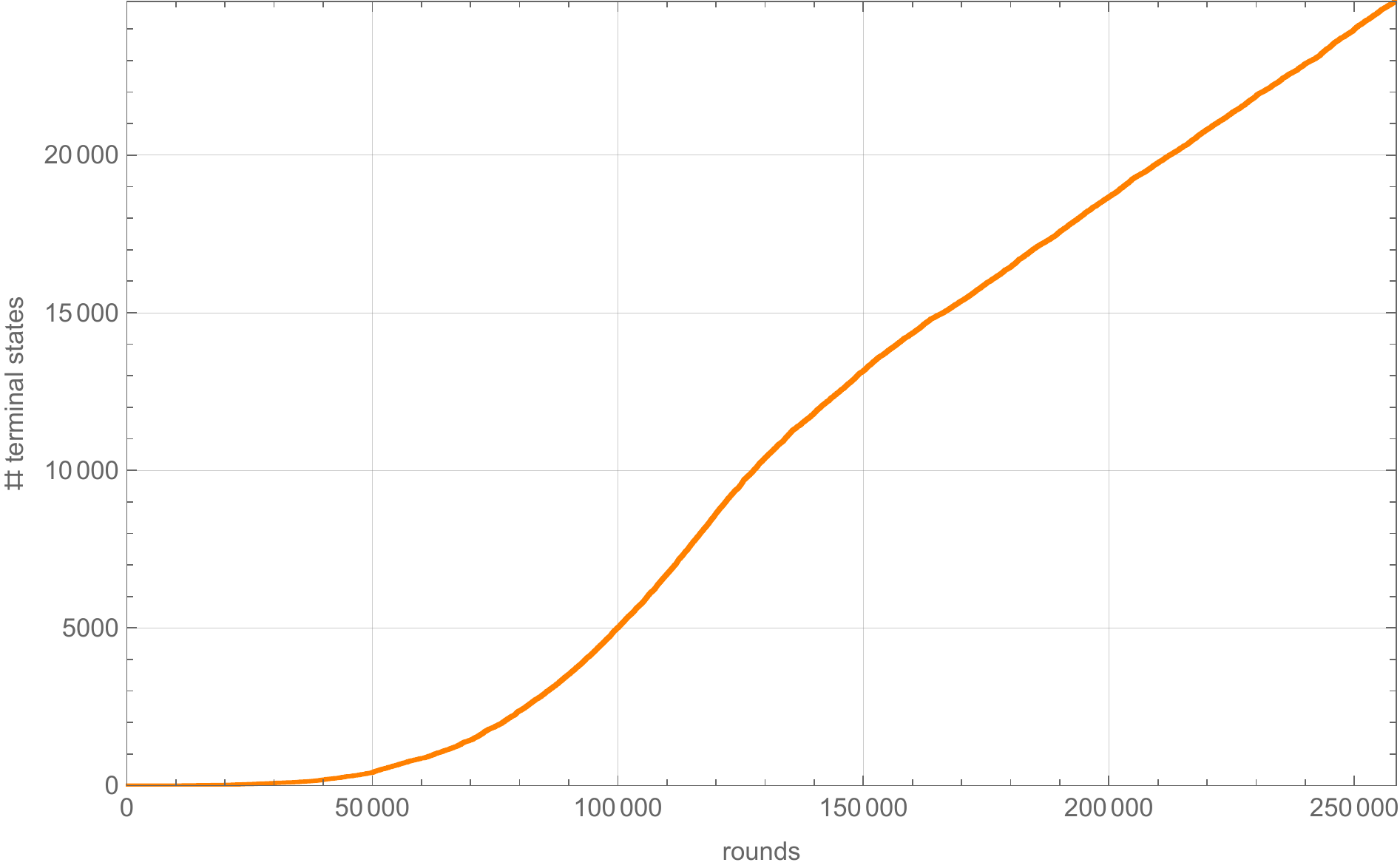}}\qquad
        \subfloat[\sf Terminal fraction vs episode number.]{\includegraphics[width=0.3\textwidth]{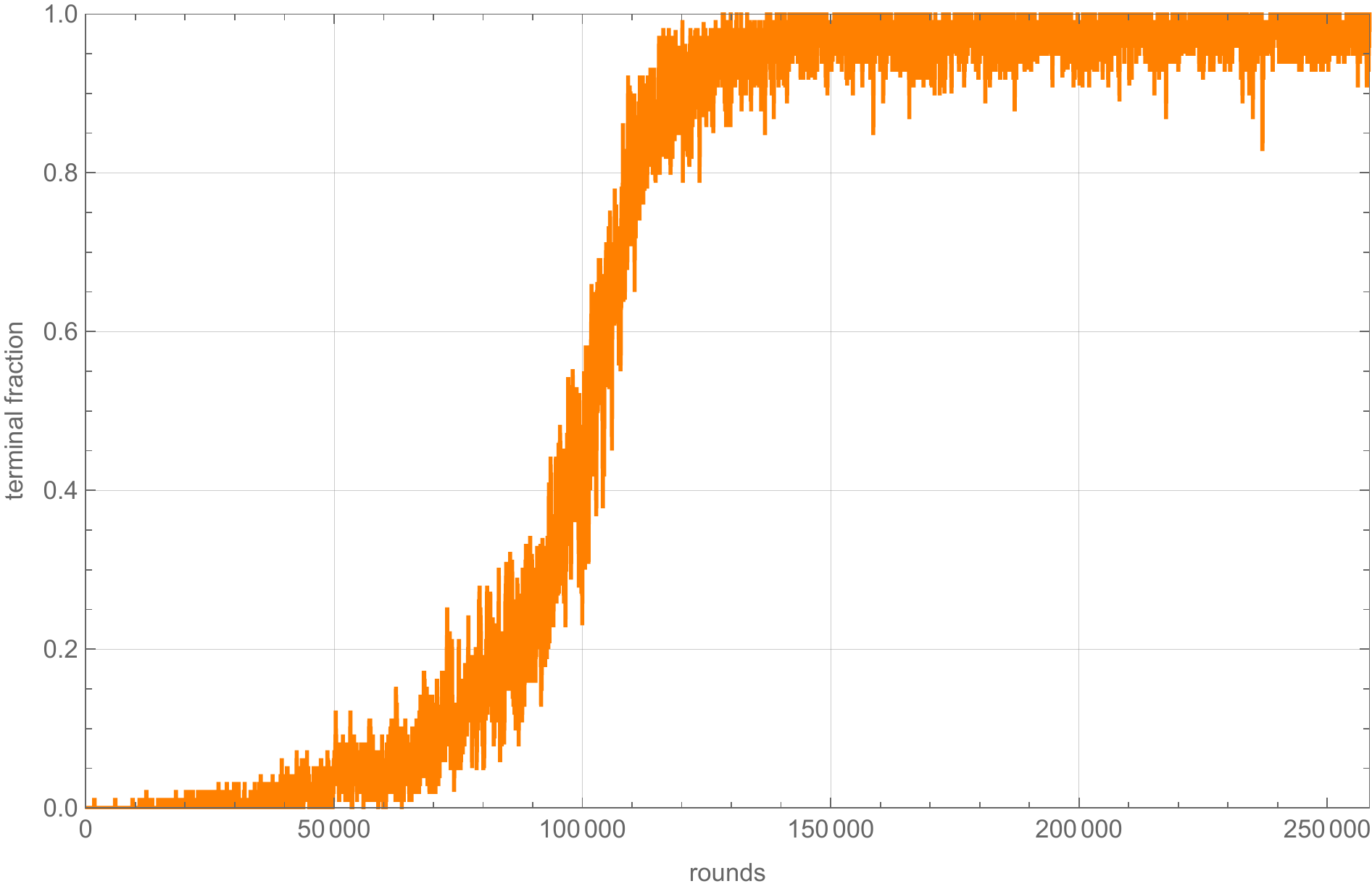}}%
        \caption{\sf Training metrics for a monad environment on the triple tri-linear CY with $(r_B,r_C)=(6,2)$.}
        \label{figMonadTTLinear62Training}
\end{figure}

For this manifold the actor-critic algorithm proves more efficient. $258,500$ rounds of training take about a day on a single CPU and training measurements are shown in Fig.~\ref{figMonadTTLinear62Training}. Only a tiny fraction of about $10^{-13}$ of the environment's states has been sampled during training.
Similar to the previous example, Fig.~\ref{figMonadTTLinear62Training}(f) shows a dramatic increase of the terminal fraction to a value close to~$1$. At the same time, the average episode length decreases to a value of about $22.5$. After removing redundancies, $12819$ terminal states are found during training. Out of these, $7638$ models have $27$ families, no anti-families and have vanishing first and third cohomologies as required by stability.

As before, the trained network can be used to find further terminal states. After running $1000$ episodes with the trained network, removing redundancies and checking for anti-families, $174$ models are found, taking the total to $7812$. This data set has been included as ancillary material in the arXiv submission. 

Performing stability checks for these models is computationally intense, due to the higher Picard number. For this reason, we have refrained from performing stability checks for the entire list of $7812$ models. Instead we have randomly selected $100$ models from this list and have checked injection of line bundle with entries in the range $\pm 3$ into $V$ and $V^*$. Of these, $7$ pass the test and these models are listed in Appendix~\ref{apptriptri}. This result points to about $\cO(500)$ stable models in the entire set. 
	
\section{Conclusions}\label{secConclusion}
In this paper, we have demonstrated that reinforcement learning (RL) can be used to engineer geometric string backgrounds with prescribed properties and that it provides an efficient method to explore large sets of geometries which defy systematic scanning.
	
All results have been obtained with modest computing resources, specifically a single CPU with running times from a few minutes to at most a day. A characteristic feature in all the studied examples has been the explosive increase of the fraction of terminal episodes to a value close to $1$ during training. This means that the trained networks leads to a terminal state from virtually any starting point in the environment within a small number of steps,  in line with the estimate~\eqref{eq:straightpath}. In this way the environment splits into basins of attractions for the various terminal states, a structure most clearly illustrated by Fig.~\ref{figLineBundleDomains} which shows the basins of attraction of a simple two-dimensional line bundle environment.	Finding all terminal states then amounts to finding one starting point in each basin of attraction. If basins of attraction are roughly the same size this scales with the number of terminal states rather than the total number of states in the environment, as would be the case for a systematic scan. 

This feature explains why RL can be efficient at exploring large spaces and we have shown RL is capable of exploring the sizeable environments realised by monad bundle environments on Calabi-Yau (CY) three-folds. In the process, many physically promising heterotic string compactifications are found, most of them not previously known. Specifically, the RL system has found about $80$ new potential standard models on the bicubic CY and a few hundred potential standard models on the triple tri-linear CY.  In either case, the environment is too large and the desirable states are too sparse to find these models by systematic scans or by random search. 

There are many extensions of the present work. The present method can be used for a systematic analysis of an entire class of string models. The class of heterotic monad bundles leading to $SO(10)$ GUT theories would be an interesting starting point. It is based on a relatively small number of CY manifolds with large freely-acting symmetry group, basically the manifolds listed in Table~\ref{tab:CYs}, which fits well with our manifold by manifold approach. But it is likely that suitable variants of our set-up can be used for other classes of models, such as heterotic models with $SU(5)$ bundles or F-theory models. 

The number of heterotic flux choices is controlled by the Picard number $h=h^{1,1}(X)$ of the CY manifold or, more generally, by the CY Hodge numbers. Systematic scans become impossible for large $h$ and, as our examples have shown, may not even be feasible for small $h$. Can RL be used to explore the full space of string models, including at large $h$? As we have seen, the number of monad bundles is of the order of $10^{h(r_B+r_C-1)}$ (where $r_B+r_B-1\geq 5$ for realistic models) whereas the number of models with the correct chiral spectrum is much smaller, probably roughly of the order $10^h$. Hence, RL searches should scale with $10^h$, provided most basins of attraction are sufficiently large. Our successful searches provide modest support that this is so, but clearly this has to be checked for $h>3$. Even then, finding and storing $10^h$ models is not feasible for even moderately large values of $h$. What is required is a more sophisticated RL system with a reward function which includes refined model properties, such as the full spectrum or Yukawa couplings, which selects on a smaller class of models. For some of these properties, this will require theoretical progress which allows for computations which are sufficiently fast for the purposes of RL. For properties such as the full spectrum or bundle stability analytical formulae for cohomology dimensions might well prove to be the crucial ingredient. It is conceivable that combining such theoretical progress with RL methods and substantial computing power allows for a systematic exploration of the entire string landscape.
	
\section*{Acknowledgements}
A.~C. is supported by a Stephen Hawking Fellowship, EPSRC grant EP/T016280/1, and T.~R.~H is supported by an STFC studentship. 

\newpage	
\appendix
\section{Example models with $(r_B,r_C)=(6,2)$ on the bicubic}\label{appbicubic62}

The table below contains the monad bundles with ranks $(r_B,r_C)=(6,2)$ obtained on the bicubic manifold $X$, leading to models with $27$ families and no anti-families. For these bundles equivariance with respect to the $\IZ_3\times\IZ_3$ action on the manifold has been checked at the level of index divisibility for the line bundles in $B$ and $C$. A number of necessary stability checks have been carried out as explained in Section~\ref{secResTTL62}. 

A number of these models have repeated line bundles in $B$ and $C$, and as a result, have an overlapping moduli space with models with lower $r_B$ and $r_C$. These have been left in the list as they are still valid $(r_B,r_C)=(6,2)$ models on the bicubic. These models also indicate that one could in fact complete only a single search over the bicubic with sufficiently large $r_B$ and $r_C$, where $r_B-r_C=4$, to find all $SO(10)$ monad models. In this set up, the lower rank models will appear with repeated line bundles in $B$ and $C$.

In order to save space we use the notation $\cO(k_1,k_2)$ instead of $\cO_X(k_1,k_2)$. All the line bundles are to be understood on the bicubic threefold. 

\begin{center}
\begin{longtable}{||c|c||} 
 \hline
\varstr{14pt}{7pt}$ B$ & $C$\\
 \hline\hline
\varstr{14pt}{7pt} $~~~~~~~~\mathcal{O}$(0,1)$\oplus\mathcal{O}$(0,1)$\oplus\mathcal{O}$(0,1)$\oplus\mathcal{O}$(1,-1)$\oplus\mathcal{O}$(1,1)$\oplus\mathcal{O}$(1,1)~~~~~~~&$~~~~\mathcal{O}$(1,2)$\oplus\mathcal{O}$(2,2)~~~~\\ \hline
\varstr{14pt}{7pt} $\mathcal{O}$(-1,1)$\oplus\mathcal{O}$(0,1)$\oplus\mathcal{O}$(0,1)$\oplus\mathcal{O}$(0,1)$\oplus\mathcal{O}$(2,-1)$\oplus\mathcal{O}$(2,1)&$\mathcal{O}$(1,2)$\oplus\mathcal{O}$(2,2)\\ \hline
\varstr{14pt}{7pt} $\mathcal{O}$(0,1)$\oplus\mathcal{O}$(0,1)$\oplus\mathcal{O}$(0,1)$\oplus\mathcal{O}$(1,-1)$\oplus\mathcal{O}$(1,1)$\oplus\mathcal{O}$(1,2)&$\mathcal{O}$(1,4)$\oplus\mathcal{O}$(2,1)\\ \hline
\varstr{14pt}{7pt} $\mathcal{O}$(0,1)$\oplus\mathcal{O}$(0,1)$\oplus\mathcal{O}$(0,1)$\oplus\mathcal{O}$(1,0)$\oplus\mathcal{O}$(1,0)$\oplus\mathcal{O}$(1,0)&$\mathcal{O}$(1,1)$\oplus\mathcal{O}$(2,2)\\ \hline
\varstr{14pt}{7pt} $\mathcal{O}$(-1,1)$\oplus\mathcal{O}$(-1,1)$\oplus\mathcal{O}$(1,1)$\oplus\mathcal{O}$(1,1)$\oplus\mathcal{O}$(1,1)$\oplus\mathcal{O}$(2,-1)&$\mathcal{O}$(1,2)$\oplus\mathcal{O}$(2,2)\\ \hline
\varstr{14pt}{7pt} $\mathcal{O}$(0,1)$\oplus\mathcal{O}$(0,1)$\oplus\mathcal{O}$(0,1)$\oplus\mathcal{O}$(1,-2)$\oplus\mathcal{O}$(1,2)$\oplus\mathcal{O}$(2,1)&$\mathcal{O}$(2,2)$\oplus\mathcal{O}$(2,2)\\ \hline
\varstr{14pt}{7pt} $\mathcal{O}$(-2,1)$\oplus\mathcal{O}$(1,0)$\oplus\mathcal{O}$(1,0)$\oplus\mathcal{O}$(1,0)$\oplus\mathcal{O}$(1,2)$\oplus\mathcal{O}$(2,1)&$\mathcal{O}$(2,2)$\oplus\mathcal{O}$(2,2)\\ \hline
\varstr{14pt}{7pt} $\mathcal{O}$(0,1)$\oplus\mathcal{O}$(0,1)$\oplus\mathcal{O}$(0,1)$\oplus\mathcal{O}$(0,3)$\oplus\mathcal{O}$(1,-2)$\oplus\mathcal{O}$(1,1)&$\mathcal{O}$(1,1)$\oplus\mathcal{O}$(1,4)\\ \hline
\varstr{14pt}{7pt} $\mathcal{O}$(-2,2)$\oplus\mathcal{O}$(1,0)$\oplus\mathcal{O}$(1,0)$\oplus\mathcal{O}$(1,0)$\oplus\mathcal{O}$(1,1)$\oplus\mathcal{O}$(1,1)&$\mathcal{O}$(1,2)$\oplus\mathcal{O}$(2,2)\\ \hline
\varstr{14pt}{7pt} $\mathcal{O}$(-1,1)$\oplus\mathcal{O}$(-1,2)$\oplus\mathcal{O}$(1,0)$\oplus\mathcal{O}$(1,0)$\oplus\mathcal{O}$(1,0)$\oplus\mathcal{O}$(2,1)&$\mathcal{O}$(1,2)$\oplus\mathcal{O}$(2,2)\\ \hline
\varstr{14pt}{7pt} $\mathcal{O}$(-1,2)$\oplus\mathcal{O}$(0,1)$\oplus\mathcal{O}$(0,1)$\oplus\mathcal{O}$(0,1)$\oplus\mathcal{O}$(2,-2)$\oplus\mathcal{O}$(2,1)&$\mathcal{O}$(1,2)$\oplus\mathcal{O}$(2,2)\\ \hline
\varstr{14pt}{7pt} $\mathcal{O}$(0,1)$\oplus\mathcal{O}$(0,1)$\oplus\mathcal{O}$(0,1)$\oplus\mathcal{O}$(0,3)$\oplus\mathcal{O}$(1,-2)$\oplus\mathcal{O}$(2,1)&$\mathcal{O}$(1,4)$\oplus\mathcal{O}$(2,1)\\ \hline
\varstr{14pt}{7pt} $\mathcal{O}$(0,2)$\oplus\mathcal{O}$(0,2)$\oplus\mathcal{O}$(0,2)$\oplus\mathcal{O}$(1,-2)$\oplus\mathcal{O}$(1,-1)$\oplus\mathcal{O}$(1,1)&$\mathcal{O}$(1,2)$\oplus\mathcal{O}$(2,2)\\ \hline
\varstr{14pt}{7pt} $\mathcal{O}$(-1,2)$\oplus\mathcal{O}$(-1,2)$\oplus\mathcal{O}$(1,-1)$\oplus\mathcal{O}$(1,-1)$\oplus\mathcal{O}$(1,1)$\oplus\mathcal{O}$(2,1)&$\mathcal{O}$(1,2)$\oplus\mathcal{O}$(2,2)\\ \hline
\varstr{14pt}{7pt} $\mathcal{O}$(-2,1)$\oplus\mathcal{O}$(0,3)$\oplus\mathcal{O}$(1,0)$\oplus\mathcal{O}$(1,0)$\oplus\mathcal{O}$(1,0)$\oplus\mathcal{O}$(1,1)&$\mathcal{O}$(1,1)$\oplus\mathcal{O}$(1,4)\\ \hline
\varstr{14pt}{7pt} $\mathcal{O}$(-1,1)$\oplus\mathcal{O}$(-1,1)$\oplus\mathcal{O}$(1,1)$\oplus\mathcal{O}$(1,1)$\oplus\mathcal{O}$(1,1)$\oplus\mathcal{O}$(2,-1)&$\mathcal{O}$(1,2)$\oplus\mathcal{O}$(2,2)\\ \hline
\varstr{14pt}{7pt} $\mathcal{O}$(-1,1)$\oplus\mathcal{O}$(0,1)$\oplus\mathcal{O}$(0,1)$\oplus\mathcal{O}$(0,1)$\oplus\mathcal{O}$(2,-1)$\oplus\mathcal{O}$(2,1)&$\mathcal{O}$(1,2)$\oplus\mathcal{O}$(2,2)\\ \hline
\varstr{14pt}{7pt} $\mathcal{O}$(-1,1)$\oplus\mathcal{O}$(-1,1)$\oplus\mathcal{O}$(-1,1)$\oplus\mathcal{O}$(2,0)$\oplus\mathcal{O}$(2,0)$\oplus\mathcal{O}$(2,0)&$\mathcal{O}$(1,1)$\oplus\mathcal{O}$(2,2)\\ \hline
\end{longtable}
\end{center}

\vspace{-50pt}
\section{Example models with $(r_B,r_C)=(7,3)$ on the bicubic}\label{appbicubic73}
\vspace{-4pt}
The following table contains the monad bundles with ranks $(r_B,r_C)=(7,3)$ obtained on the bicubic manifold. Some of these bundles have a common line bundle in $B$ and $C$ and are, therefore, have an overlapping moduli space with models where $(r_B,r_C)=(6,2)$. A small number of models have two common line bundles in $B$ and $C$ and have an overlapping moduli space with models where $(r_B,r_C)=(5,1)$. These models have been left in this list for the same reasons as discussed in Appendix \ref{appbicubic62}.
\vspace{-14pt}

\begin{center}
\begin{longtable}{||c|c||} 
 \hline
 \varstr{12pt}{7pt}$B$ & $C$\\
 \hline\hline
\varstr{14pt}{7pt}$~~~~\mathcal{O}$(-1,2)$\oplus\mathcal{O}$(0,1)$\oplus\mathcal{O}$(0,1)$\oplus\mathcal{O}$(0,1)$\oplus\mathcal{O}$(1,1)$\oplus\mathcal{O}$(2,-2)$\oplus\mathcal{O}$(2,1)~~~~&~~~~$\mathcal{O}$(1,1)$\oplus\mathcal{O}$(1,2)$\oplus\mathcal{O}$(2,2)~~~~\\ \hline 
\varstr{14pt}{7pt}$\mathcal{O}$(0,1)$\oplus\mathcal{O}$(0,1)$\oplus\mathcal{O}$(0,1)$\oplus\mathcal{O}$(1,-1)$\oplus\mathcal{O}$(1,1)$\oplus\mathcal{O}$(1,1)$\oplus\mathcal{O}$(1,2)&$\mathcal{O}$(1,2)$\oplus\mathcal{O}$(1,2)$\oplus\mathcal{O}$(2,2)\\ \hline 
\varstr{14pt}{7pt}$\mathcal{O}$(0,1)$\oplus\mathcal{O}$(0,1)$\oplus\mathcal{O}$(0,1)$\oplus\mathcal{O}$(1,-1)$\oplus\mathcal{O}$(1,1)$\oplus\mathcal{O}$(1,1)$\oplus\mathcal{O}$(1,1)&$\mathcal{O}$(1,1)$\oplus\mathcal{O}$(1,2)$\oplus\mathcal{O}$(2,2)\\ \hline 
\varstr{14pt}{7pt}$\mathcal{O}$(-2,2)$\oplus\mathcal{O}$(1,0)$\oplus\mathcal{O}$(1,0)$\oplus\mathcal{O}$(1,0)$\oplus\mathcal{O}$(1,1)$\oplus\mathcal{O}$(1,1)$\oplus\mathcal{O}$(2,1)&$\mathcal{O}$(1,2)$\oplus\mathcal{O}$(2,1)$\oplus\mathcal{O}$(2,2)\\ \hline 
\varstr{14pt}{7pt}$\mathcal{O}$(0,1)$\oplus\mathcal{O}$(0,1)$\oplus\mathcal{O}$(0,1)$\oplus\mathcal{O}$(1,0)$\oplus\mathcal{O}$(1,0)$\oplus\mathcal{O}$(1,0)$\oplus\mathcal{O}$(2,5)&$\mathcal{O}$(1,1)$\oplus\mathcal{O}$(2,2)$\oplus\mathcal{O}$(2,5)\\ \hline 
\varstr{14pt}{7pt}$\mathcal{O}$(0,1)$\oplus\mathcal{O}$(0,1)$\oplus\mathcal{O}$(0,1)$\oplus\mathcal{O}$(1,0)$\oplus\mathcal{O}$(1,0)$\oplus\mathcal{O}$(1,0)$\oplus\mathcal{O}$(1,1)&$\mathcal{O}$(1,1)$\oplus\mathcal{O}$(1,1)$\oplus\mathcal{O}$(2,2)\\ \hline 
\varstr{14pt}{7pt}$\mathcal{O}$(-1,1)$\oplus\mathcal{O}$(0,1)$\oplus\mathcal{O}$(0,1)$\oplus\mathcal{O}$(0,1)$\oplus\mathcal{O}$(1,1)$\oplus\mathcal{O}$(2,-1)$\oplus\mathcal{O}$(2,1)&$\mathcal{O}$(1,1)$\oplus\mathcal{O}$(1,2)$\oplus\mathcal{O}$(2,2)\\ \hline 
\varstr{14pt}{7pt}$\mathcal{O}$(-1,1)$\oplus\mathcal{O}$(0,1)$\oplus\mathcal{O}$(0,1)$\oplus\mathcal{O}$(0,1)$\oplus\mathcal{O}$(1,2)$\oplus\mathcal{O}$(2,-1)$\oplus\mathcal{O}$(2,1)&$\mathcal{O}$(1,2)$\oplus\mathcal{O}$(1,2)$\oplus\mathcal{O}$(2,2)\\ \hline 
\varstr{14pt}{7pt}$\mathcal{O}$(-2,1)$\oplus\mathcal{O}$(1,0)$\oplus\mathcal{O}$(1,0)$\oplus\mathcal{O}$(1,0)$\oplus\mathcal{O}$(1,1)$\oplus\mathcal{O}$(1,2)$\oplus\mathcal{O}$(2,1)&$\mathcal{O}$(1,1)$\oplus\mathcal{O}$(2,2)$\oplus\mathcal{O}$(2,2)\\ \hline 
\varstr{14pt}{7pt}$\mathcal{O}$(0,1)$\oplus\mathcal{O}$(0,1)$\oplus\mathcal{O}$(0,1)$\oplus\mathcal{O}$(1,-2)$\oplus\mathcal{O}$(1,1)$\oplus\mathcal{O}$(1,2)$\oplus\mathcal{O}$(2,1)&$\mathcal{O}$(1,1)$\oplus\mathcal{O}$(2,2)$\oplus\mathcal{O}$(2,2)\\ \hline 
\varstr{14pt}{7pt}$\mathcal{O}$(0,2)$\oplus\mathcal{O}$(0,2)$\oplus\mathcal{O}$(0,2)$\oplus\mathcal{O}$(1,-2)$\oplus\mathcal{O}$(1,-1)$\oplus\mathcal{O}$(1,1)$\oplus\mathcal{O}$(1,1)&$\mathcal{O}$(1,1)$\oplus\mathcal{O}$(1,2)$\oplus\mathcal{O}$(2,2)\\ \hline 
\varstr{14pt}{7pt}$\mathcal{O}$(-1,1)$\oplus\mathcal{O}$(-1,2)$\oplus\mathcal{O}$(1,0)$\oplus\mathcal{O}$(1,0)$\oplus\mathcal{O}$(1,0)$\oplus\mathcal{O}$(1,2)$\oplus\mathcal{O}$(2,1)&$\mathcal{O}$(1,2)$\oplus\mathcal{O}$(1,2)$\oplus\mathcal{O}$(2,2)\\ \hline 
\varstr{14pt}{7pt}$\mathcal{O}$(-1,2)$\oplus\mathcal{O}$(-1,2)$\oplus\mathcal{O}$(1,0)$\oplus\mathcal{O}$(1,0)$\oplus\mathcal{O}$(1,0)$\oplus\mathcal{O}$(1,1)$\oplus\mathcal{O}$(1,1)&$\mathcal{O}$(1,2)$\oplus\mathcal{O}$(1,2)$\oplus\mathcal{O}$(1,2)\\ \hline 
\varstr{14pt}{7pt}$\mathcal{O}$(0,2)$\oplus\mathcal{O}$(0,2)$\oplus\mathcal{O}$(0,2)$\oplus\mathcal{O}$(1,-2)$\oplus\mathcal{O}$(1,-1)$\oplus\mathcal{O}$(1,1)$\oplus\mathcal{O}$(1,2)&$\mathcal{O}$(1,2)$\oplus\mathcal{O}$(1,2)$\oplus\mathcal{O}$(2,2)\\ \hline 
\varstr{14pt}{7pt}$\mathcal{O}$(-2,2)$\oplus\mathcal{O}$(1,0)$\oplus\mathcal{O}$(1,0)$\oplus\mathcal{O}$(1,0)$\oplus\mathcal{O}$(1,1)$\oplus\mathcal{O}$(1,1)$\oplus\mathcal{O}$(1,2)&$\mathcal{O}$(1,2)$\oplus\mathcal{O}$(1,2)$\oplus\mathcal{O}$(2,2)\\ \hline 
\varstr{14pt}{7pt}$\mathcal{O}$(-1,1)$\oplus\mathcal{O}$(1,0)$\oplus\mathcal{O}$(1,0)$\oplus\mathcal{O}$(1,0)$\oplus\mathcal{O}$(1,1)$\oplus\mathcal{O}$(1,1)$\oplus\mathcal{O}$(1,2)&$\mathcal{O}$(1,2)$\oplus\mathcal{O}$(2,1)$\oplus\mathcal{O}$(2,2)\\ \hline 
\varstr{14pt}{7pt}$\mathcal{O}$(-2,2)$\oplus\mathcal{O}$(-1,1)$\oplus\mathcal{O}$(1,1)$\oplus\mathcal{O}$(1,2)$\oplus\mathcal{O}$(2,-1)$\oplus\mathcal{O}$(2,-1)$\oplus\mathcal{O}$(2,1)&$\mathcal{O}$(1,2)$\oplus\mathcal{O}$(2,1)$\oplus\mathcal{O}$(2,2)\\ \hline 
\varstr{14pt}{7pt}$\mathcal{O}$(0,1)$\oplus\mathcal{O}$(0,1)$\oplus\mathcal{O}$(0,1)$\oplus\mathcal{O}$(1,0)$\oplus\mathcal{O}$(1,0)$\oplus\mathcal{O}$(1,0)$\oplus\mathcal{O}$(1,2)&$\mathcal{O}$(1,1)$\oplus\mathcal{O}$(1,2)$\oplus\mathcal{O}$(2,2)\\ \hline 
\varstr{14pt}{7pt}$\mathcal{O}$(-1,2)$\oplus\mathcal{O}$(-1,2)$\oplus\mathcal{O}$(1,-1)$\oplus\mathcal{O}$(1,-1)$\oplus\mathcal{O}$(1,1)$\oplus\mathcal{O}$(1,2)$\oplus\mathcal{O}$(2,1)&$\mathcal{O}$(1,2)$\oplus\mathcal{O}$(1,2)$\oplus\mathcal{O}$(2,2)\\ \hline 
\varstr{14pt}{7pt}$\mathcal{O}$(-1,1)$\oplus\mathcal{O}$(-1,1)$\oplus\mathcal{O}$(1,1)$\oplus\mathcal{O}$(1,1)$\oplus\mathcal{O}$(1,1)$\oplus\mathcal{O}$(2,-1)$\oplus\mathcal{O}$(2,1)&$\mathcal{O}$(1,2)$\oplus\mathcal{O}$(2,1)$\oplus\mathcal{O}$(2,2)\\ \hline 
\varstr{14pt}{7pt}$\mathcal{O}$(-1,2)$\oplus\mathcal{O}$(-1,2)$\oplus\mathcal{O}$(1,-1)$\oplus\mathcal{O}$(1,-1)$\oplus\mathcal{O}$(1,1)$\oplus\mathcal{O}$(1,1)$\oplus\mathcal{O}$(2,1)&$\mathcal{O}$(1,1)$\oplus\mathcal{O}$(1,2)$\oplus\mathcal{O}$(2,2)\\ \hline 
\varstr{14pt}{7pt}$\mathcal{O}$(0,2)$\oplus\mathcal{O}$(0,2)$\oplus\mathcal{O}$(0,2)$\oplus\mathcal{O}$(1,-2)$\oplus\mathcal{O}$(1,-1)$\oplus\mathcal{O}$(1,1)$\oplus\mathcal{O}$(2,2)&$\mathcal{O}$(1,2)$\oplus\mathcal{O}$(2,2)$\oplus\mathcal{O}$(2,2)\\ \hline 
\varstr{14pt}{7pt}$\mathcal{O}$(-1,1)$\oplus\mathcal{O}$(1,0)$\oplus\mathcal{O}$(1,0)$\oplus\mathcal{O}$(1,0)$\oplus\mathcal{O}$(1,1)$\oplus\mathcal{O}$(1,1)$\oplus\mathcal{O}$(1,4)&$\mathcal{O}$(1,4)$\oplus\mathcal{O}$(2,1)$\oplus\mathcal{O}$(2,2)\\ \hline 
\varstr{14pt}{7pt}$\mathcal{O}$(0,1)$\oplus\mathcal{O}$(0,1)$\oplus\mathcal{O}$(0,1)$\oplus\mathcal{O}$(1,0)$\oplus\mathcal{O}$(1,0)$\oplus\mathcal{O}$(1,0)$\oplus\mathcal{O}$(2,4)&$\mathcal{O}$(1,1)$\oplus\mathcal{O}$(2,2)$\oplus\mathcal{O}$(2,4)\\ \hline 
\varstr{14pt}{7pt}$\mathcal{O}$(0,1)$\oplus\mathcal{O}$(0,1)$\oplus\mathcal{O}$(0,1)$\oplus\mathcal{O}$(1,-2)$\oplus\mathcal{O}$(1,2)$\oplus\mathcal{O}$(1,2)$\oplus\mathcal{O}$(2,1)&$\mathcal{O}$(1,2)$\oplus\mathcal{O}$(2,2)$\oplus\mathcal{O}$(2,2)\\ \hline 
\varstr{14pt}{7pt}$\mathcal{O}$(-1,2)$\oplus\mathcal{O}$(-1,2)$\oplus\mathcal{O}$(1,0)$\oplus\mathcal{O}$(1,0)$\oplus\mathcal{O}$(1,0)$\oplus\mathcal{O}$(1,2)$\oplus\mathcal{O}$(2,-1)&$\mathcal{O}$(1,2)$\oplus\mathcal{O}$(1,2)$\oplus\mathcal{O}$(2,1)\\ \hline 
\varstr{14pt}{7pt}$\mathcal{O}$(-2,1)$\oplus\mathcal{O}$(1,0)$\oplus\mathcal{O}$(1,0)$\oplus\mathcal{O}$(1,0)$\oplus\mathcal{O}$(1,2)$\oplus\mathcal{O}$(2,1)$\oplus\mathcal{O}$(2,2)&$\mathcal{O}$(2,2)$\oplus\mathcal{O}$(2,2)$\oplus\mathcal{O}$(2,2)\\ \hline 
\varstr{14pt}{7pt}$\mathcal{O}$(-1,1)$\oplus\mathcal{O}$(-1,1)$\oplus\mathcal{O}$(1,1)$\oplus\mathcal{O}$(1,1)$\oplus\mathcal{O}$(1,1)$\oplus\mathcal{O}$(2,-1)$\oplus\mathcal{O}$(2,2)&$\mathcal{O}$(1,2)$\oplus\mathcal{O}$(2,2)$\oplus\mathcal{O}$(2,2)\\ \hline 
\varstr{14pt}{7pt}$\mathcal{O}$(-2,2)$\oplus\mathcal{O}$(1,0)$\oplus\mathcal{O}$(1,0)$\oplus\mathcal{O}$(1,0)$\oplus\mathcal{O}$(1,1)$\oplus\mathcal{O}$(1,1)$\oplus\mathcal{O}$(1,2)&$\mathcal{O}$(1,1)$\oplus\mathcal{O}$(1,4)$\oplus\mathcal{O}$(2,1)\\ \hline 
\varstr{14pt}{7pt}$\mathcal{O}$(-1,1)$\oplus\mathcal{O}$(-1,2)$\oplus\mathcal{O}$(1,0)$\oplus\mathcal{O}$(1,0)$\oplus\mathcal{O}$(1,0)$\oplus\mathcal{O}$(1,1)$\oplus\mathcal{O}$(2,1)&$\mathcal{O}$(1,1)$\oplus\mathcal{O}$(1,2)$\oplus\mathcal{O}$(2,2)\\ \hline 
\varstr{14pt}{7pt}$\mathcal{O}$(0,1)$\oplus\mathcal{O}$(0,1)$\oplus\mathcal{O}$(0,1)$\oplus\mathcal{O}$(1,0)$\oplus\mathcal{O}$(1,0)$\oplus\mathcal{O}$(1,0)$\oplus\mathcal{O}$(1,4)&$\mathcal{O}$(1,1)$\oplus\mathcal{O}$(1,4)$\oplus\mathcal{O}$(2,2)\\ \hline 
\varstr{14pt}{7pt}$\mathcal{O}$(-1,2)$\oplus\mathcal{O}$(1,-1)$\oplus\mathcal{O}$(1,0)$\oplus\mathcal{O}$(1,0)$\oplus\mathcal{O}$(1,0)$\oplus\mathcal{O}$(1,2)$\oplus\mathcal{O}$(1,2)&$\mathcal{O}$(1,2)$\oplus\mathcal{O}$(2,1)$\oplus\mathcal{O}$(2,2)\\ \hline 
\varstr{14pt}{7pt}$\mathcal{O}$(-1,2)$\oplus\mathcal{O}$(1,-1)$\oplus\mathcal{O}$(1,-1)$\oplus\mathcal{O}$(1,1)$\oplus\mathcal{O}$(1,1)$\oplus\mathcal{O}$(1,1)$\oplus\mathcal{O}$(1,2)&$\mathcal{O}$(1,2)$\oplus\mathcal{O}$(2,1)$\oplus\mathcal{O}$(2,2)\\ \hline 
\varstr{14pt}{7pt}$\mathcal{O}$(0,1)$\oplus\mathcal{O}$(0,1)$\oplus\mathcal{O}$(0,1)$\oplus\mathcal{O}$(1,-1)$\oplus\mathcal{O}$(1,1)$\oplus\mathcal{O}$(1,1)$\oplus\mathcal{O}$(2,2)&$\mathcal{O}$(1,2)$\oplus\mathcal{O}$(2,2)$\oplus\mathcal{O}$(2,2)\\ \hline 
\varstr{14pt}{7pt}$\mathcal{O}$(-2,2)$\oplus\mathcal{O}$(1,0)$\oplus\mathcal{O}$(1,0)$\oplus\mathcal{O}$(1,0)$\oplus\mathcal{O}$(1,1)$\oplus\mathcal{O}$(1,1)$\oplus\mathcal{O}$(1,1)&$\mathcal{O}$(1,1)$\oplus\mathcal{O}$(1,2)$\oplus\mathcal{O}$(2,2)\\ \hline 
\varstr{14pt}{7pt}$\mathcal{O}$(0,1)$\oplus\mathcal{O}$(0,1)$\oplus\mathcal{O}$(0,1)$\oplus\mathcal{O}$(1,-1)$\oplus\mathcal{O}$(1,1)$\oplus\mathcal{O}$(1,1)$\oplus\mathcal{O}$(2,1)&$\mathcal{O}$(1,2)$\oplus\mathcal{O}$(2,1)$\oplus\mathcal{O}$(2,2)\\ \hline 
\varstr{14pt}{7pt}$\mathcal{O}$(-1,2)$\oplus\mathcal{O}$(-1,2)$\oplus\mathcal{O}$(1,-1)$\oplus\mathcal{O}$(1,-1)$\oplus\mathcal{O}$(1,1)$\oplus\mathcal{O}$(2,1)$\oplus\mathcal{O}$(2,2)&$\mathcal{O}$(1,2)$\oplus\mathcal{O}$(2,2)$\oplus\mathcal{O}$(2,2)\\ \hline 
\varstr{14pt}{7pt}$\mathcal{O}$(-2,1)$\oplus\mathcal{O}$(-1,1)$\oplus\mathcal{O}$(1,1)$\oplus\mathcal{O}$(1,2)$\oplus\mathcal{O}$(2,0)$\oplus\mathcal{O}$(2,0)$\oplus\mathcal{O}$(2,0)&$\mathcal{O}$(1,2)$\oplus\mathcal{O}$(2,1)$\oplus\mathcal{O}$(2,2)\\ \hline 
\varstr{14pt}{7pt}$\mathcal{O}$(0,1)$\oplus\mathcal{O}$(0,1)$\oplus\mathcal{O}$(0,1)$\oplus\mathcal{O}$(1,1)$\oplus\mathcal{O}$(1,1)$\oplus\mathcal{O}$(1,2)$\oplus\mathcal{O}$(2,-2)&$\mathcal{O}$(1,2)$\oplus\mathcal{O}$(2,1)$\oplus\mathcal{O}$(2,2)\\ \hline 
\varstr{14pt}{7pt}$\mathcal{O}$(-1,1)$\oplus\mathcal{O}$(-1,1)$\oplus\mathcal{O}$(1,1)$\oplus\mathcal{O}$(1,1)$\oplus\mathcal{O}$(1,1)$\oplus\mathcal{O}$(1,1)$\oplus\mathcal{O}$(2,-1)&$\mathcal{O}$(1,1)$\oplus\mathcal{O}$(1,2)$\oplus\mathcal{O}$(2,2)\\ \hline 
\varstr{14pt}{7pt}$\mathcal{O}$(-1,1)$\oplus\mathcal{O}$(-1,1)$\oplus\mathcal{O}$(1,1)$\oplus\mathcal{O}$(1,1)$\oplus\mathcal{O}$(1,1)$\oplus\mathcal{O}$(1,2)$\oplus\mathcal{O}$(2,-1)&$\mathcal{O}$(1,2)$\oplus\mathcal{O}$(1,2)$\oplus\mathcal{O}$(2,2)\\ \hline 
\varstr{14pt}{7pt} $\mathcal{O}$(0,1)$\oplus\mathcal{O}$(0,1)$\oplus\mathcal{O}$(0,1)$\oplus\mathcal{O}$(1,-1)$\oplus\mathcal{O}$(1,2)$\oplus\mathcal{O}$(1,2)$\oplus\mathcal{O}$(2,-1)&$\mathcal{O}$(1,2)$\oplus\mathcal{O}$(2,1)$\oplus\mathcal{O}$(2,2)\\ \hline 
\varstr{14pt}{7pt}$\mathcal{O}$(-1,2)$\oplus\mathcal{O}$(-1,2)$\oplus\mathcal{O}$(1,-1)$\oplus\mathcal{O}$(1,1)$\oplus\mathcal{O}$(1,1)$\oplus\mathcal{O}$(1,1)$\oplus\mathcal{O}$(2,-1)&$\mathcal{O}$(1,2)$\oplus\mathcal{O}$(1,2)$\oplus\mathcal{O}$(2,1)\\ \hline 
\varstr{14pt}{7pt}$\mathcal{O}$(-1,2)$\oplus\mathcal{O}$(0,1)$\oplus\mathcal{O}$(0,1)$\oplus\mathcal{O}$(0,1)$\oplus\mathcal{O}$(1,2)$\oplus\mathcal{O}$(2,-2)$\oplus\mathcal{O}$(2,1)&$\mathcal{O}$(1,2)$\oplus\mathcal{O}$(1,2)$\oplus\mathcal{O}$(2,2)\\ \hline 
\varstr{14pt}{7pt}$\mathcal{O}$(-2,2)$\oplus\mathcal{O}$(-1,2)$\oplus\mathcal{O}$(1,1)$\oplus\mathcal{O}$(1,1)$\oplus\mathcal{O}$(1,1)$\oplus\mathcal{O}$(2,-1)$\oplus\mathcal{O}$(2,-1)&$\mathcal{O}$(1,2)$\oplus\mathcal{O}$(1,2)$\oplus\mathcal{O}$(2,1)\\ \hline
\varstr{14pt}{7pt}$\mathcal{O}$(-2,1)$\oplus\mathcal{O}$(1,0)$\oplus\mathcal{O}$(1,0)$\oplus\mathcal{O}$(1,0)$\oplus\mathcal{O}$(1,2)$\oplus\mathcal{O}$(1,2)$\oplus\mathcal{O}$(2,1)&$\mathcal{O}$(1,2)$\oplus\mathcal{O}$(2,2)$\oplus\mathcal{O}$(2,2)\\ \hline 
\varstr{14pt}{7pt}$\mathcal{O}$(-1,2)$\oplus\mathcal{O}$(0,1)$\oplus\mathcal{O}$(0,1)$\oplus\mathcal{O}$(0,1)$\oplus\mathcal{O}$(1,1)$\oplus\mathcal{O}$(2,-2)$\oplus\mathcal{O}$(2,1)&$\mathcal{O}$(1,1)$\oplus\mathcal{O}$(1,2)$\oplus\mathcal{O}$(2,2)\\ \hline 
\varstr{14pt}{7pt}$\mathcal{O}$(-2,2)$\oplus\mathcal{O}$(1,0)$\oplus\mathcal{O}$(1,0)$\oplus\mathcal{O}$(1,0)$\oplus\mathcal{O}$(1,1)$\oplus\mathcal{O}$(1,1)$\oplus\mathcal{O}$(2,2)&$\mathcal{O}$(1,2)$\oplus\mathcal{O}$(2,2)$\oplus\mathcal{O}$(2,2)\\ \hline 
\varstr{14pt}{7pt}$\mathcal{O}$(0,2)$\oplus\mathcal{O}$(0,2)$\oplus\mathcal{O}$(0,2)$\oplus\mathcal{O}$(1,-2)$\oplus\mathcal{O}$(1,0)$\oplus\mathcal{O}$(1,0)$\oplus\mathcal{O}$(1,0)&$\mathcal{O}$(1,1)$\oplus\mathcal{O}$(1,1)$\oplus\mathcal{O}$(2,2)\\ \hline 
\varstr{14pt}{7pt}$\mathcal{O}$(0,1)$\oplus\mathcal{O}$(0,1)$\oplus\mathcal{O}$(0,1)$\oplus\mathcal{O}$(1,-1)$\oplus\mathcal{O}$(1,1)$\oplus\mathcal{O}$(1,2)$\oplus\mathcal{O}$(1,2)&$\mathcal{O}$(1,2)$\oplus\mathcal{O}$(1,4)$\oplus\mathcal{O}$(2,1)\\ \hline 
\varstr{14pt}{7pt}$\mathcal{O}$(0,2)$\oplus\mathcal{O}$(0,2)$\oplus\mathcal{O}$(0,2)$\oplus\mathcal{O}$(1,-1)$\oplus\mathcal{O}$(1,0)$\oplus\mathcal{O}$(1,0)$\oplus\mathcal{O}$(1,0)&$\mathcal{O}$(1,2)$\oplus\mathcal{O}$(1,2)$\oplus\mathcal{O}$(2,1)\\ \hline 
\varstr{14pt}{7pt}$\mathcal{O}$(0,1)$\oplus\mathcal{O}$(0,1)$\oplus\mathcal{O}$(0,1)$\oplus\mathcal{O}$(1,-1)$\oplus\mathcal{O}$(1,1)$\oplus\mathcal{O}$(1,2)$\oplus\mathcal{O}$(3,3)&$\mathcal{O}$(1,4)$\oplus\mathcal{O}$(2,1)$\oplus\mathcal{O}$(3,3)\\ \hline 
\varstr{14pt}{7pt}$\mathcal{O}$(-2,1)$\oplus\mathcal{O}$(0,3)$\oplus\mathcal{O}$(1,0)$\oplus\mathcal{O}$(1,0)$\oplus\mathcal{O}$(1,0)$\oplus\mathcal{O}$(1,1)$\oplus\mathcal{O}$(1,1)&$\mathcal{O}$(1,1)$\oplus\mathcal{O}$(1,1)$\oplus\mathcal{O}$(1,4)\\ \hline 
\varstr{14pt}{7pt}$\mathcal{O}$(-1,2)$\oplus\mathcal{O}$(-1,2)$\oplus\mathcal{O}$(1,-1)$\oplus\mathcal{O}$(1,0)$\oplus\mathcal{O}$(1,0)$\oplus\mathcal{O}$(1,0)$\oplus\mathcal{O}$(2,1)&$\mathcal{O}$(1,1)$\oplus\mathcal{O}$(1,1)$\oplus\mathcal{O}$(2,2)\\ \hline 
\varstr{14pt}{7pt}$\mathcal{O}$(-1,1)$\oplus\mathcal{O}$(-1,2)$\oplus\mathcal{O}$(1,0)$\oplus\mathcal{O}$(1,0)$\oplus\mathcal{O}$(1,0)$\oplus\mathcal{O}$(2,1)$\oplus\mathcal{O}$(2,1)&$\mathcal{O}$(1,2)$\oplus\mathcal{O}$(2,1)$\oplus\mathcal{O}$(2,2)\\ \hline 
\varstr{14pt}{7pt}$\mathcal{O}$(0,1)$\oplus\mathcal{O}$(0,1)$\oplus\mathcal{O}$(0,1)$\oplus\mathcal{O}$(1,0)$\oplus\mathcal{O}$(1,0)$\oplus\mathcal{O}$(1,0)$\oplus\mathcal{O}$(1,4)&$\mathcal{O}$(1,1)$\oplus\mathcal{O}$(1,5)$\oplus\mathcal{O}$(2,1)\\ \hline 
\varstr{14pt}{7pt}$\mathcal{O}$(0,1)$\oplus\mathcal{O}$(0,1)$\oplus\mathcal{O}$(0,1)$\oplus\mathcal{O}$(1,-2)$\oplus\mathcal{O}$(1,2)$\oplus\mathcal{O}$(2,1)$\oplus\mathcal{O}$(2,2)&$\mathcal{O}$(2,2)$\oplus\mathcal{O}$(2,2)$\oplus\mathcal{O}$(2,2)\\ \hline 
\varstr{14pt}{7pt}$\mathcal{O}$(-1,1)$\oplus\mathcal{O}$(-1,2)$\oplus\mathcal{O}$(1,0)$\oplus\mathcal{O}$(1,0)$\oplus\mathcal{O}$(1,0)$\oplus\mathcal{O}$(2,1)$\oplus\mathcal{O}$(2,2)&$\mathcal{O}$(1,2)$\oplus\mathcal{O}$(2,2)$\oplus\mathcal{O}$(2,2)\\ \hline 
\varstr{14pt}{7pt}$\mathcal{O}$(0,1)$\oplus\mathcal{O}$(0,1)$\oplus\mathcal{O}$(0,1)$\oplus\mathcal{O}$(1,0)$\oplus\mathcal{O}$(1,0)$\oplus\mathcal{O}$(1,0)$\oplus\mathcal{O}$(2,2)&$\mathcal{O}$(1,1)$\oplus\mathcal{O}$(2,2)$\oplus\mathcal{O}$(2,2)\\ \hline 
\varstr{14pt}{7pt}$\mathcal{O}$(-1,2)$\oplus\mathcal{O}$(-1,2)$\oplus\mathcal{O}$(1,0)$\oplus\mathcal{O}$(1,0)$\oplus\mathcal{O}$(1,0)$\oplus\mathcal{O}$(1,1)$\oplus\mathcal{O}$(2,-1)&$\mathcal{O}$(1,1)$\oplus\mathcal{O}$(1,2)$\oplus\mathcal{O}$(2,1)\\ \hline 
\varstr{14pt}{7pt}$\mathcal{O}$(-2,2)$\oplus\mathcal{O}$(-1,1)$\oplus\mathcal{O}$(1,1)$\oplus\mathcal{O}$(1,1)$\oplus\mathcal{O}$(1,1)$\oplus\mathcal{O}$(2,-1)$\oplus\mathcal{O}$(2,1)&$\mathcal{O}$(1,2)$\oplus\mathcal{O}$(1,2)$\oplus\mathcal{O}$(2,2)\\ \hline 
\varstr{14pt}{7pt}$\mathcal{O}$(0,1)$\oplus\mathcal{O}$(0,1)$\oplus\mathcal{O}$(0,1)$\oplus\mathcal{O}$(1,0)$\oplus\mathcal{O}$(1,0)$\oplus\mathcal{O}$(1,0)$\oplus\mathcal{O}$(3,3)&$\mathcal{O}$(1,1)$\oplus\mathcal{O}$(2,2)$\oplus\mathcal{O}$(3,3)\\ \hline 
\varstr{14pt}{7pt}$\mathcal{O}$(0,1)$\oplus\mathcal{O}$(0,1)$\oplus\mathcal{O}$(0,1)$\oplus\mathcal{O}$(1,-1)$\oplus\mathcal{O}$(1,1)$\oplus\mathcal{O}$(1,1)$\oplus\mathcal{O}$(1,2)&$\mathcal{O}$(1,1)$\oplus\mathcal{O}$(1,4)$\oplus\mathcal{O}$(2,1)\\ \hline 
\varstr{14pt}{7pt}$\mathcal{O}$(-1,2)$\oplus\mathcal{O}$(0,1)$\oplus\mathcal{O}$(0,1)$\oplus\mathcal{O}$(0,1)$\oplus\mathcal{O}$(1,2)$\oplus\mathcal{O}$(2,-1)$\oplus\mathcal{O}$(2,-1)&$\mathcal{O}$(1,2)$\oplus\mathcal{O}$(1,2)$\oplus\mathcal{O}$(2,1)\\ \hline 
\varstr{14pt}{7pt}$\mathcal{O}$(-1,1)$\oplus\mathcal{O}$(0,1)$\oplus\mathcal{O}$(0,1)$\oplus\mathcal{O}$(0,1)$\oplus\mathcal{O}$(2,-1)$\oplus\mathcal{O}$(2,1)$\oplus\mathcal{O}$(2,1)&$\mathcal{O}$(1,2)$\oplus\mathcal{O}$(2,1)$\oplus\mathcal{O}$(2,2)\\ \hline 
\varstr{14pt}{7pt}$\mathcal{O}$(0,1)$\oplus\mathcal{O}$(0,1)$\oplus\mathcal{O}$(0,1)$\oplus\mathcal{O}$(1,-1)$\oplus\mathcal{O}$(1,1)$\oplus\mathcal{O}$(1,2)$\oplus\mathcal{O}$(2,2)&$\mathcal{O}$(1,4)$\oplus\mathcal{O}$(2,1)$\oplus\mathcal{O}$(2,2)\\ \hline 
\varstr{14pt}{7pt}$\mathcal{O}$(-1,1)$\oplus\mathcal{O}$(0,1)$\oplus\mathcal{O}$(0,1)$\oplus\mathcal{O}$(0,1)$\oplus\mathcal{O}$(2,-1)$\oplus\mathcal{O}$(2,1)$\oplus\mathcal{O}$(2,2)&$\mathcal{O}$(1,2)$\oplus\mathcal{O}$(2,2)$\oplus\mathcal{O}$(2,2)\\ \hline 
\varstr{14pt}{7pt}$\mathcal{O}$(0,2)$\oplus\mathcal{O}$(0,2)$\oplus\mathcal{O}$(0,2)$\oplus\mathcal{O}$(1,-2)$\oplus\mathcal{O}$(1,-1)$\oplus\mathcal{O}$(1,1)$\oplus\mathcal{O}$(2,1)&$\mathcal{O}$(1,2)$\oplus\mathcal{O}$(2,1)$\oplus\mathcal{O}$(2,2)\\ \hline 
\varstr{14pt}{7pt}$\mathcal{O}$(-1,1)$\oplus\mathcal{O}$(-1,1)$\oplus\mathcal{O}$(1,1)$\oplus\mathcal{O}$(1,1)$\oplus\mathcal{O}$(1,1)$\oplus\mathcal{O}$(1,2)$\oplus\mathcal{O}$(2,-1)&$\mathcal{O}$(1,1)$\oplus\mathcal{O}$(1,4)$\oplus\mathcal{O}$(2,1) \\ \hline
\end{longtable}
\end{center}

\vspace{-41pt}
\section{Example models on the triple tri-linear CY}\label{apptriptri}
\vspace{-12pt}
The  table contains $7$ heterotic monad bundles on the triple trilinear manifold. These have been obtained from a sample of $100$ bundles randomly selected out of the total of 7812 bundles found during the RL scan, after imposing a limited number of computationally intense stability checks. It so happens that all of these models have shared line bundles in $B$ and $C$. These models have been left in this list for the same reasons as discussed in Appendix \ref{appbicubic62}. The full data set of 7812 monad bundles has been included in the arXiv submission in an auxiliary file. 
\vspace{-14pt}

\begin{center}
\begin{longtable}{||c|c||} 
 \hline
\varstr{12pt}{7pt} $B$ & $C$\\
 \hline\hline
 \varstr{14pt}{7pt}$~~~~\mathcal{O}$(-1,1,0)$\oplus\mathcal{O}$(0,-1,1)$\oplus\mathcal{O}$(0,-1,1)$\oplus\mathcal{O}$(1,1,0)$\oplus\mathcal{O}$(1,1,0)$\oplus\mathcal{O}$(1,3,1)~~~~&$~~~~\mathcal{O}$(1,1,2)$\oplus\mathcal{O}$(1,3,1)~~~~\\ \hline 
\varstr{14pt}{7pt}$\mathcal{O}$(-1,0,2)$\oplus\mathcal{O}$(0,1,-1)$\oplus\mathcal{O}$(0,1,2)$\oplus\mathcal{O}$(1,-1,0)$\oplus\mathcal{O}$(1,0,-1)$\oplus\mathcal{O}$(1,1,2)&$\mathcal{O}$(1,1,2)$\oplus\mathcal{O}$(1,1,2)\\ \hline 
\varstr{14pt}{7pt}$\mathcal{O}$(-1,0,2)$\oplus\mathcal{O}$(-1,1,0)$\oplus\mathcal{O}$(1,-1,1)$\oplus\mathcal{O}$(1,0,2)$\oplus\mathcal{O}$(1,1,-3)$\oplus\mathcal{O}$(1,2,2)&$\mathcal{O}$(1,1,2)$\oplus\mathcal{O}$(1,2,2)\\ \hline 
\varstr{14pt}{7pt}$\mathcal{O}$(-1,0,1)$\oplus\mathcal{O}$(0,-2,1)$\oplus\mathcal{O}$(0,2,1)$\oplus\mathcal{O}$(1,1,-1)$\oplus\mathcal{O}$(1,1,-1)$\oplus\mathcal{O}$(1,2,1)&$\mathcal{O}$(1,2,1)$\oplus\mathcal{O}$(1,2,1)\\ \hline 
\varstr{14pt}{7pt}$\mathcal{O}$(-1,1,1)$\oplus\mathcal{O}$(-1,1,1)$\oplus\mathcal{O}$(1,-2,0)$\oplus\mathcal{O}$(1,0,-1)$\oplus\mathcal{O}$(1,2,0)$\oplus\mathcal{O}$(1,2,1)&$\mathcal{O}$(1,2,1)$\oplus\mathcal{O}$(1,2,1)\\ \hline 
\varstr{14pt}{7pt}$\mathcal{O}$(-1,1,1)$\oplus\mathcal{O}$(0,-1,2)$\oplus\mathcal{O}$(0,1,-1)$\oplus\mathcal{O}$(1,0,-2)$\oplus\mathcal{O}$(1,0,2)$\oplus\mathcal{O}$(1,2,4)&$\mathcal{O}$(1,1,2)$\oplus\mathcal{O}$(1,2,4)\\ \hline 
\varstr{14pt}{7pt}$\mathcal{O}$(-1,0,2)$\oplus\mathcal{O}$(-1,0,2)$\oplus\mathcal{O}$(1,0,-1)$\oplus\mathcal{O}$(1,0,2)$\oplus\mathcal{O}$(1,1,-3)$\oplus\mathcal{O}$(1,1,2)&$\mathcal{O}$(1,1,2)$\oplus\mathcal{O}$(1,1,2)\\
\hline
\end{longtable}
\end{center}

\bibliographystyle{utcaps}
\providecommand{\href}[2]{#2}\begingroup\raggedright\endgroup

\end{document}